\documentclass{ab}

\usepackage{graphicx}
\usepackage{latexsym}

\def\degr{\hbox{$^\circ$}}

\begin{document}

\title{
Study of the Power Beam Pattern of RATAN-600 \\
During the Deep RZF Survey  \\
(1998--2003)}

\author{E.K.~Majorova$^{1}$,  N.N.~Bursov$^{1}$}
\institute{Special Astrophysical Observatory Russian Academy of Sciences,
N. Arkhyz, KChR, 369167, Russia}

\offprints{E.K.Majorova, \email{len@sao.ru}}
 
\titlerunning{Study of the Power Beam Pattern of RATAN-600}

\authorrunning{Majorova et al.}
 
%\received{July 10, 2007}%
\date{Received: April 4, 2007/Revised:April  10, 2007}
%\revised{August 17, 2007}%

\abstract{
This paper proposes a method for constructing an experimental
power beam pattern (PB) of RATAN-600 based on the sample of NVSS
sources observed in the process of a deep sky survey near local
zenith. The data obtained from observations of radio sources at
$\lambda7.6$ cm in nine bands of the survey  (the 2002 and 2003
sets) are used to construct vertical PB of the telescope at
rather large offsets from the central horizontal section of the
PB ($\pm36'$). The experimental PBs obtained using different
methods are compared and the root-mean-square deviations of the
experimental PB from the corresponding computed PB are
determined. The stability of the power beam pattern in its central
part ($\pm6'$) during the RATAN-600 Zenith Field (RZF) survey (1998--2003)
and the accuracies of the fluxes of the sources observed within
the framework of this survey and included into the
RZF catalog are estimated [\cite{p1:Majorova_n}].
}
\maketitle

\section{INTRODUCTION}

This paper continues a series  of works dedicated to the study of
the power beam pattern of the radio telescope RATAN-600. The PB was analyzed numerically
[\cite{e1:Majorova_n,e2:Majorova_n,e3:Majorova_n,e4:Majorova_n,e5:Majorova_n,g1:Majorova_n,k1:Majorova_n,mm1:Majorova_n}],
using methods of optical modeling
[\cite{g2:Majorova_n,e6:Majorova_n,k2:Majorova_n}], and also by
observing bright cosmic sources
[\cite{t1:Majorova_n,t2:Majorova_n,mm2:Majorova_n}]. These studies
give a comprehensive understanding of the PB
in terms of intensity and polarization. The final reconciliation of the
computed and observed PBs was carried out in our
 papers [\cite{mm1:Majorova_n,mm2:Majorova_n}]. We showed
that the additional allowance for diffraction effects in the
aperture of the primary mirror and for the finite vertical size
of its reflective elements [\cite{mm1:Majorova_n}] brings about a
good agreement between the computed and measured PB within a large
solid angle down to a level of  0.5\% of the maximum beam
intensity [\cite{mm2:Majorova_n}]. We perform numerical simulations
to study the effect of the error distribution in the settings of
the reflective elements of the primary mirror on the structure of
the PB in distant sections. These computations were
experimentally corroborated in a series of observations of
reference sources in 2001 and 2002 [\cite{mm2:Majorova_n}].

The power beam pattern of RATAN-600 requires such a detailed study
because of its complex structure and because the shape of the
PB depends significantly on the elevation of the
source observed. In observations performed with a single sector
of the antenna  the PB varies from the
knife-edge beam (in observations at the horizon) to a virtually
pencil beam --- in observations of
near-zenith sources. Near the local zenith the power beam pattern has a narrow
 beam whose half-power beam width (HPBW) is determined by the horizontal
aperture of the sector, and   extended scattered background
whose intensity decreases as  1/y in the vertical plane. The
extent of this background in the vertical plane depends on the
vertical size of the reflective panel and may be as large as 100
HPBW. This scattered background is concentrated in the sky in two
sectors located symmetrically with respect to the central section
with the summits at the maximum of the PB and with angular sizes
of $90^{o}$. This ``fan'' of  scattered background of the PB
gradually ``shrinks'' (the angular size of the sectors decreases)
with decreasing elevation at which the antenna is focused (the
elevation $H$ of the source observed) and at low elevations it
transforms into a narrow structure extended in the vertical
direction --- the  PB transform into knife-edge beam.

Temirova~[\cite{t1:Majorova_n,t2:Majorova_n}] and Majorova and
Trushkin~[\cite{mm2:Majorova_n}] studied experimentally the PB of
RATAN-600 using the radio-astronomical method described by
Kuz'min and Solomonovich~[\cite{ks:Majorova_n}]. This method
consists in measuring the antenna temperatures of a cosmic source
as it passes through different sections of the PB. The so-called
``point''\, sources are measured whose angular sizes are much
smaller than the width of the PB. In this case, the response of
the antenna at the output of the radiometer coincides with the PB
in the given horizontal direction up to a constant factor. During
observations the transit curves of the source are recorded while passing
through various horizontal sections of the PB shifted by $dH$ in
declination with respect to the central section. The PB is
normalized to the maximum of the antenna temperature of the
source in the central section ($dH=0$). The dependences of the
normalized antenna temperature of the source on  $dH$ give the
vertical power  patterns  $F_{v}$ of the radiotelescope.

This method allows the two-dimensional PB to be measured with
high accuracy. However, to measure the power  pattern within a
sufficiently large solid angle, one has to observe many transits
of a bright cosmic source across a fixed PB. The time required to
perform such a study increases with the degree of detail the
power pattern is to be measured. Therefore such measurements
of the PB cannot be made too often. At the same time, the deep
sky surveys performed with RATAN-600 over many years require good
knowledge of the PB within large solid angles. The more sensitive
is the survey, the more reliable must be the information about
the state of the  beam pattern the time the observations
are performed. Good knowledge of the shape of the power beam pattern
 in distant sections and of the vertical PB of the
 telescope are needed to determine the fluxes from the
sources and estimate the declinations of the latter. In addition,
information on the stability of the PB in time is of great
importance.

Note that the fluxes of individual sources crossing the central
section of the PB are determined using the  ``reference''\,
sources with well-known fluxes and with declinations close to
that of the source studied. The information on the central
section of the PB is quite sufficient when performing
observations in this mode. When performing deep sky surveys one
has to know the PB at least within the declination band of the
survey.

The power beam pattern can be computed using the available codes
[\cite{mm1:Majorova_n}] or based on the results of experimental
studies performed in  2001 -- 2002 [\cite{mm2:Majorova_n}].
However, the $F_{v}(dH)$ dependences can also be inferred from
the observational data of the survey proper. In this paper we
propose to use to this end the transit curves of the sources of
the RZF survey  with known fluxes whose
declinations differ from that of the central sky section of the
survey. One can select a sufficient number of ``point''\, sources
with well-known coordinates and fluxes from a single 24-hour sky
section with more than 30000 radio sources passing within the
envelope of the reflective panel when observed at $\lambda7.6$ cm
to construct the two-dimensional power beam pattern of the radio
telescope [\cite{co1:Majorova_n,co2:Majorova_n}].

This method of the PB measurement may be less accurate compared
to the technique described by Kuz'min and
Solomonovivh~[\cite{ks:Majorova_n}], because it requires a good
knowledge of the fluxes of the calibration sources. However, it
allows one to estimate the state of the power  pattern at the
time of survey observations and does not require additional
antenna time. Moreover, a comparison of the PB so measured with
the power beam determined using other methods allows one
to estimate the measurement errors at the survey time.

In this paper we analyze the PB at the time of the deep zenith
sky survey performed with RATAN-600 [\cite{p1:Majorova_n}]. We
chose most of our reference sources among those steep-spectrum
sources from the  NVSS catalog that fall within the sky region of
the survey.

\section{OBSERVATIONS}

The RZF survey
 was carried out within the framework of the program
``The Genetic Code of the Universe''\, aimed at the study of
spatial variations of cosmic background radiation and weak radio
sources within the band of the [\cite{p2:Majorova_n,p3:Majorova_n}]
survey. Observations were performed from  1998 through 2003 in the
Northern sector of RATAN-600 at the elevation of  3С84
($H=87^o41'$) in the wavelength interval $\lambda=1.0\div55$ cm.
The measurement equipment employed consisted of the radiometric
complex of antenna feed No.1.
A total of eight observing sets
were run. During the  (1998.10--1999.02), (1999.03),
(1999.10--1999.11), (1999.12--2000.03), and (2001.01--03,
2001.10--11) sets observations were made in the $0^h\le R.A.<
24^h$ survey band at the declination of 3С84
($Dec_{2000}=41^{o}30'42''$), and during the (2001.12 -- 2002.03)
and (2002.12--2003.03) sets, in nine sections  $12'$ apart so
that the total band of the  $\lambda7.6$-cm survey was $0^h \le
R.A._{2000.0}< 24^h$, $Dec_{2000.0}=41^o30'42'' \pm60'$.

In this paper we focus on the study of the power beam pattern at
$\lambda7.6$ cm. The sensitivity of the receiver is maximal at
this wavelength and amounts to  2.5\,mK. Of greatest interest from
the viewpoint of studying the PB are the results
of observations made during the last two sets. For the sake of
brevity we refer to them as the 2002 (2001.12--2002.03) and 2003
(2002.12--2003.03) sets. Observational data obtained in nine
sections  (survey bands)  $\Delta\delta=12'$ apart in declination
allow the vertical power  pattern to be constructed over a
wide range of  $dH$. The long time span (from 1998 to 2003) of
the survey allows information to be obtained about the stability
of the power beam pattern in time.

To construct the PB, we selected  from the  NVSS
catalog mostly sources with steep spectra and $\lambda$\,21-cm
fluxes exceeding 16\,mJy. We converted the $\lambda$\,21-cm fluxes
into the corresponding $\lambda$\,7.6-cm fluxes using all the
available data on the fluxes of these sources at other
wavelengths as adopted from the  CATS database
[\cite{v1:Majorova_n,v2:Majorova_n}]. The $\lambda$\,7.6-cm fluxes of
the selected sources proved to be no lower than 5\,mJy. To increase
the signal-to-noise ratio, we averaged the records obtained both
over individual sets and over groups of sets. The number of
averaged single-source records in the central section of the
survey $Dec_{2000}=41^o30'42''$ ($\Delta\delta=0$) varied from 15
to 60 depending on the set. The number of averaged records did not exceed
five in sections $\Delta\delta=\pm12'n$ ($n$=1,2,3,4) apart from
the central section.
%Дальнейшая обработка
%записей сводилась к приведению их к антенным температурам и вычитанию фона.

\section{REDUCTION AND CONSTRUCTION OF THE POWER BEAM PATTERN}

When an unpolarized  ``point'' \, source crosses the central
section of the PB its antenna temperature at the
radiometer output and spectral flux density are linked by the
following relation:

\begin{equation}
 P/T_{a}=2k/S_{eff},
\label{1:Majorova_n}
\end{equation}

where $k$ is the Boltzmann constant; $S_{eff}$ is the effective
area of the antenna of the radio telescope (1000 m$^2$); $T_{a}$
is the antenna temperature (K), and $P$ is the source flux
(W/m$^2$ Hz).

When the source crosses a section that is $dH$ apart from the
central section in declination its antenna temperature should
decrease in accordance with the variation of the vertical
power pattern $F_{v}(dH)$. If the PB is
measured using a single source crossing different sections of the
power beam, its value in a section $dH$ apart from the
central section can be computed by the following formula:

\begin{equation}
  F_{v}(dH) = \frac{(P/T_{a})_{dH=0}}{(P/T_{a})_{dH}}=\frac{(T_{a})_{dH}}{(T_{a})_{dH=0}},
\label{2:Majorova_n}
\end{equation}
\noindent where $(T_{a})_{dH=0}$ and $(T_{a})_{dH}$ are the
antenna temperatures of the source in the central section and in
the section  $dH$ apart from the central section, respectively.

If the power beam  is measured using a set of sources
crossing different sections of the PB then to
compute the vertical power pattern in the  $dH$ section, one
has to know not only the antenna temperatures of the sources
(${T_{a}}^{i})_{dH=0}$, $({T_{a}}^{j})_{dH}$, but also their
fluxes $P_{i}$, $P_{j}$. In this case, the formula for $F_{v}(dH)$
acquires the following form:
\begin{equation}
    F_{v}(dH) = \frac{(P_{i}/{T_{a}}^{i})_{dH=0}}{(P_{j}/{T_{a}}^{j})_{dH}}.
\label{3:Majorova_n}
\end{equation}

In the observations performed in nine survey bands during the 2002
and 2003 sets the antenna of the radio telescope was focused on
the elevations corresponding to the following declinations:

$Dec_{n}=Dec_{3C84}\pm12'n$,  ($n=0, 1, 2, 3, 4$).

We estimated the  $dH$ offset for the source with declination
$Dec_{i}$ observed in  $n$th band of the survey from the central
section of this band as:

    $dH = Dec_{i} - Dec_{3C84}\pm12'n$,   ($n=1, 2, 3, 4$).

Sufficiently bright sources were observed in several bands of the
survey.

To construct the power beam pattern,
 we selected from all nine bands of the survey  of about 140 sources from  NVSS
catalog with the $\lambda$7.6-cm fluxes no lower than  80\,mJy,
except for three sources. A total of about 35 and 20 records for
each of these sources were averaged in the 2002 and 2003 sets,
respectively.

\begin{figure*}[htbp]
%\setcaptionmargin{15mm}
%\onelinecaptionsfalse
\centerline{
\includegraphics[angle=-90,width=0.35\textwidth,clip]{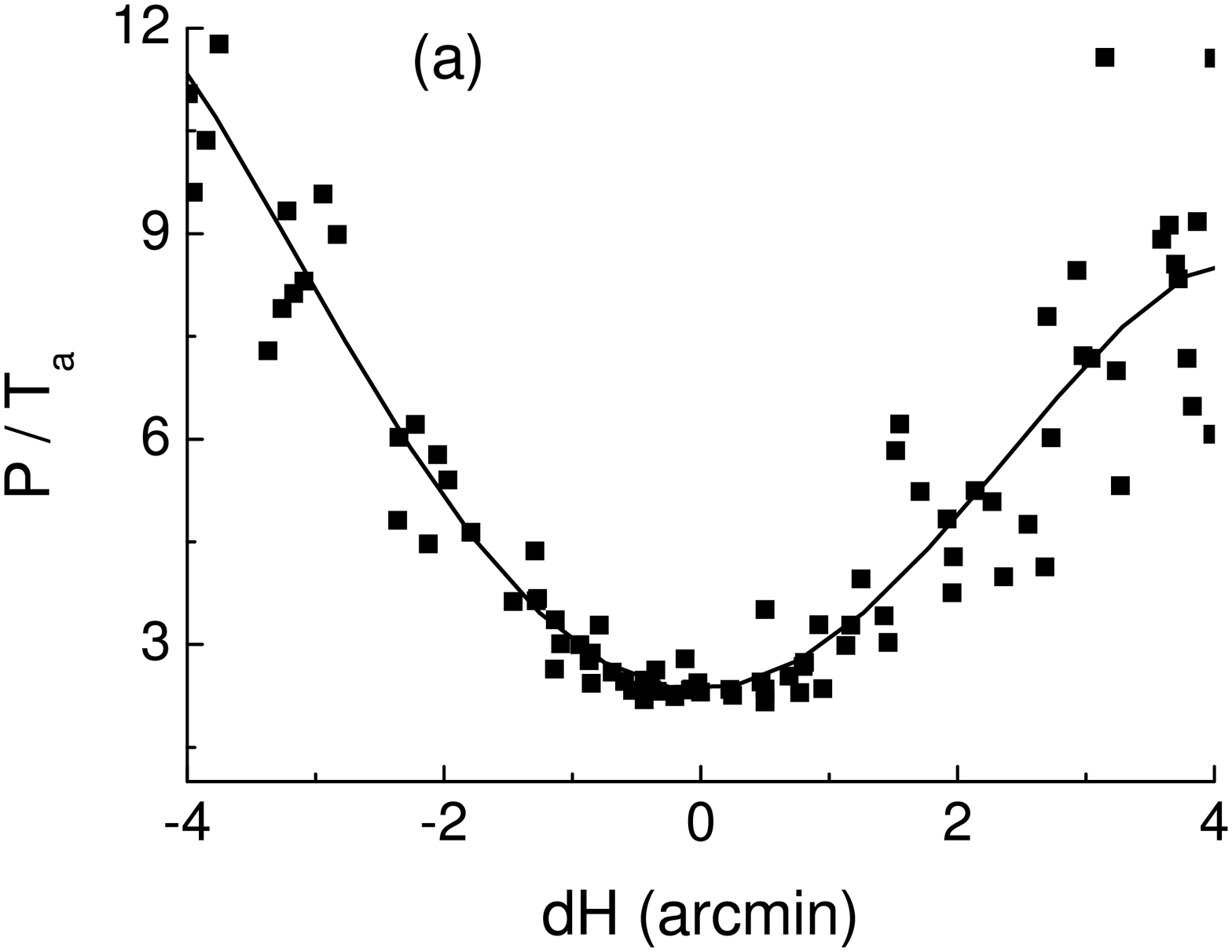}%0.45\textwidth
\includegraphics[angle=-90,width=0.35\textwidth,clip]{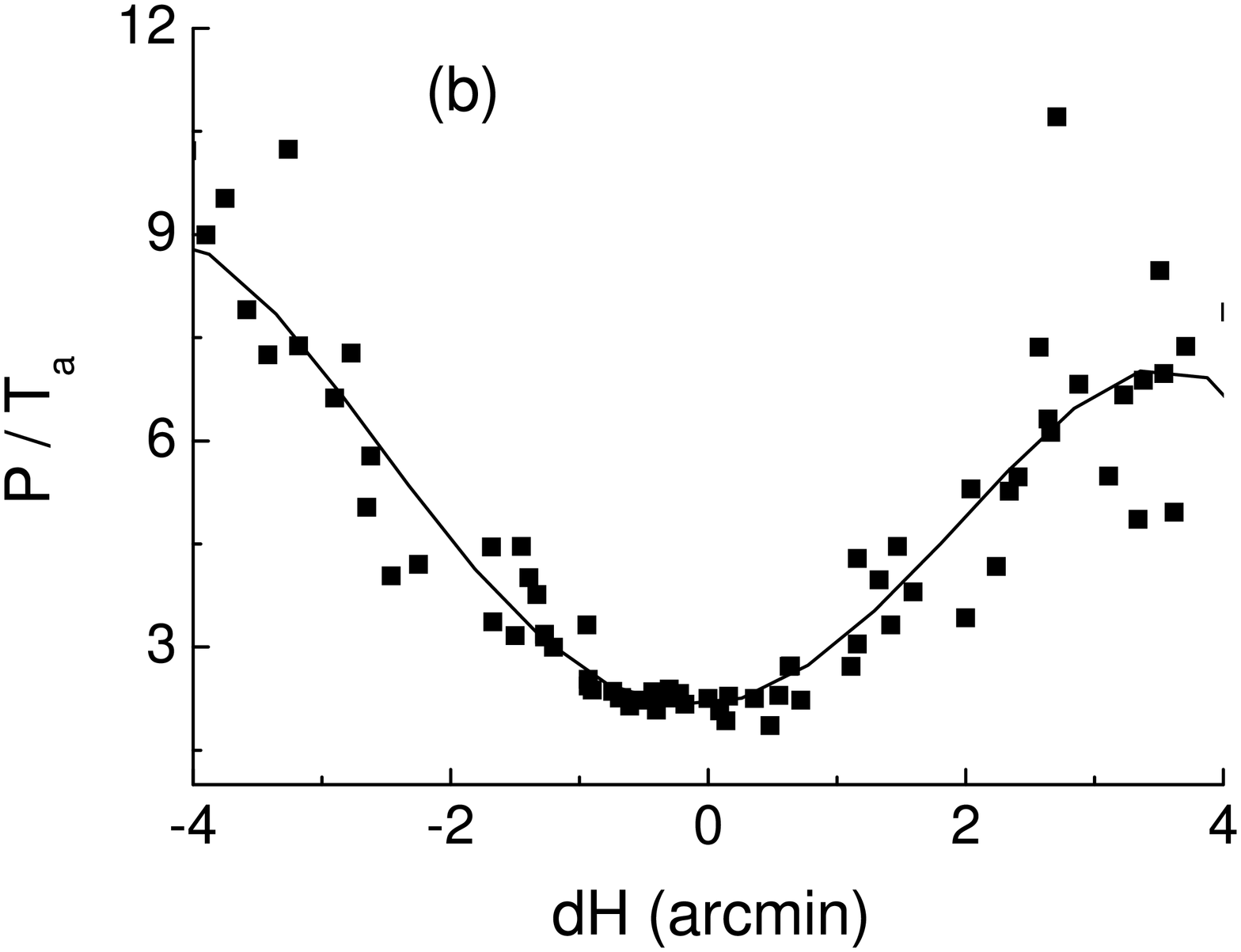}
}
%\captionstyle{normal}
\caption{ The ratio of the source flux to
its antenna temperature,  $P/T_{a}$, as a function of  $dH$. The
filled squares show the experimental values and the solid lines
show the corresponding approximating polynomials. (a) The data
for the 2002 set; (b) the data for the 2003 set. }
\label{fig1:Majorova_n}
\end{figure*}

After determining the fluxes and antenna temperatures of the
sources for each observing set we constructed the corresponding
$P/T_{a}=f(dH)$ dependences. The filled squares in
Fig.~\ref{fig1:Majorova_n} show the  $P/T_{a}$ ratios for the
sources of the sample studied located within  $dH=-4'\div4'$ ((a)
and (b) correspond to the 2002 and 2003 sets, respectively) from
the central section. We then use the least squares method to fit the
approximating curves (two- and four-degree polynomials) to the
data points obtained. These curves allowed us to determine the
average  $(P/T_{a})_{dH=0}$ ratios in the central section of the
power beam pattern. The approximating curves are shown by the solid
lines in Fig.~\ref{fig1:Majorova_n}. The approximating curves
yielded a $(P/T_{a})_{dH=0}$ ratio of  $2.3\pm0.3$ and
$2.2\pm0.3$ for the 2002 and 2003 sets, respectively. Given the
$(P/T_{a})_{dH=0}$ ratio, one can use formula
(\ref{1:Majorova_n}) to estimate the effective area of the
antenna of the radio telescope during the observing period
considered. The effective area at $\lambda$7.6 cm with the
antenna pointed to the elevation of  3С84 ($H=87^o41'$) was equal
to $S_{eff}=1200 \pm130$ and $S_{eff}=1250 \pm130$ m$^2$ during
the 2002 and 2003 sets, respectively.

To construct~~ the vertical~ PB,~~ we used the
$(P/T_{a})_{dH=0}$ ratios determined from the approximating
curves and also the antenna temperatures $({T_{a}}^{i})_{dH}$ and
fluxes $P_{i}$ of the sources. We computed the $F_{v}(dH)$ values
by formula (\ref{3:Majorova_n}). The relative standard error of
the antenna temperatures determined from the entire sample of
sources considered is equal to $\sigma_{T_{a}}=0.10\pm0.08$; the
standard error of the inferred source fluxes is equal to
$\sigma_{P}=0.16\pm0.06$, and the total standard error of
measurements is equal to  $\sigma_{\Sigma}=0.19\pm0.10$. The
total error is determined by the contributions of three
components: errors of interpolation of the spectra to the
wavelength of $\lambda7.6$ cm based on the results of other
observations; the signal-to-noise ratio, and the stability of the
parameters of the antenna system of  RATAN-600.

\section{RESULTS OF THE CONSTRUCTION OF THE EXPERIMENTAL POWER BEAM PATTERN}

\begin{figure*}[htbp]
%\setcaptionmargin{0mm}
%\onelinecaptionsfalse
\centerline{
\vbox{
\hbox{
\includegraphics[angle=-90,width=0.3\textwidth,clip]{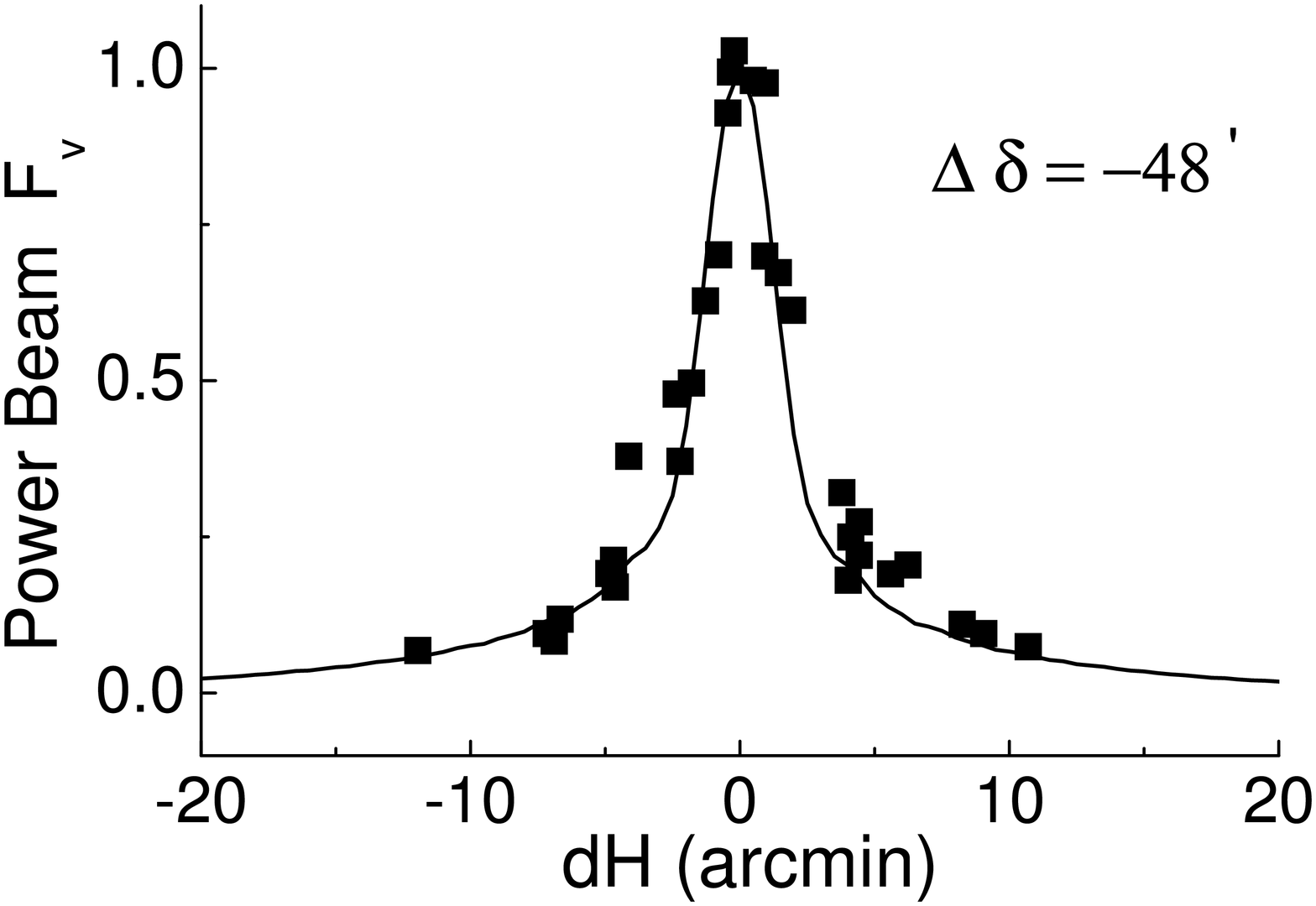}
\includegraphics[angle=-90,width=0.3\textwidth,clip]{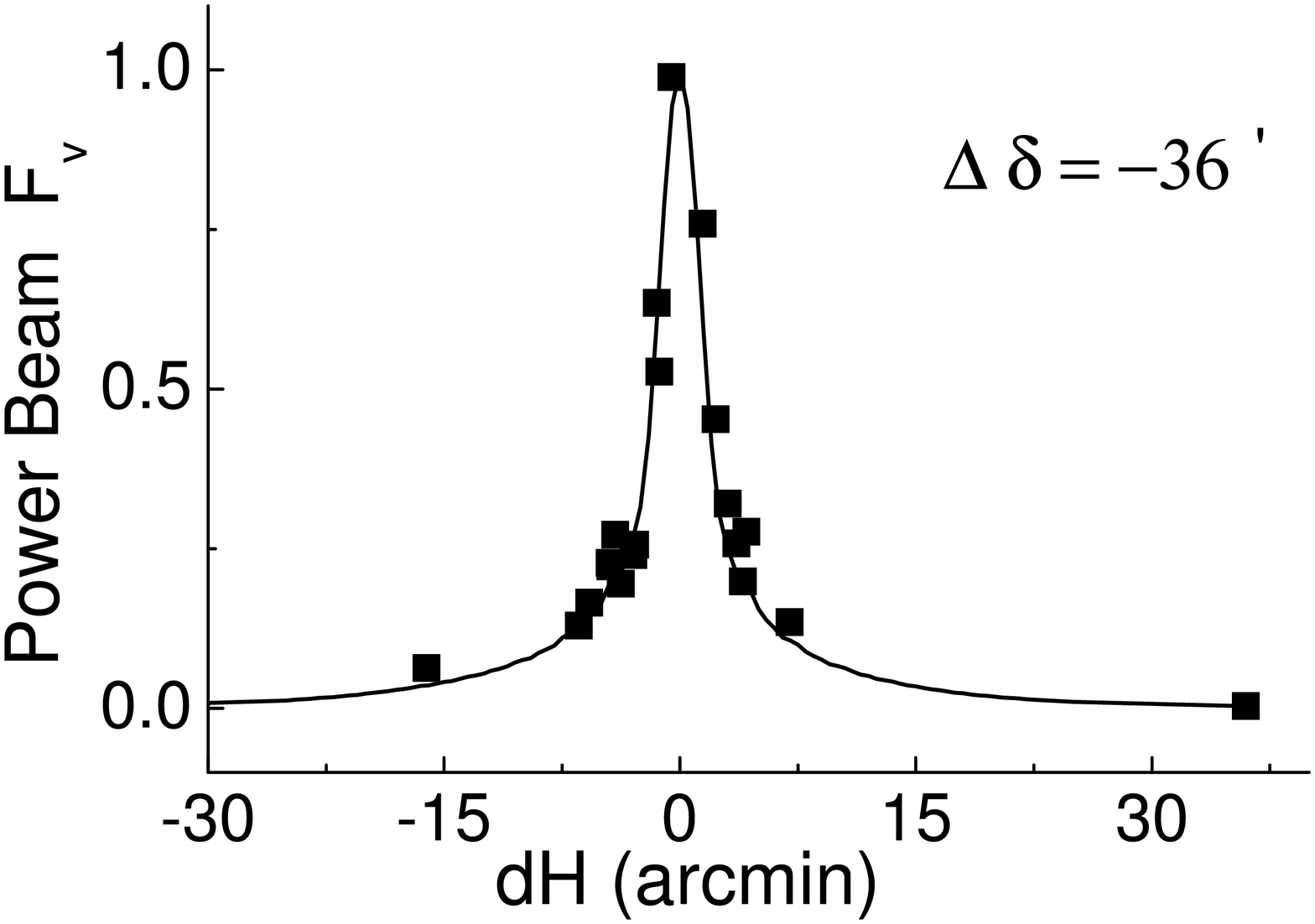}
\includegraphics[angle=-90,width=0.3\textwidth,clip]{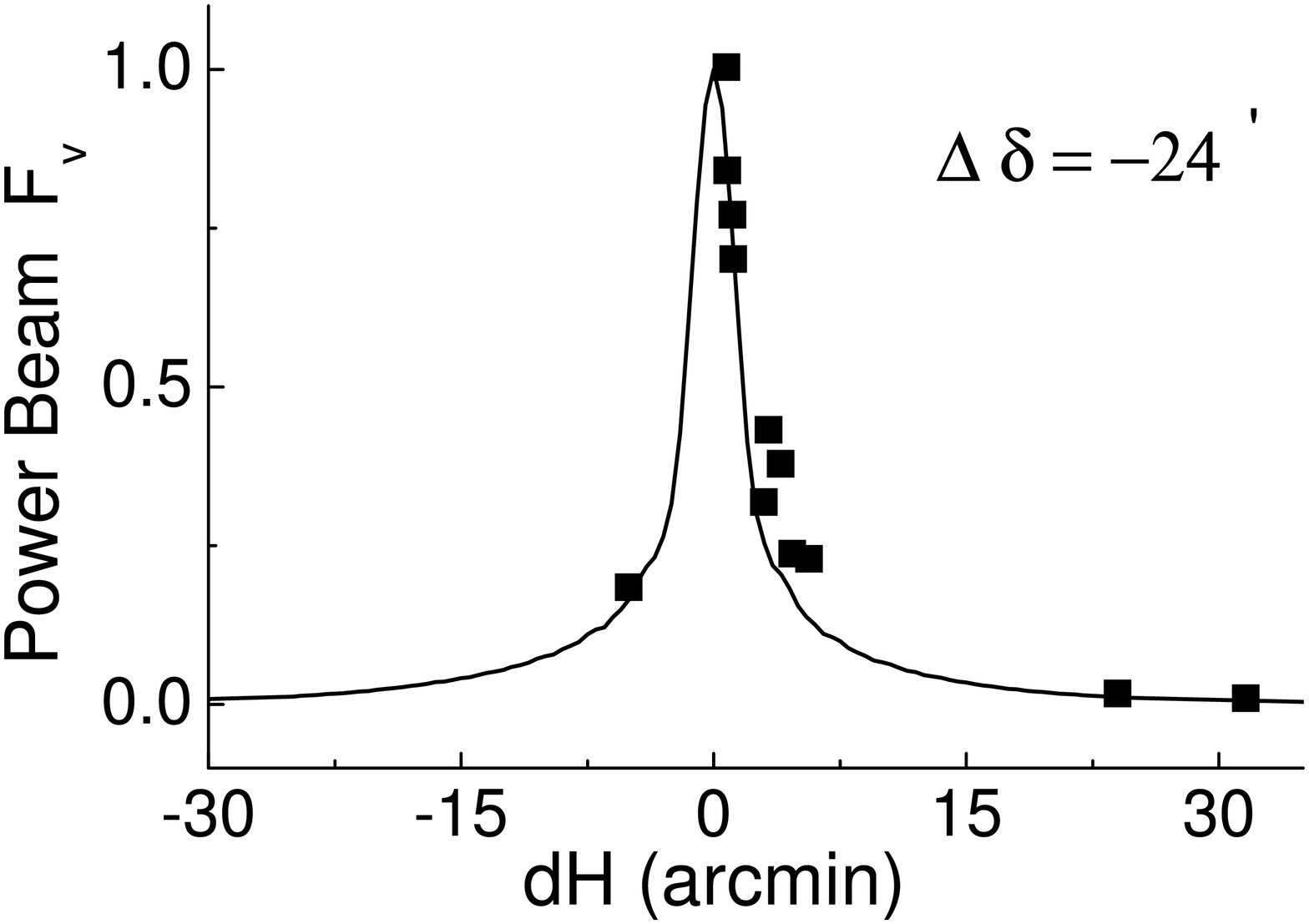}
}
\hbox{
\includegraphics[angle=-90,width=0.3\textwidth,clip]{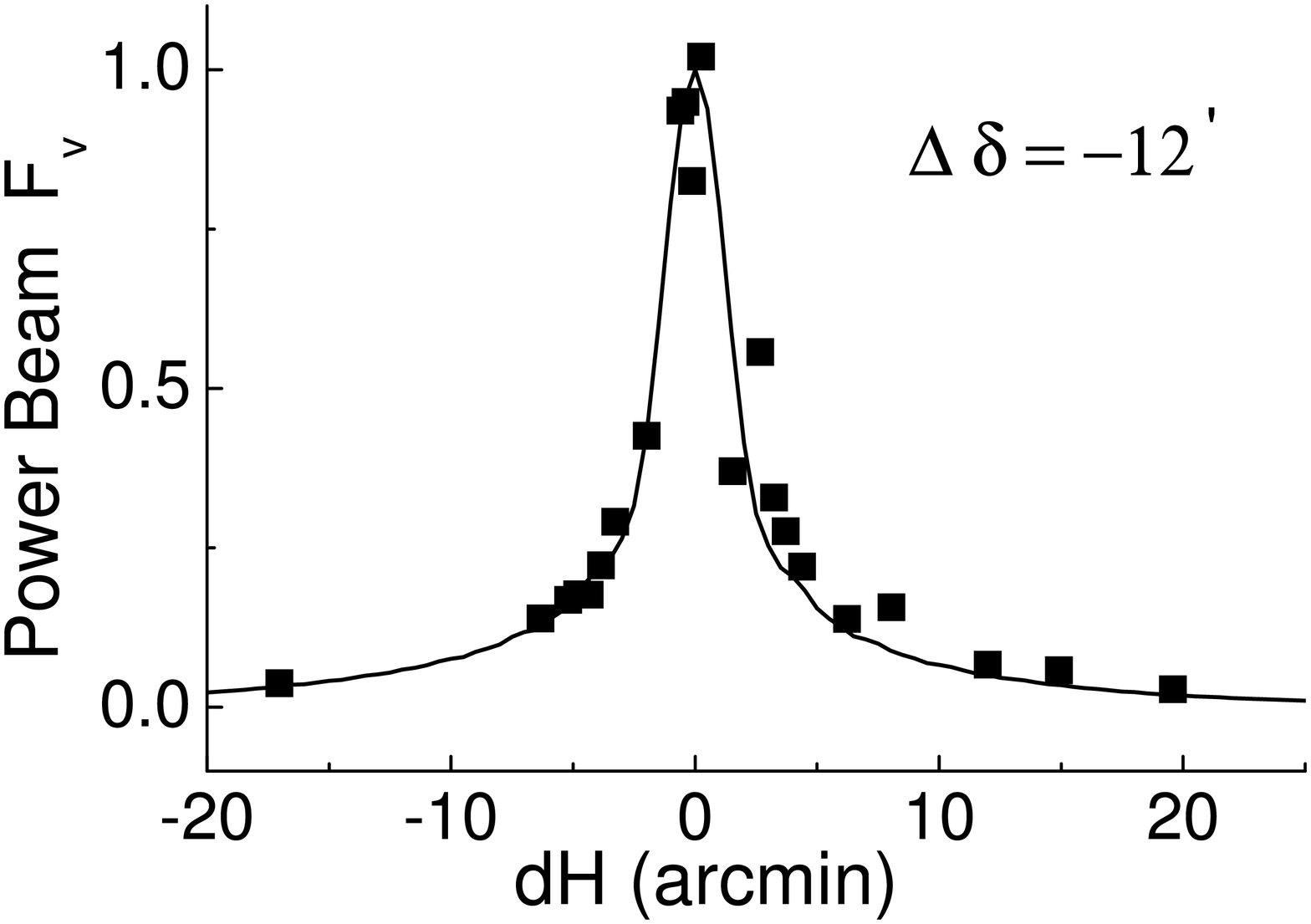}
\includegraphics[angle=-90,width=0.3\textwidth,clip]{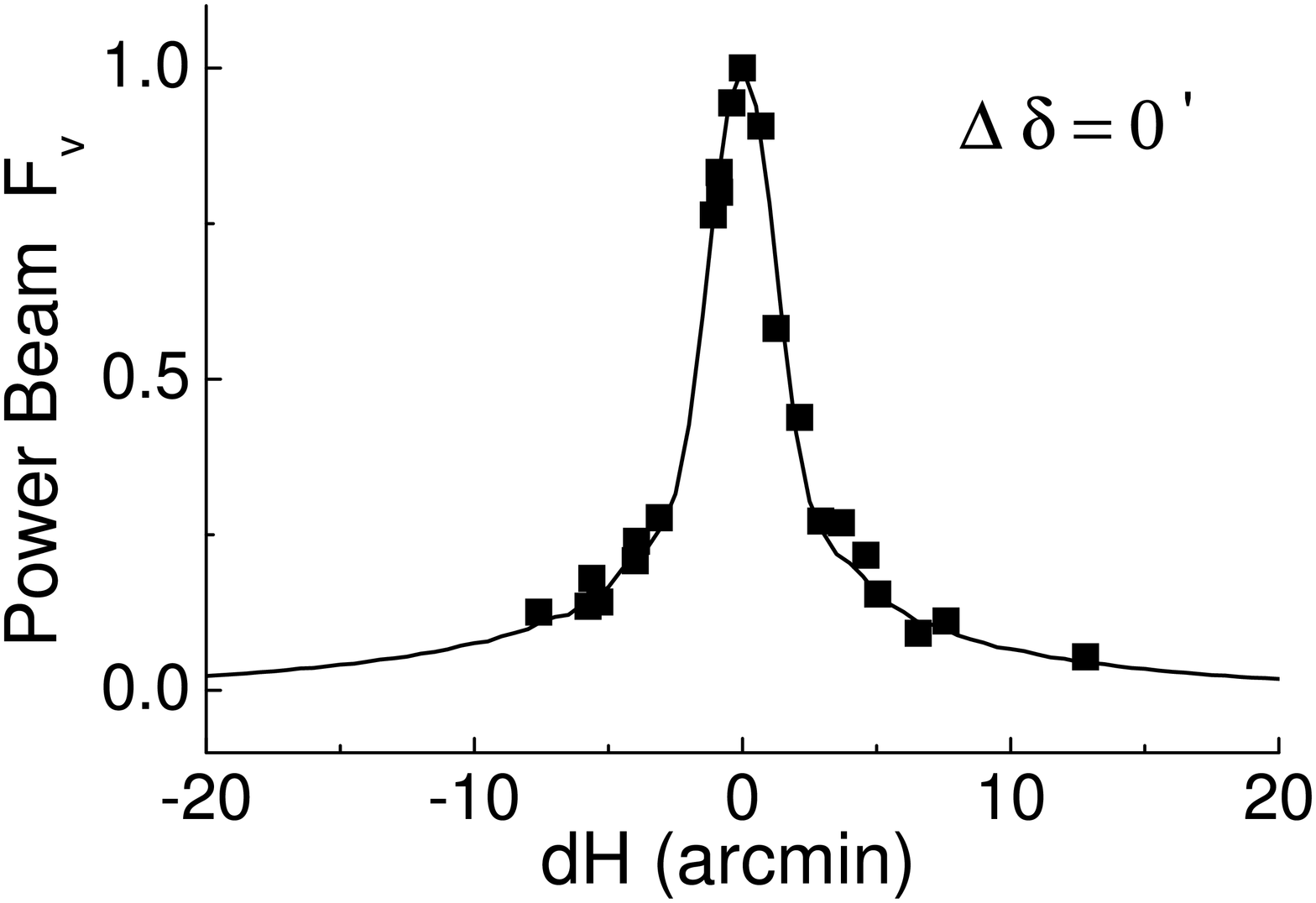}
\includegraphics[angle=-90,width=0.3\textwidth,clip]{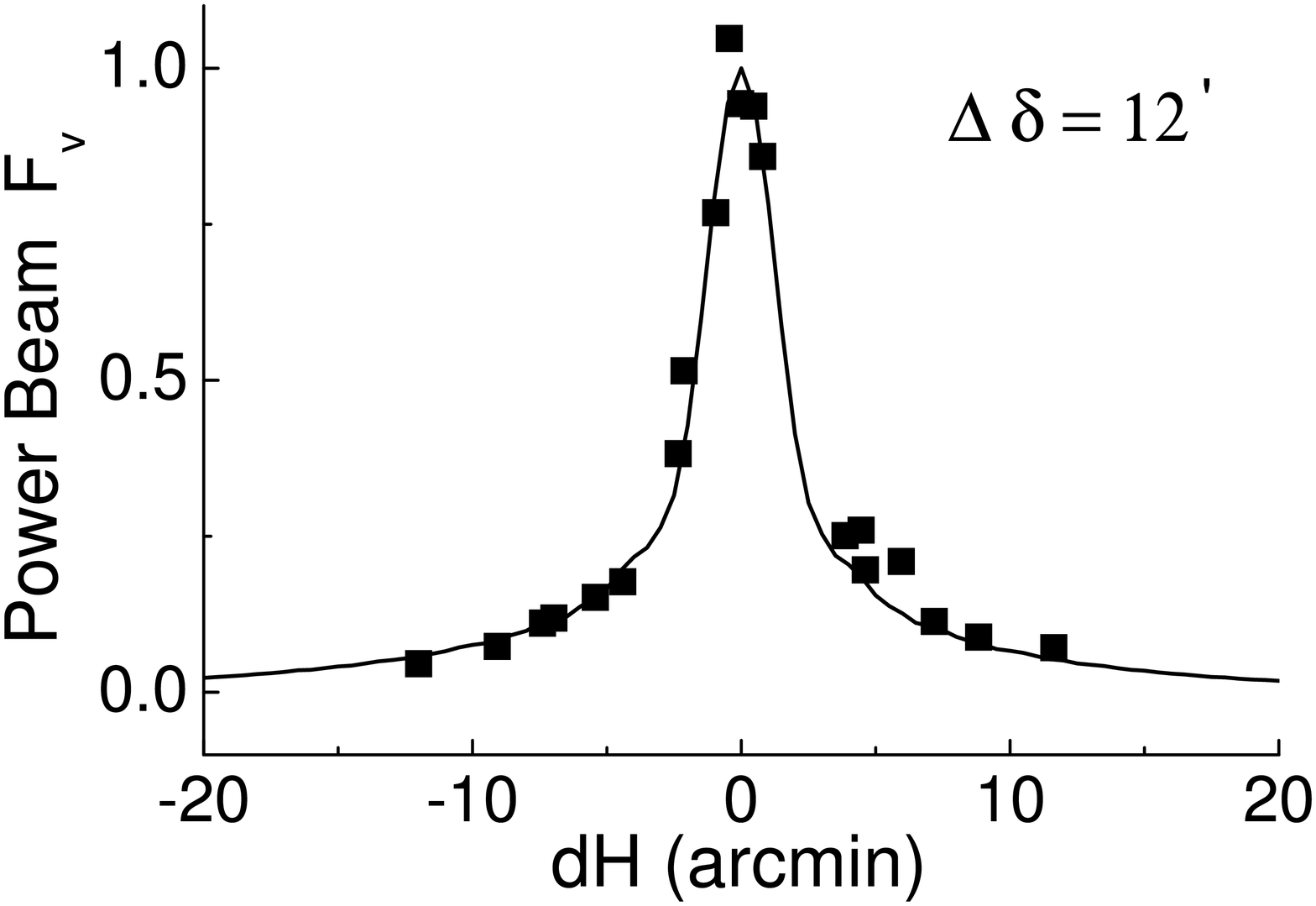}
}
\hbox{
\includegraphics[angle=-90,width=0.3\textwidth,clip]{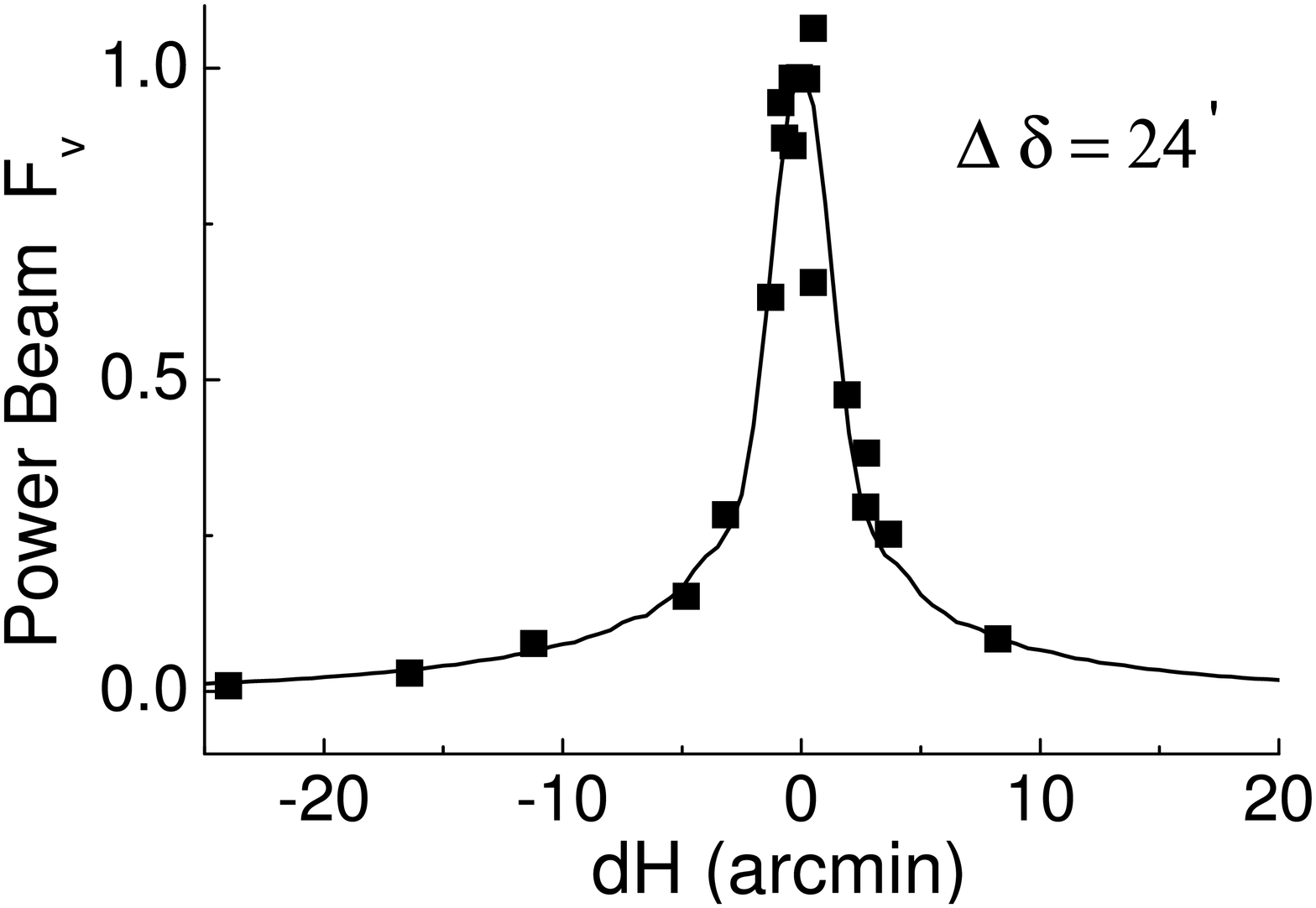}
\includegraphics[angle=-90,width=0.3\textwidth,clip]{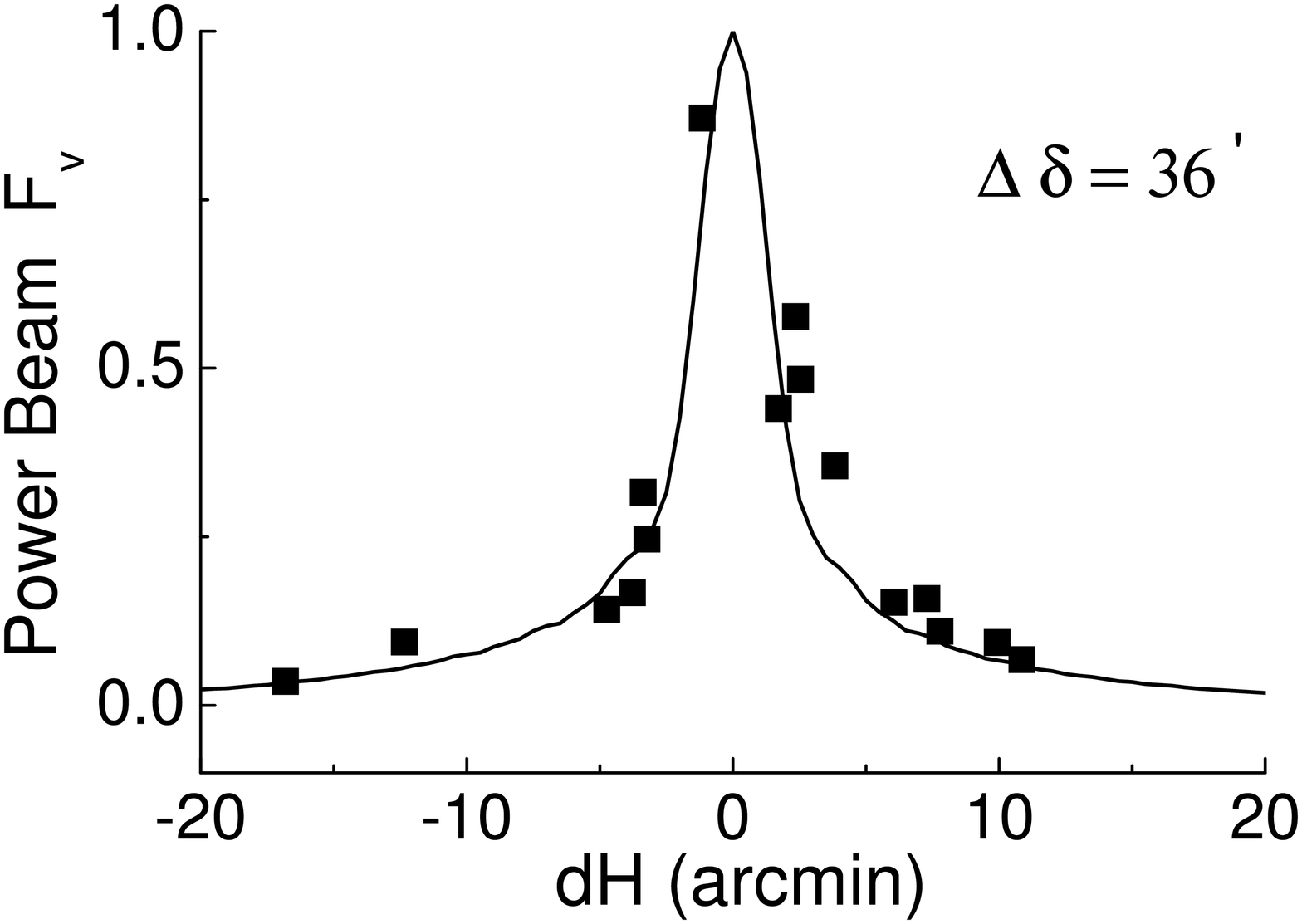}
\includegraphics[angle=-90,width=0.3\textwidth,clip]{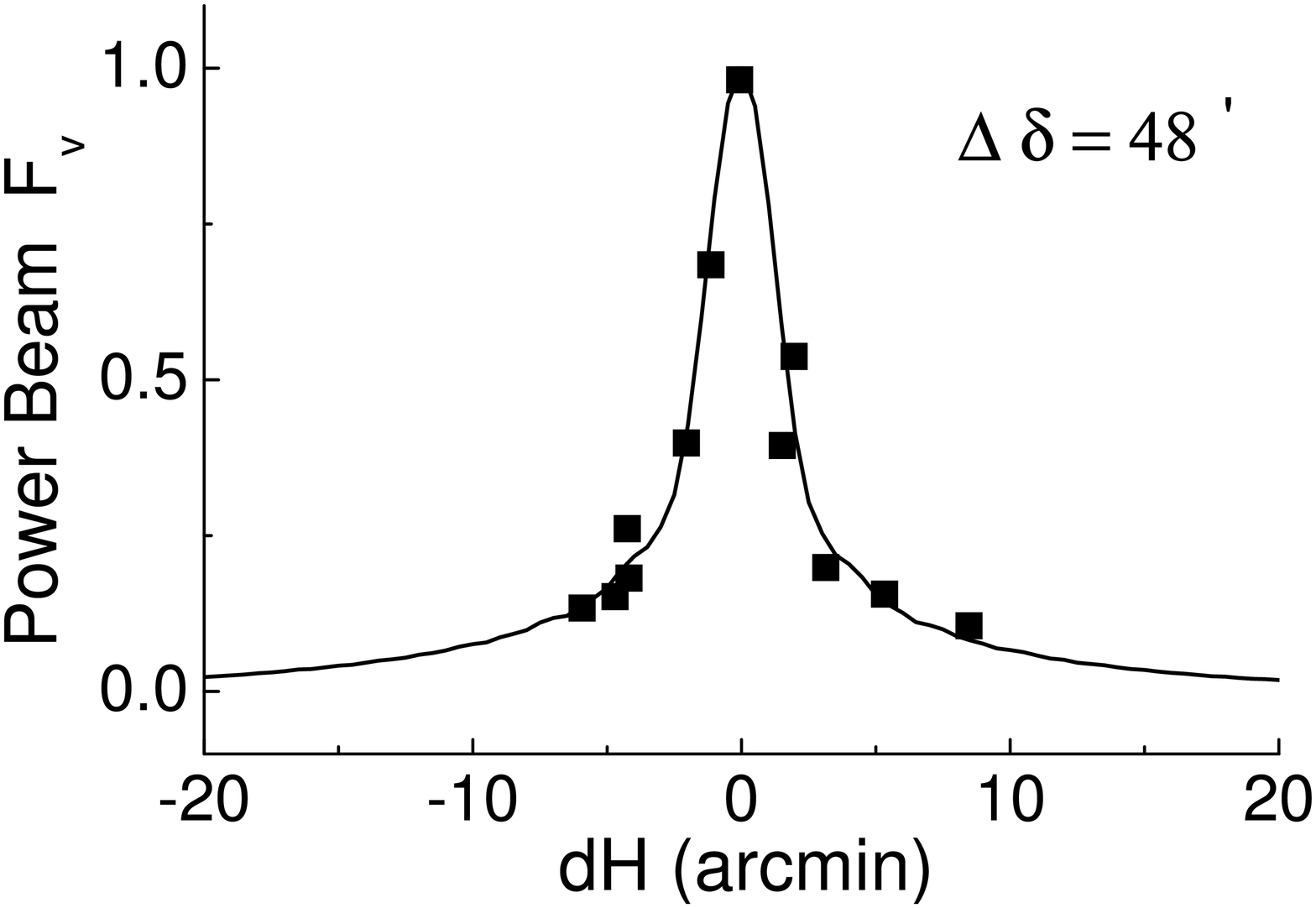}
}}}
%\captionstyle{normal}
\caption{ Vertical PB
constructed  using the results for a sample of NVSS sources
with $\lambda7.6$cm fluxes  $P > 80$\,mJy in nine bands of the RZF
survey: in the central band $Dec_{2000}=41^o30'42''$
($\Delta\delta=0'$) and in the bands
$\Delta\delta=~\pm12',~\pm24', ~\pm36,~\pm48'$ apart from the
central band. The filled squares show the experimental data
points of the power beam pattern and the the solid lines show the
computed power beam patterns. (The 2002 observing set).}
\label{fig2:Majorova_n}
\end{figure*}

\begin{figure*}[htbp]
%\onelinecaptionstrue
\centerline{
\vbox{
\hbox{
\includegraphics[angle=-90,width=0.3\textwidth,clip]{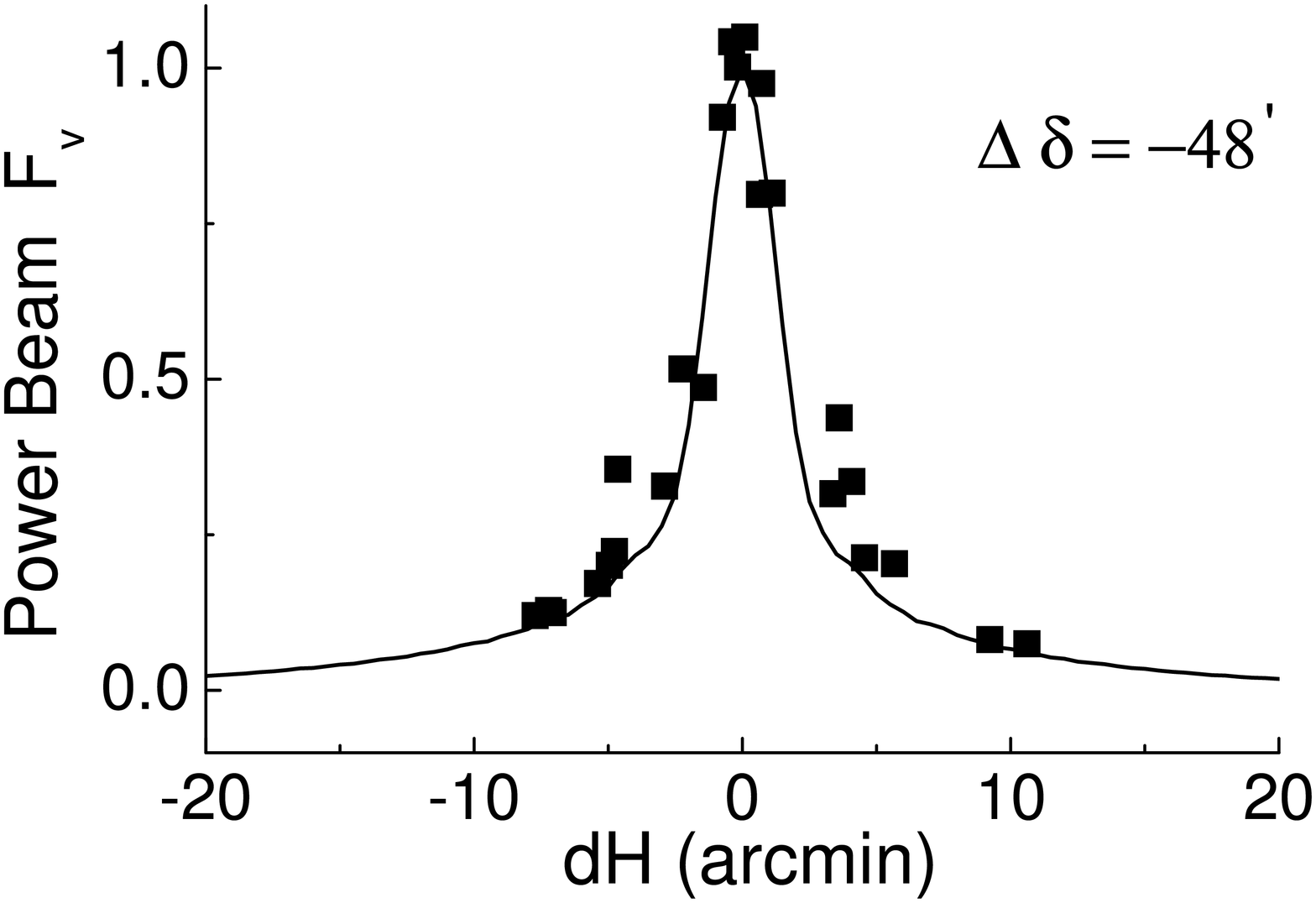}
\includegraphics[angle=-90,width=0.3\textwidth,clip]{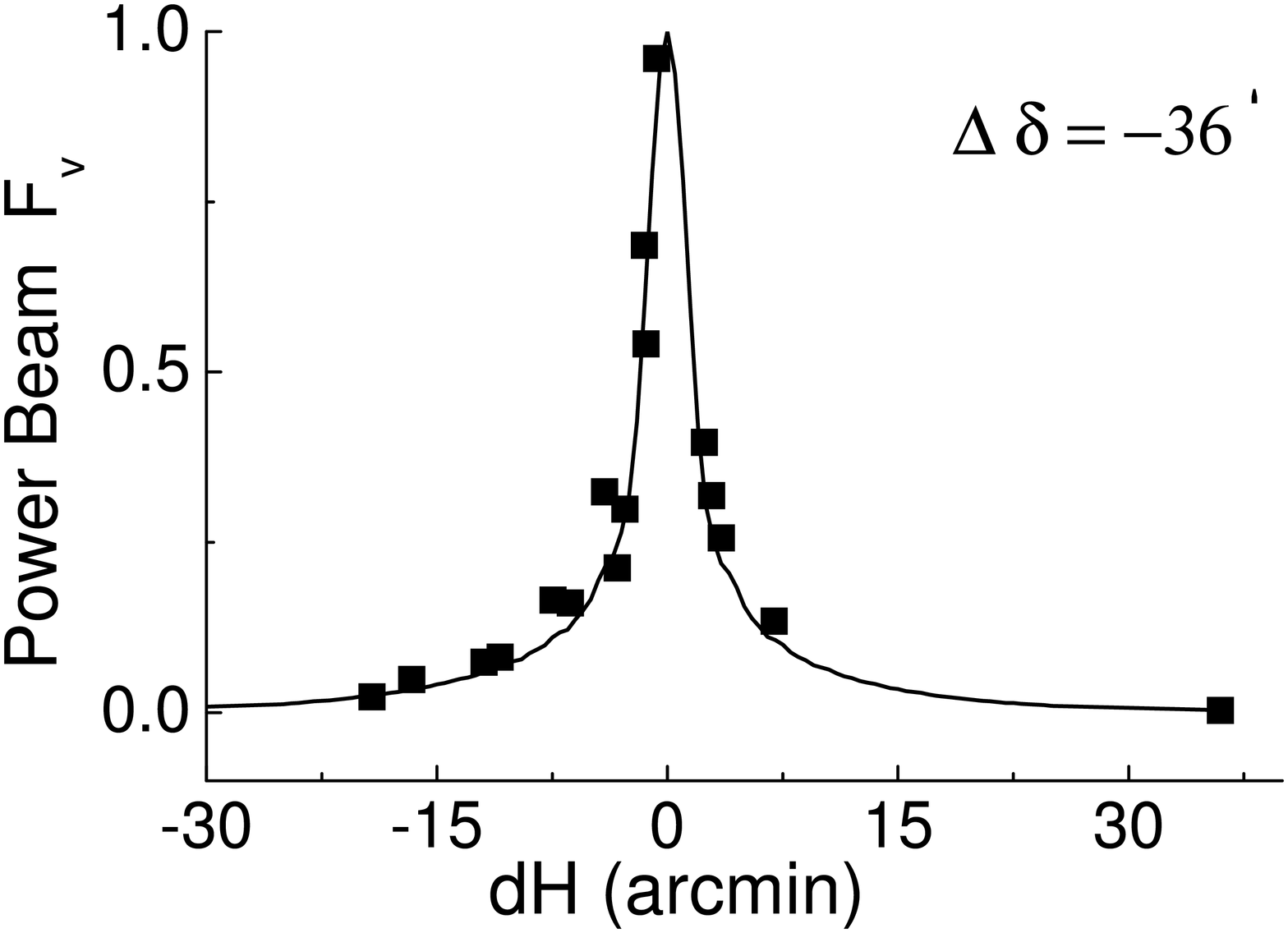}
\includegraphics[angle=-90,width=0.3\textwidth,clip]{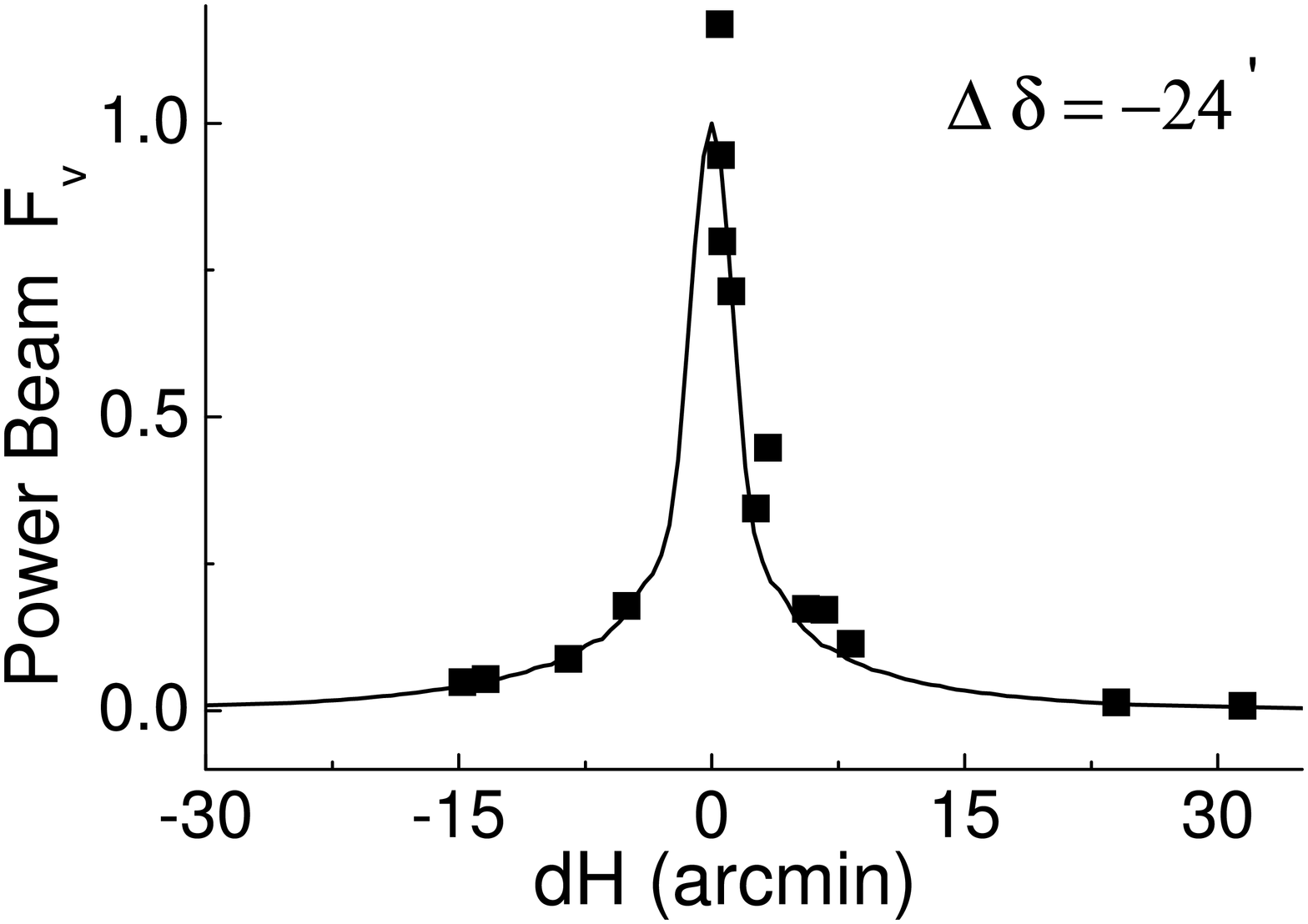}
}
\hbox{
\includegraphics[angle=-90,width=0.3\textwidth,clip]{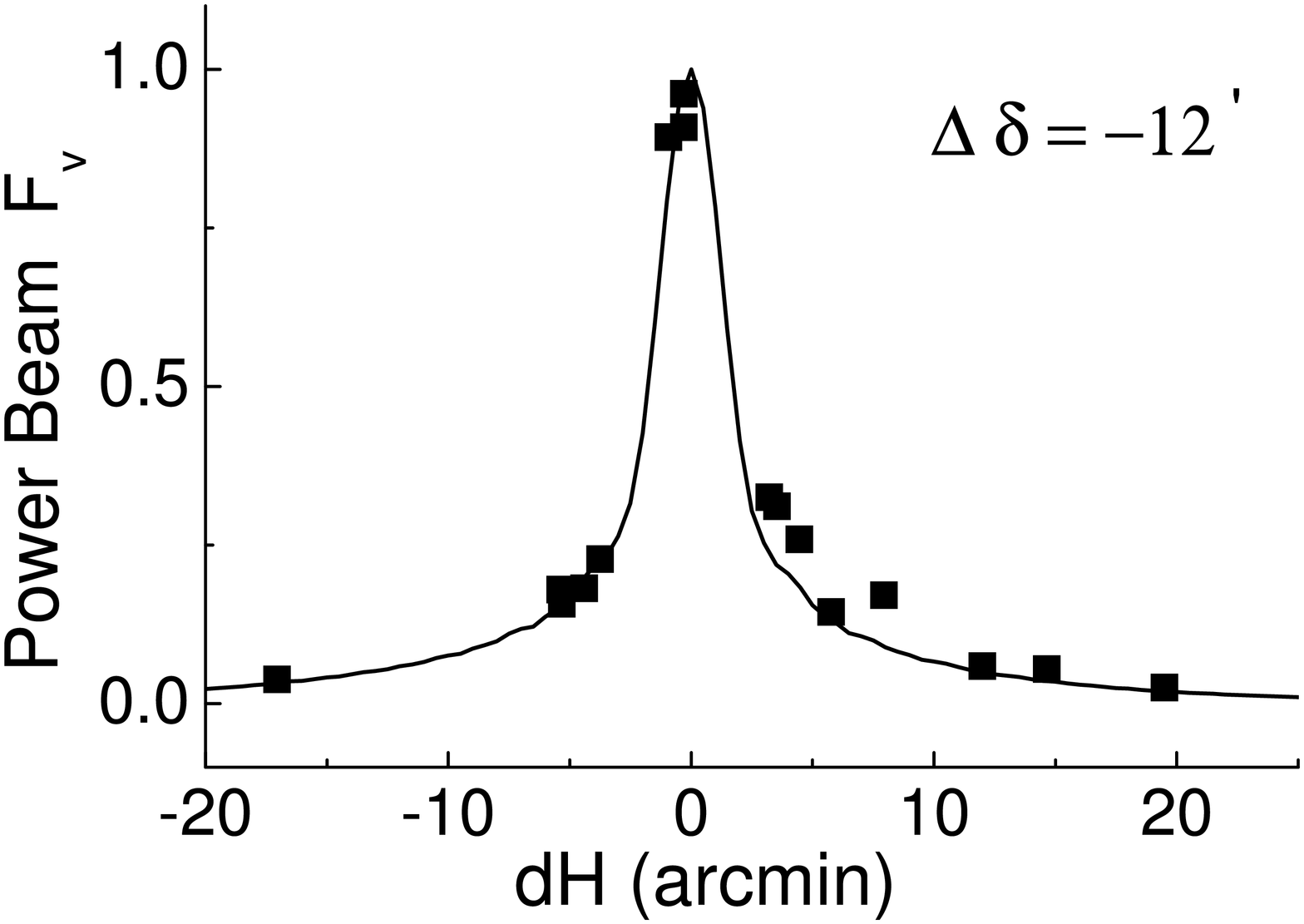}
\includegraphics[angle=-90,width=0.3\textwidth,clip]{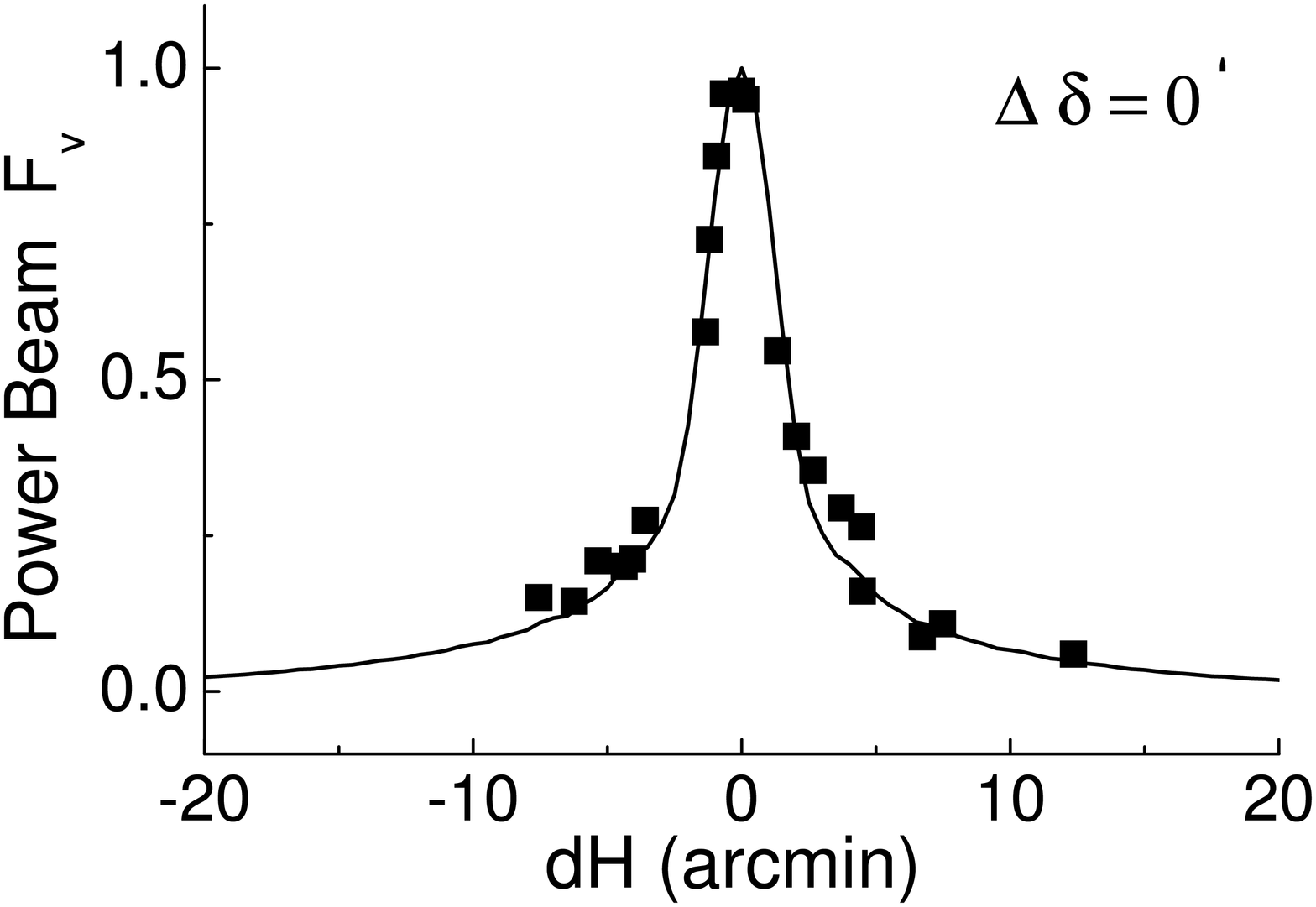}
\includegraphics[angle=-90,width=0.3\textwidth,clip]{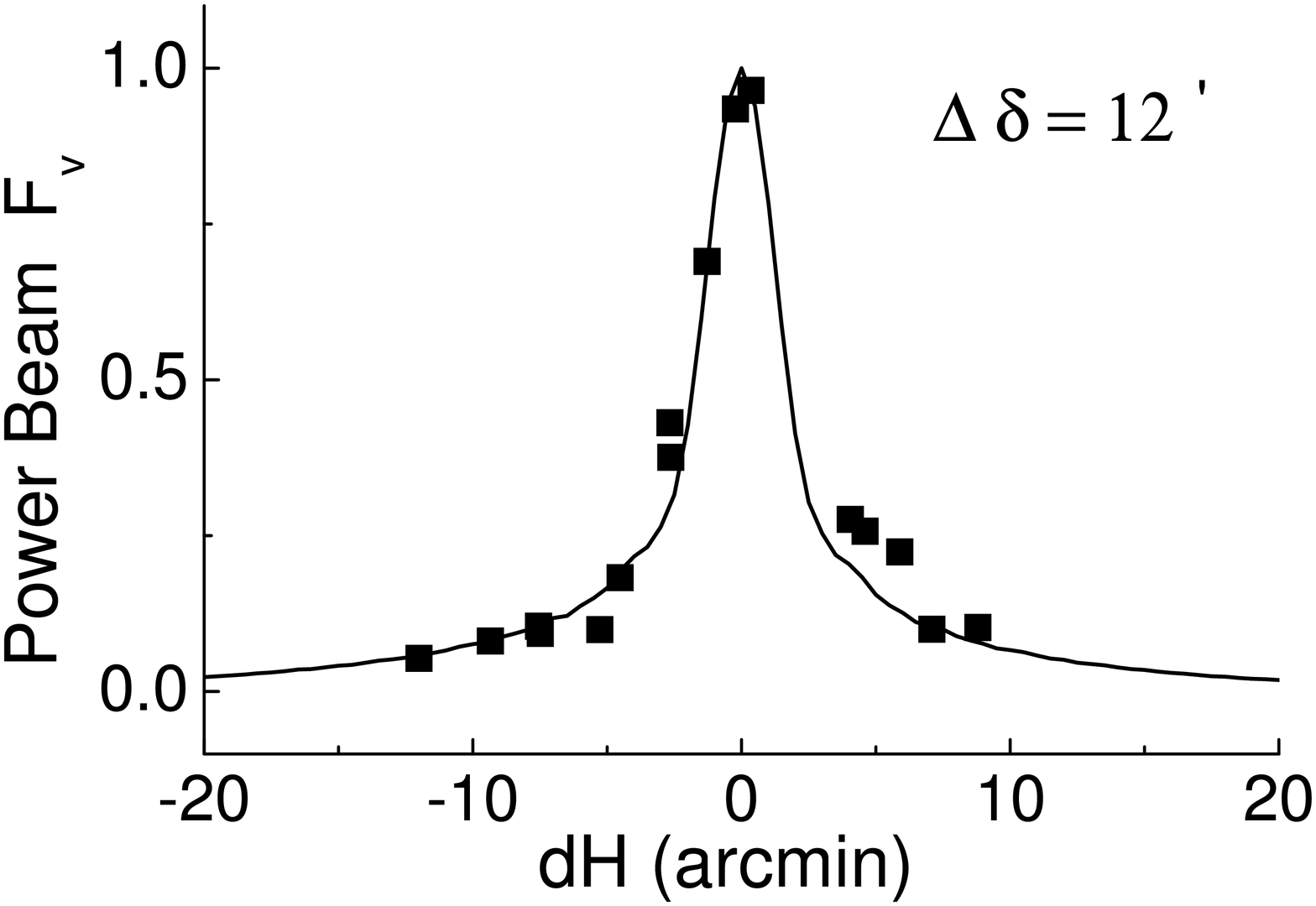}
}
\hbox{
\includegraphics[angle=-90,width=0.3\textwidth,clip]{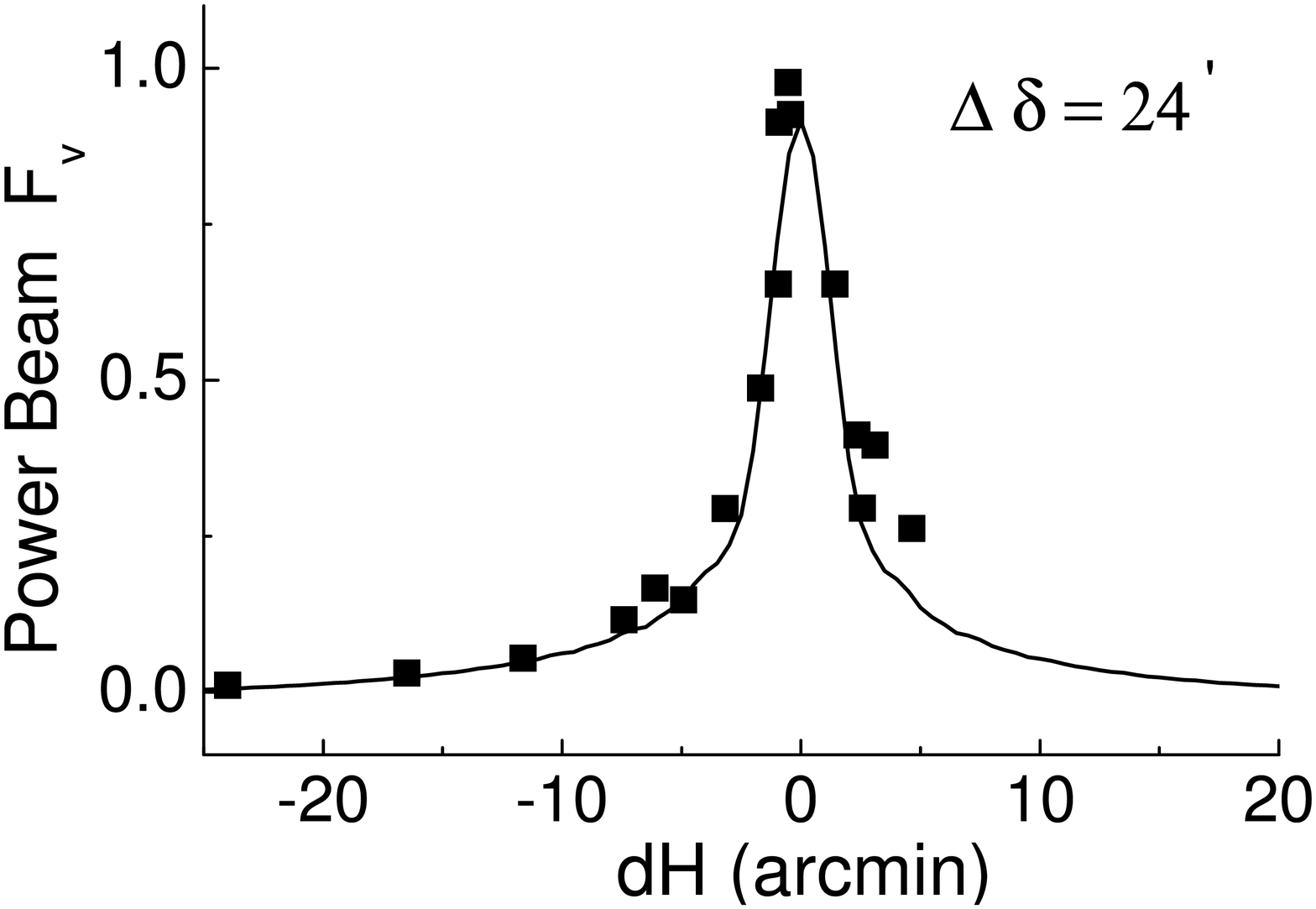}
\includegraphics[angle=-90,width=0.3\textwidth,clip]{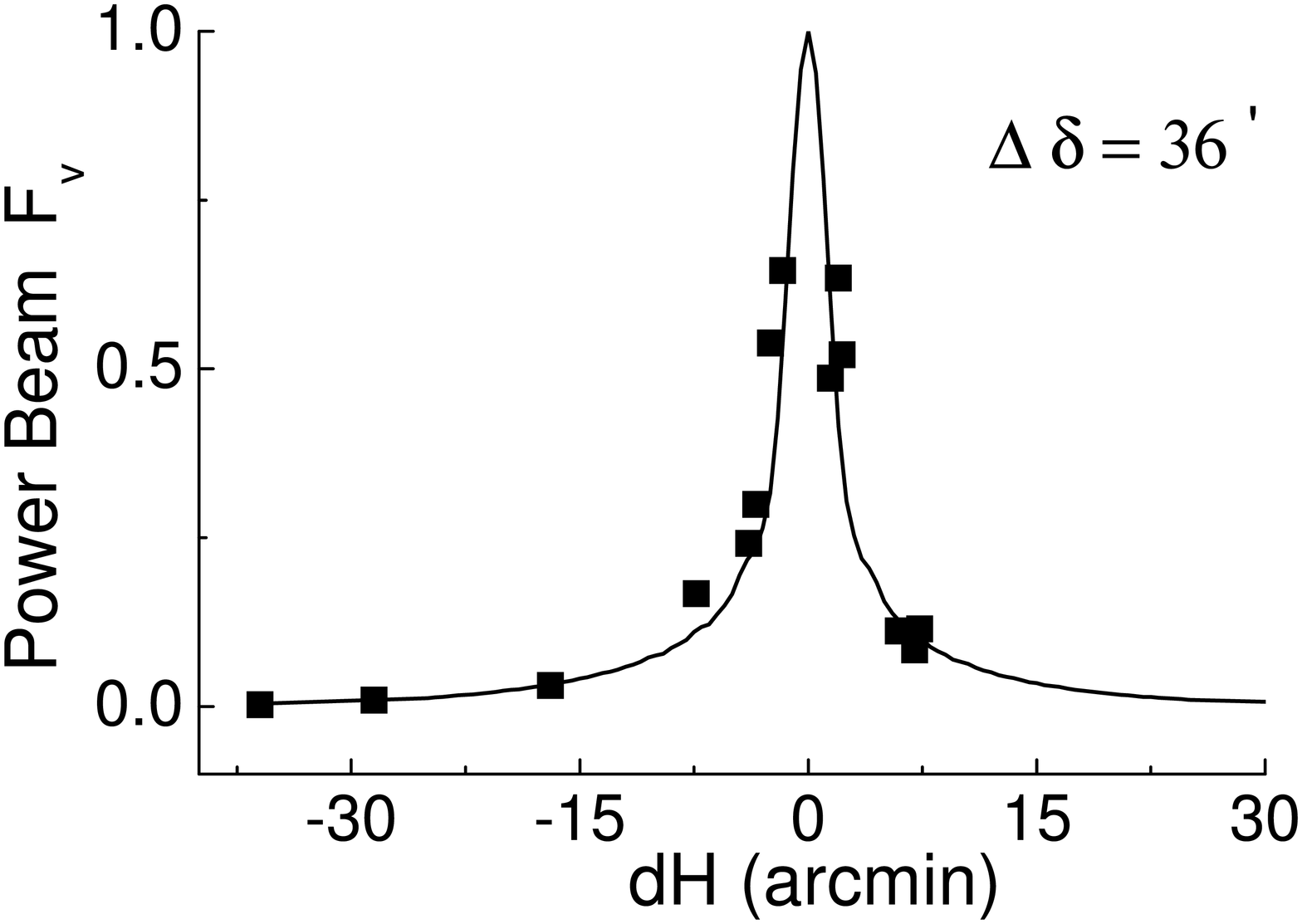}
\includegraphics[angle=-90,width=0.3\textwidth,clip]{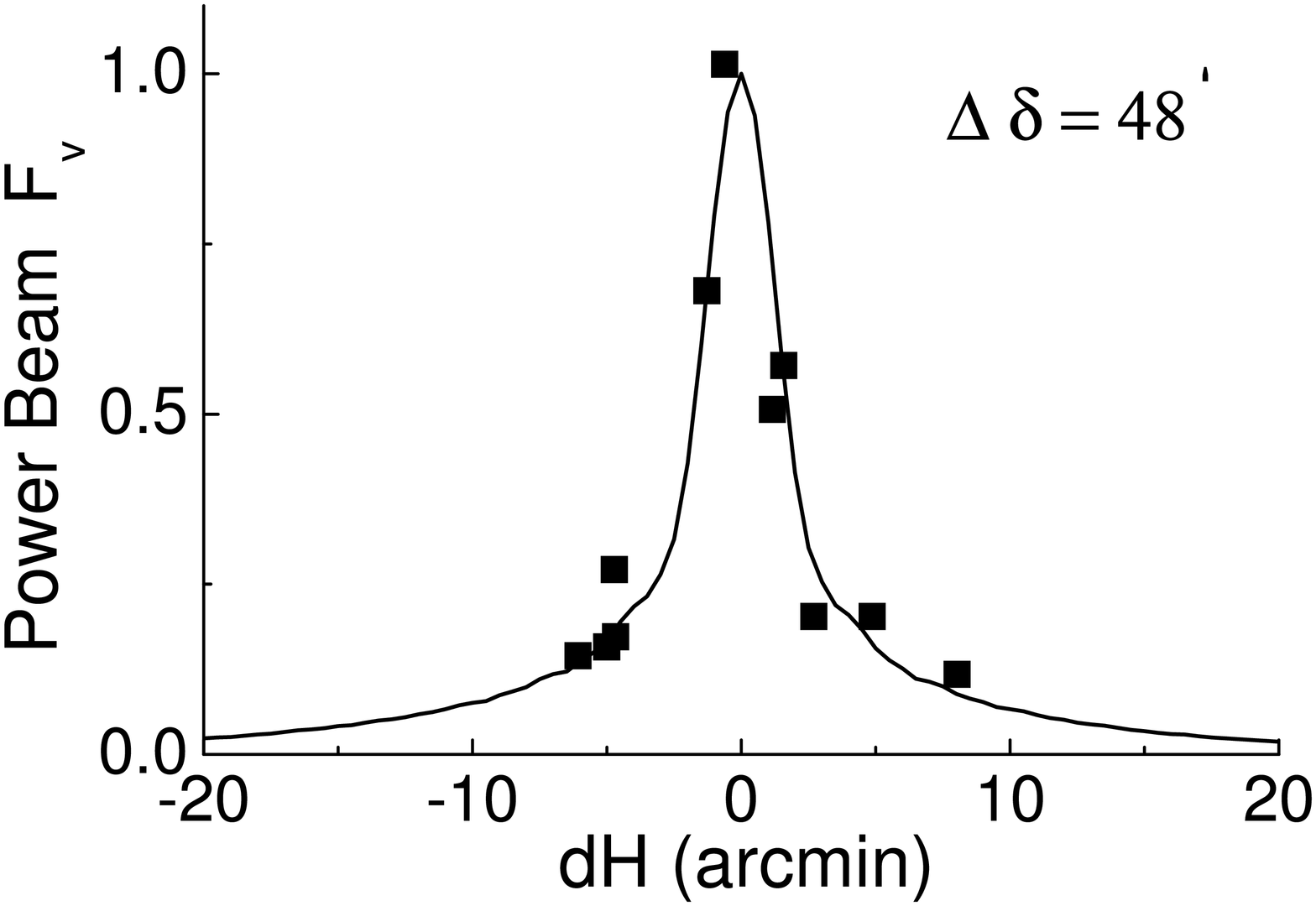}
} } }
%\setcaptionmargin{0mm}
%\captionstyle{normal}
\caption{ Same
as Fig.~\ref{fig2:Majorova_n}, but for the 2003 observing set. }
\label{fig3:Majorova_n}
\end{figure*}

\begin{figure*}[t]
%\onelinecaptionsfalse
\centerline{
\vbox{
\hbox{
\includegraphics[angle=-90,width=0.3\textwidth,clip]{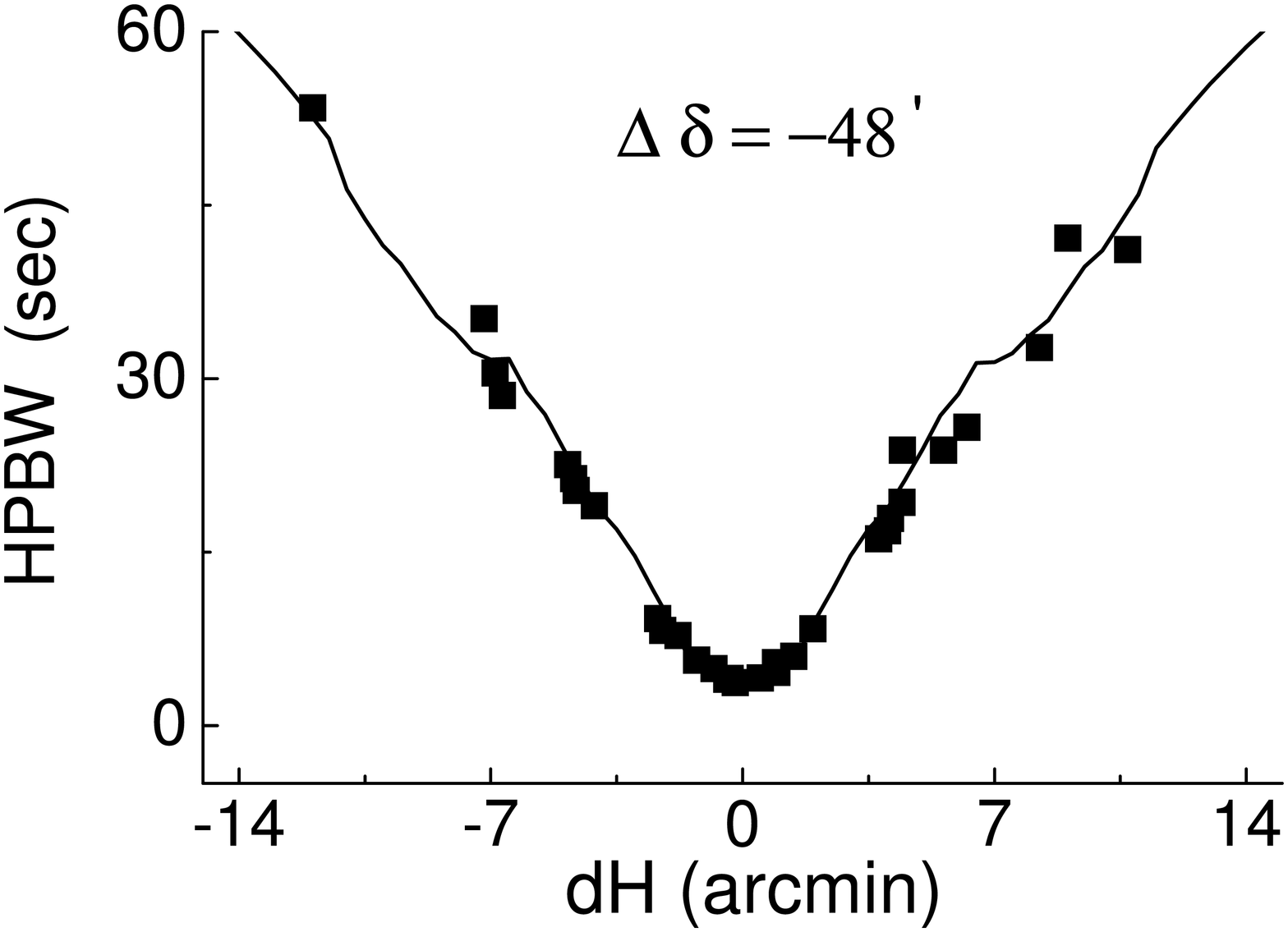}
\includegraphics[angle=-90,width=0.3\textwidth,clip]{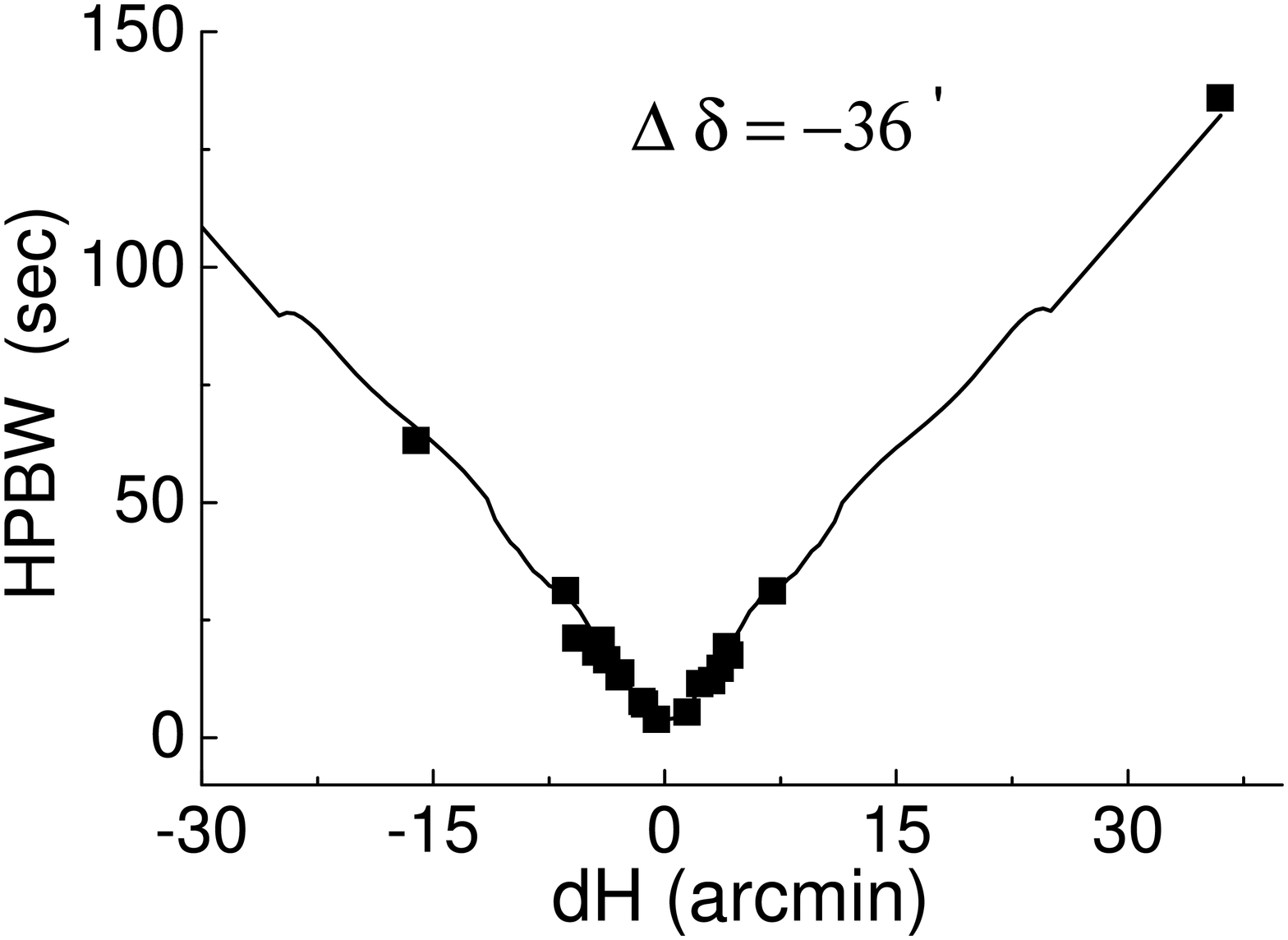}
\includegraphics[angle=-90,width=0.3\textwidth,clip]{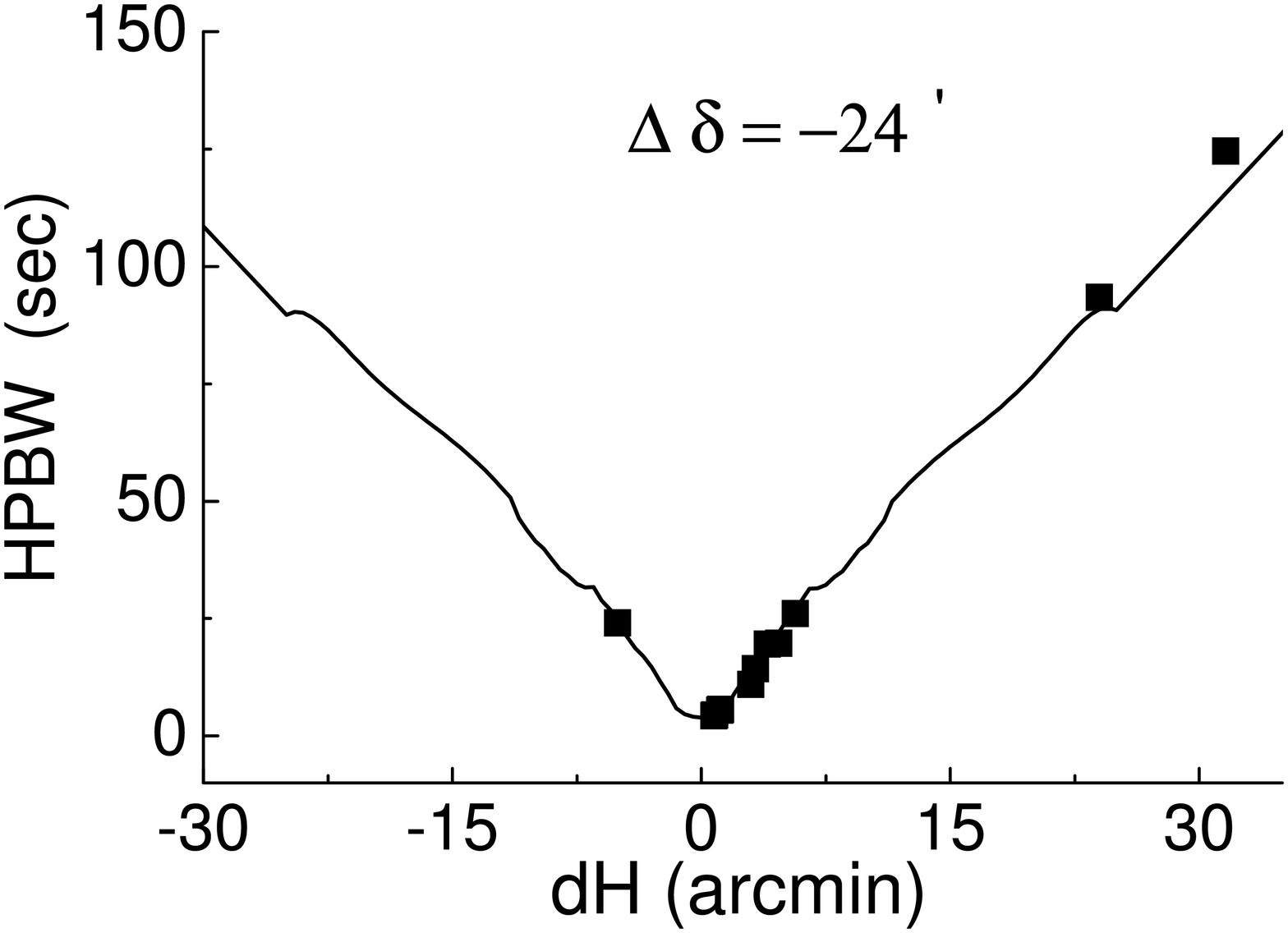}
}
\hbox{
\includegraphics[angle=-90,width=0.3\textwidth,clip]{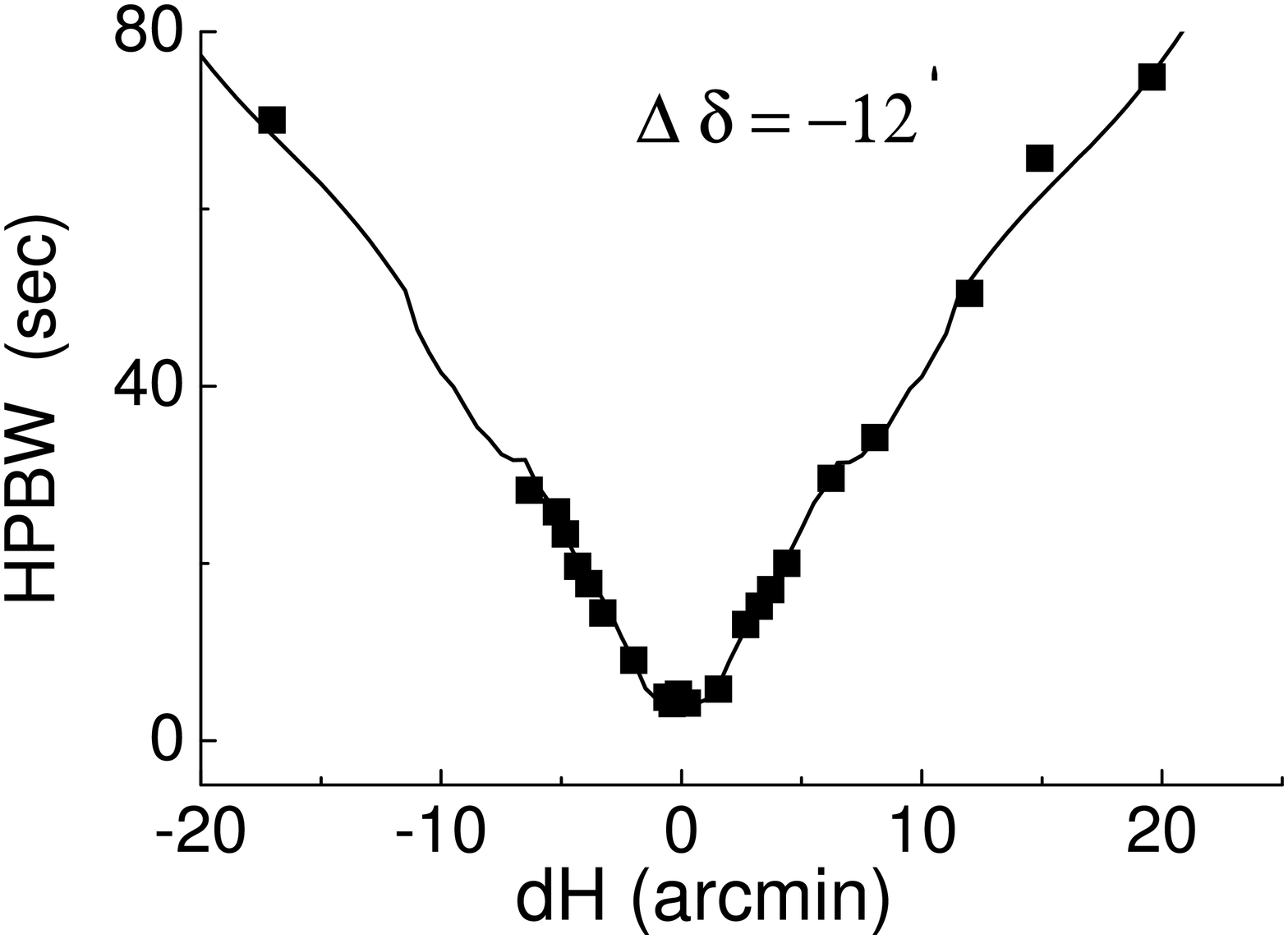}
\includegraphics[angle=-90,width=0.3\textwidth,clip]{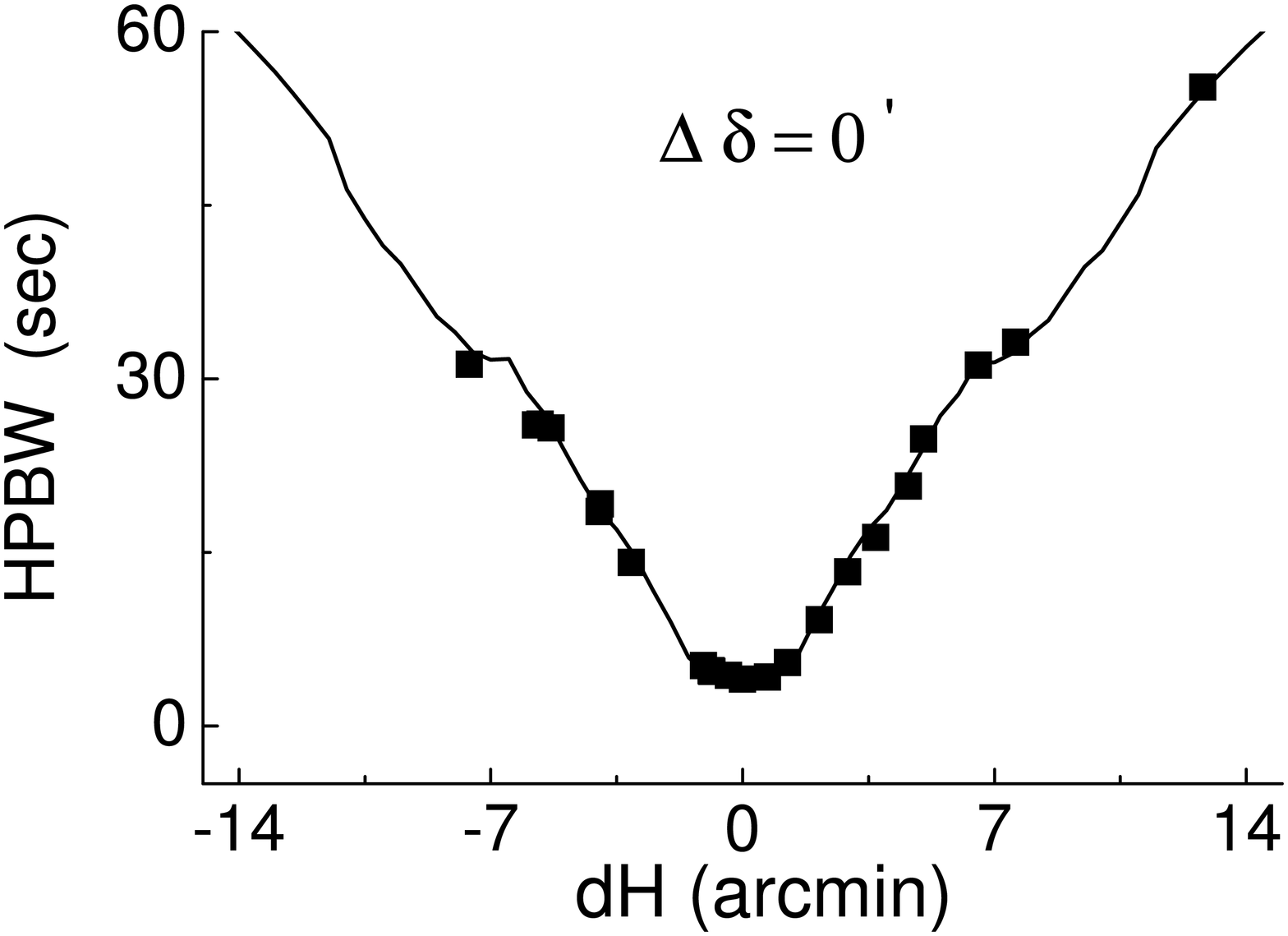}
\includegraphics[angle=-90,width=0.3\textwidth,clip]{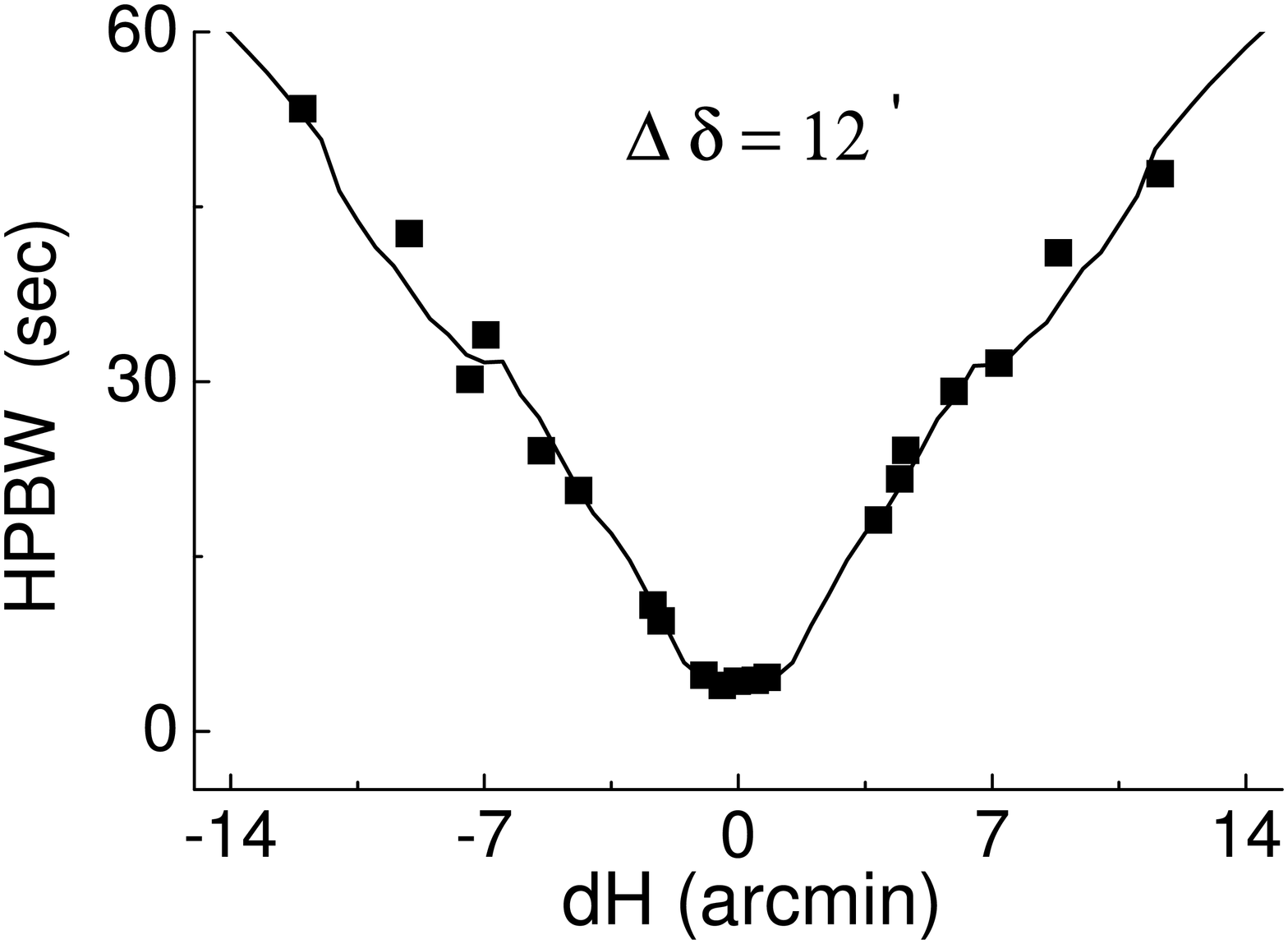}
}
\hbox{
\includegraphics[angle=-90,width=0.3\textwidth,clip]{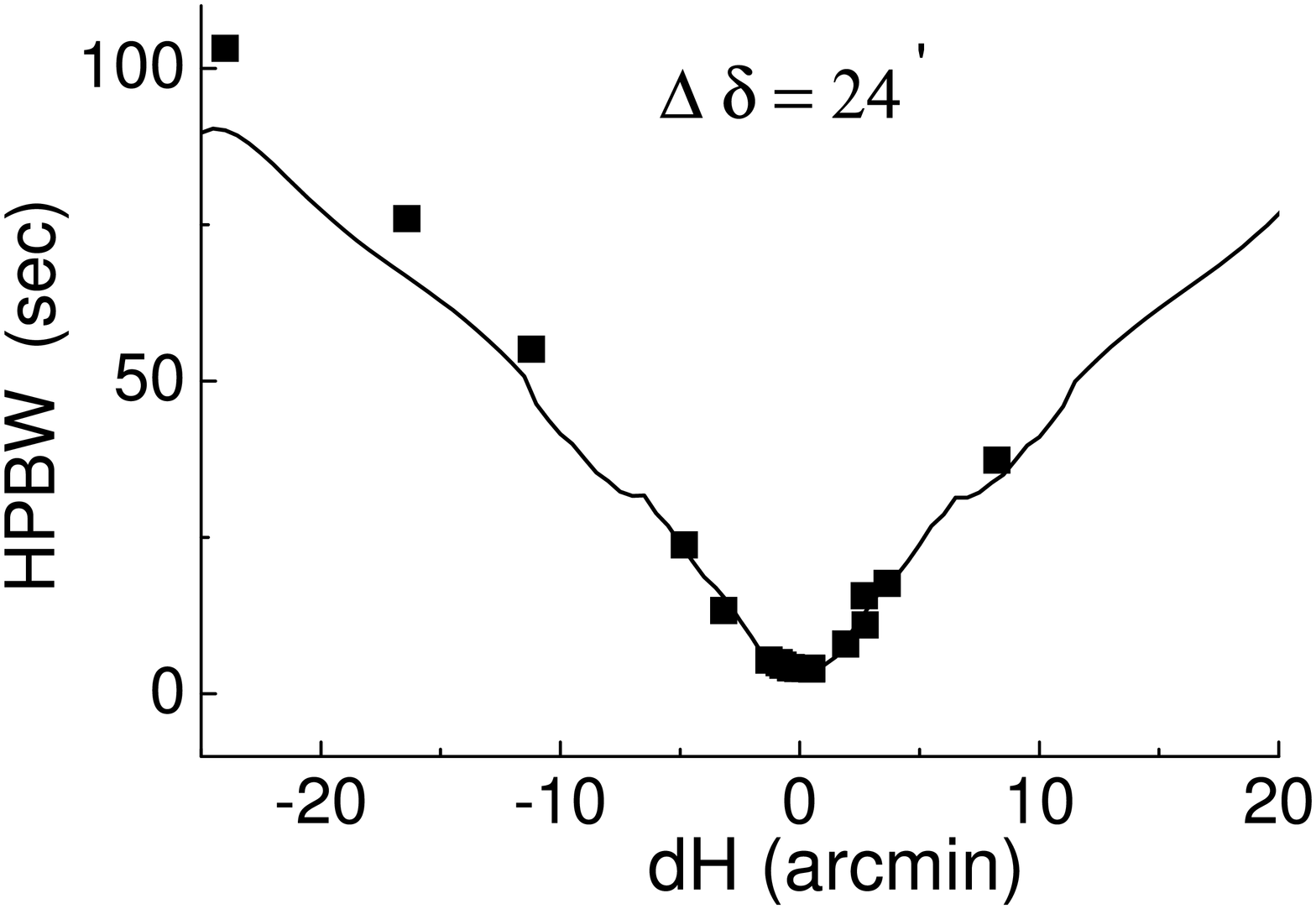}
\includegraphics[angle=-90,width=0.3\textwidth,clip]{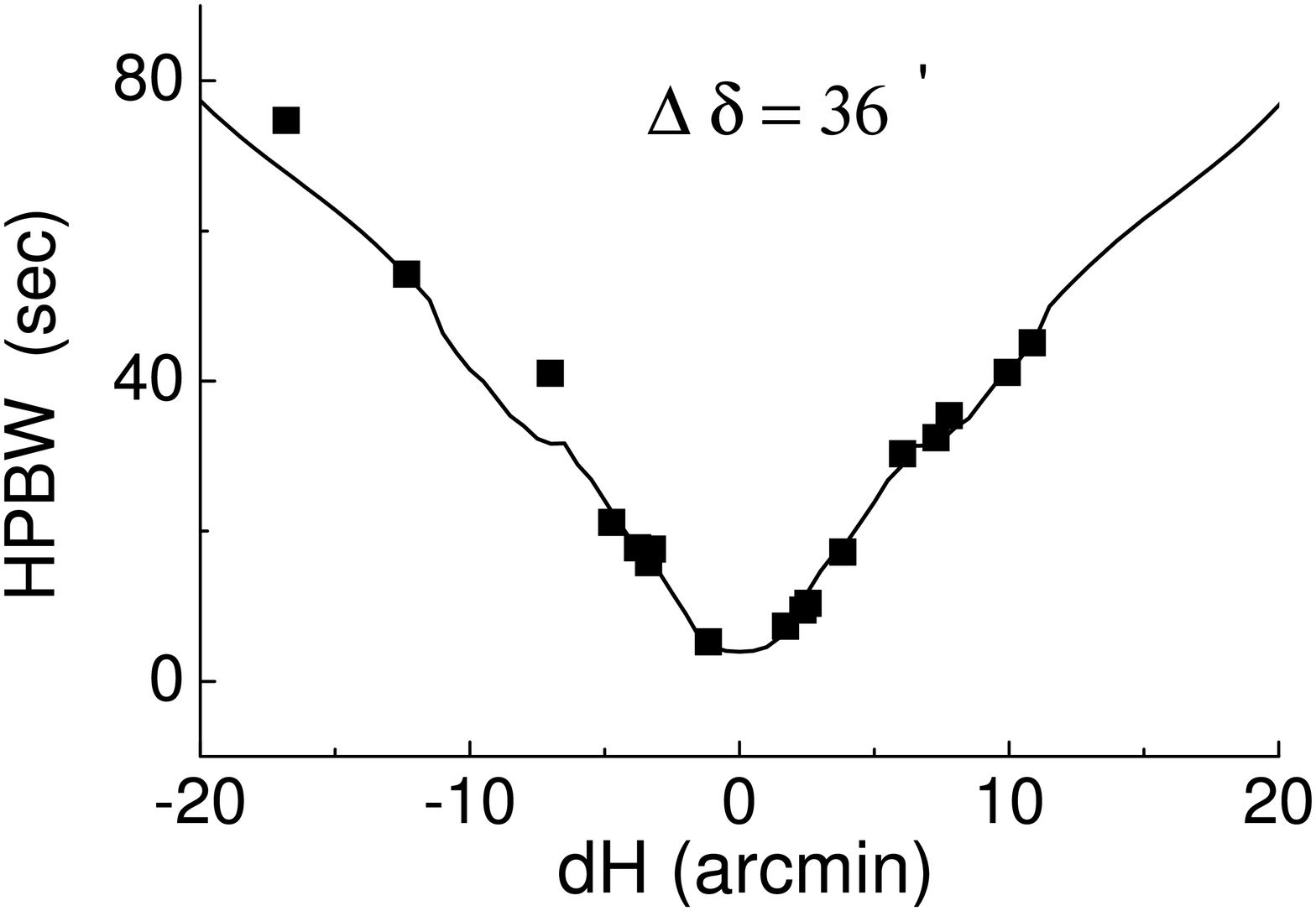}
\includegraphics[angle=-90,width=0.3\textwidth,clip]{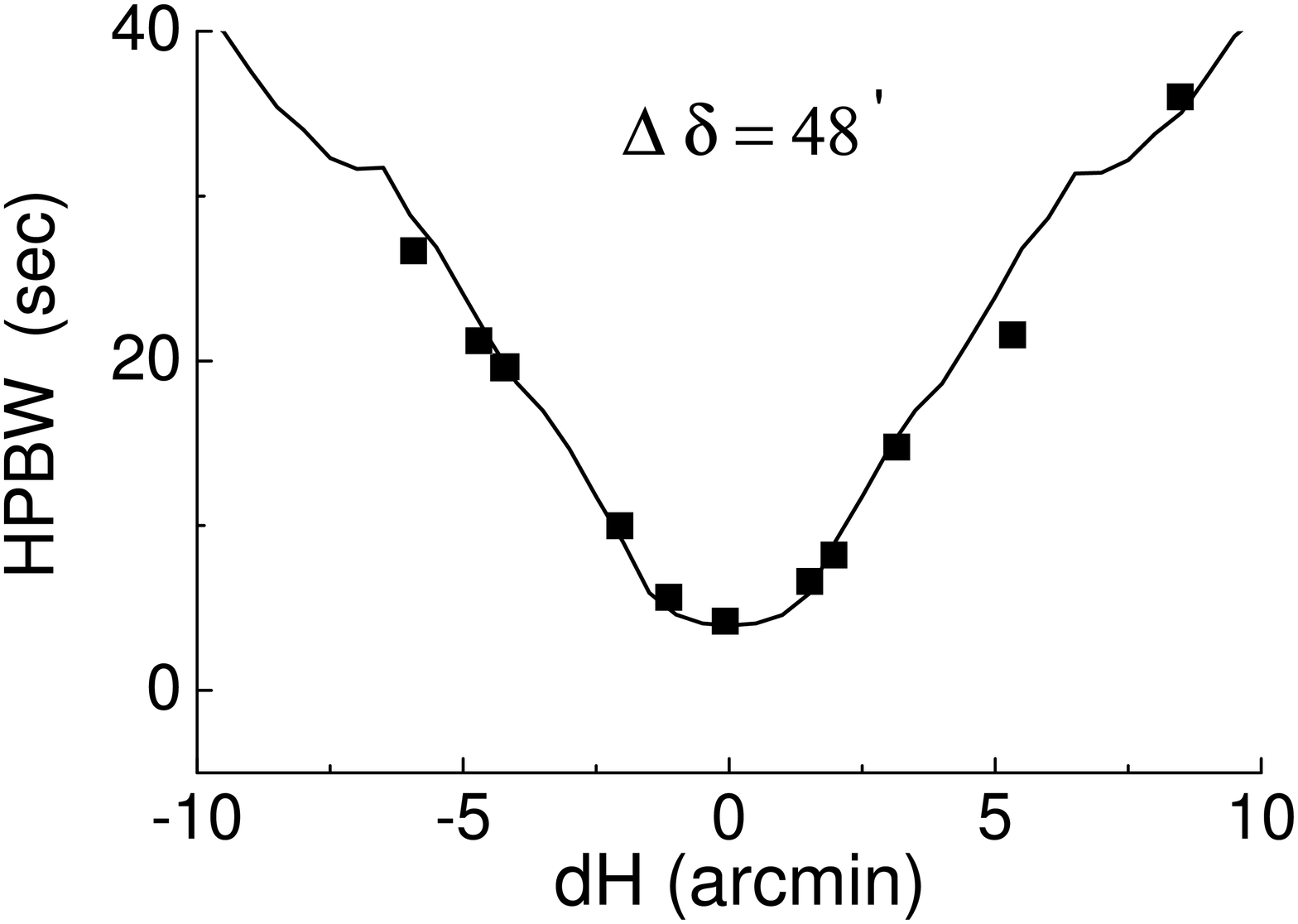}
} } }
%\setcaptionmargin{0mm}
%\captionstyle{normal}
\caption{ The
dependences of the half-widths of the  power beam pattern on  $dH$
obtained from the results of observations of  the sample of
NVSS sources with the $\lambda7.6$-cm fluxes $P > 80$\,mJy in nine
bands of the RZF survey: the central band $Dec_{2000}=41^o30'42''$
($\Delta\delta=0'$) and the bands $\Delta\delta=~\pm12',~\pm24',
~\pm36,~\pm48'$ apart from the central band. The filled squares
show the experimental data points and the solid lines show the
computed dependences. (The 2002 observing set). }
\label{fig4:Majorova_n}
\end{figure*}

\begin{figure*}[t]
%\onelinecaptionstrue
\centerline{
\vbox{
\hbox{
\includegraphics[angle=-90,width=0.3\textwidth,clip]{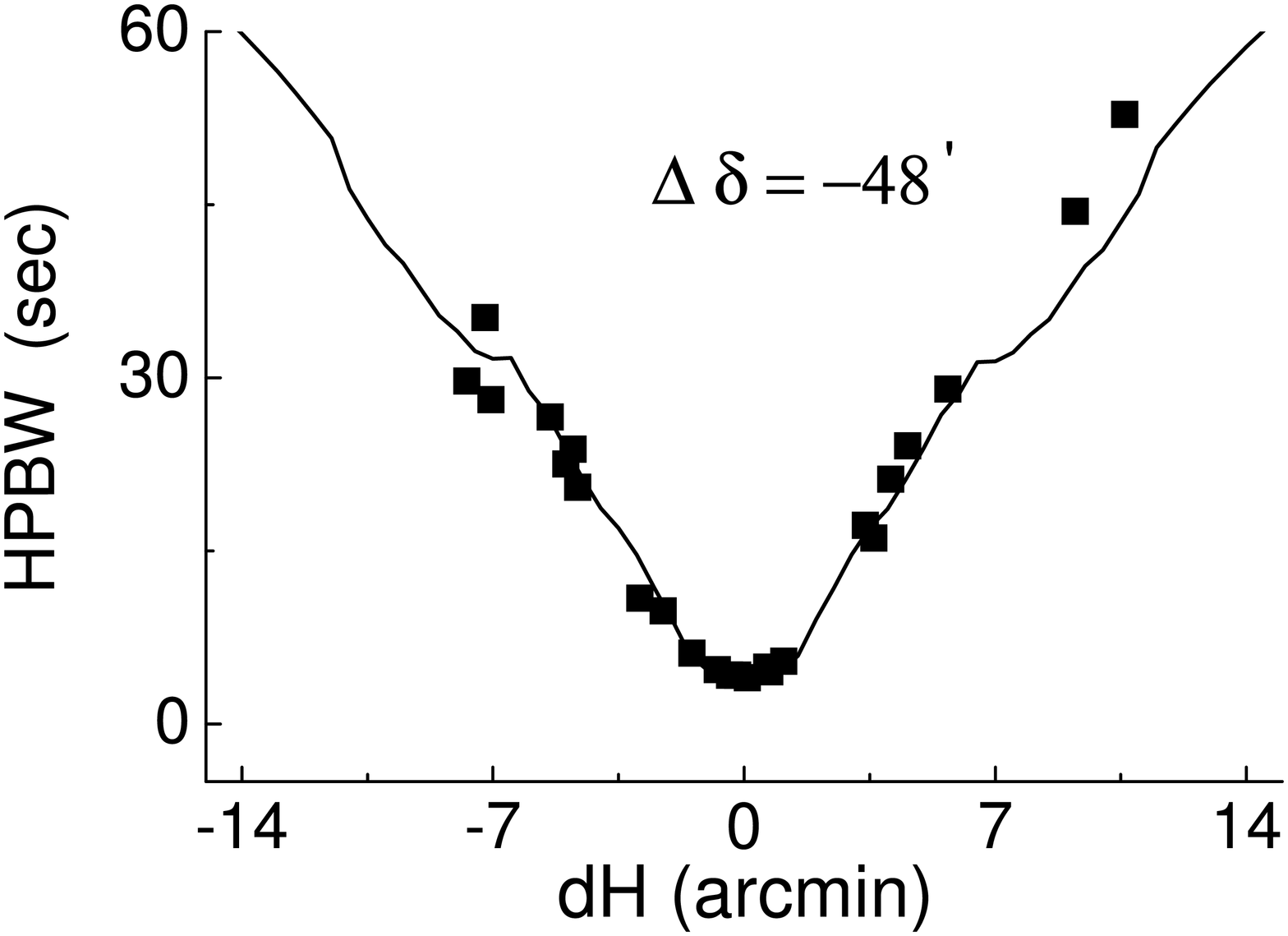}
\includegraphics[angle=-90,width=0.3\textwidth,clip]{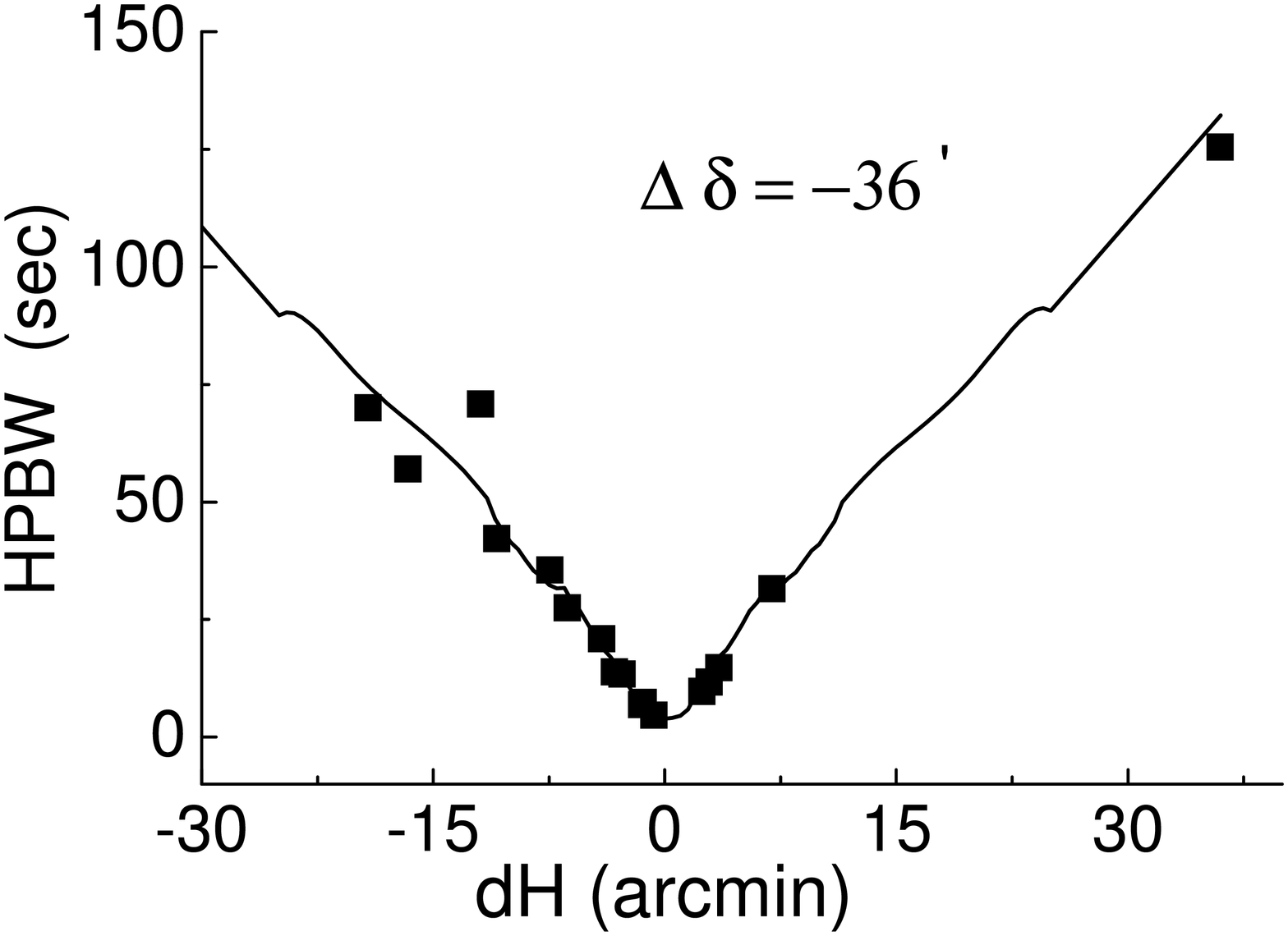}
\includegraphics[angle=-90,width=0.3\textwidth,clip]{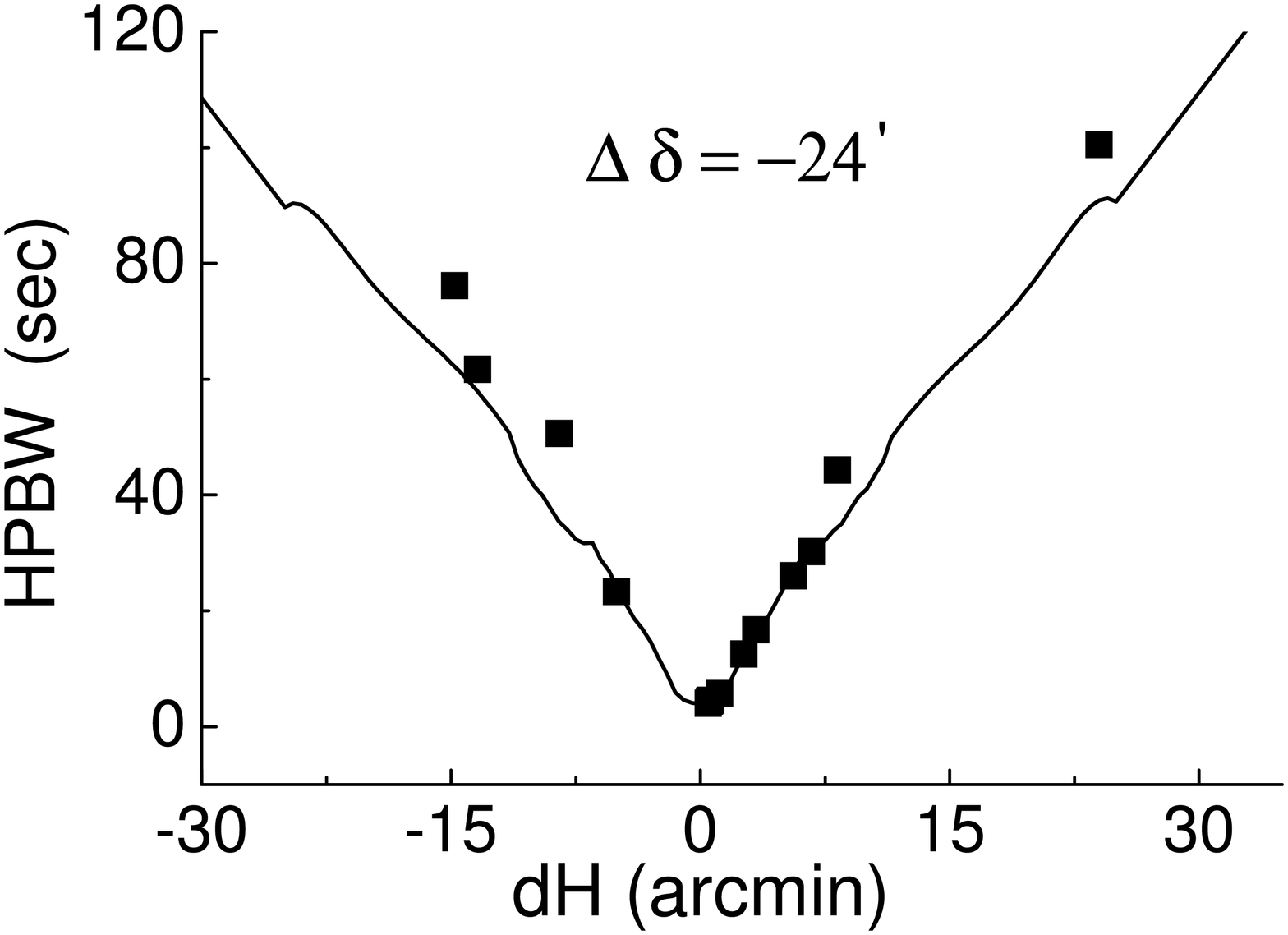}
}
\hbox{
\includegraphics[angle=-90,width=0.3\textwidth,clip]{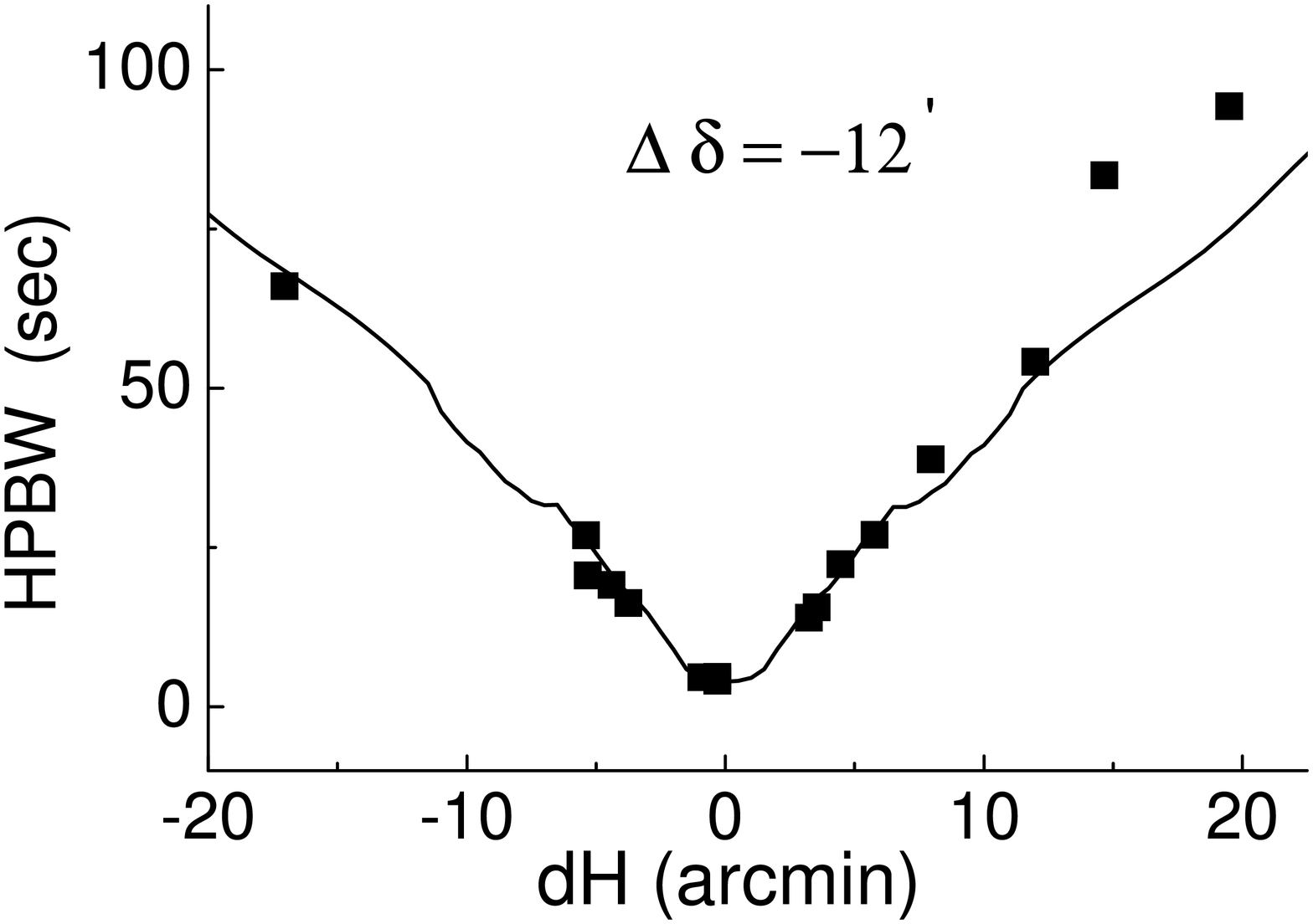}
\includegraphics[angle=-90,width=0.3\textwidth,clip]{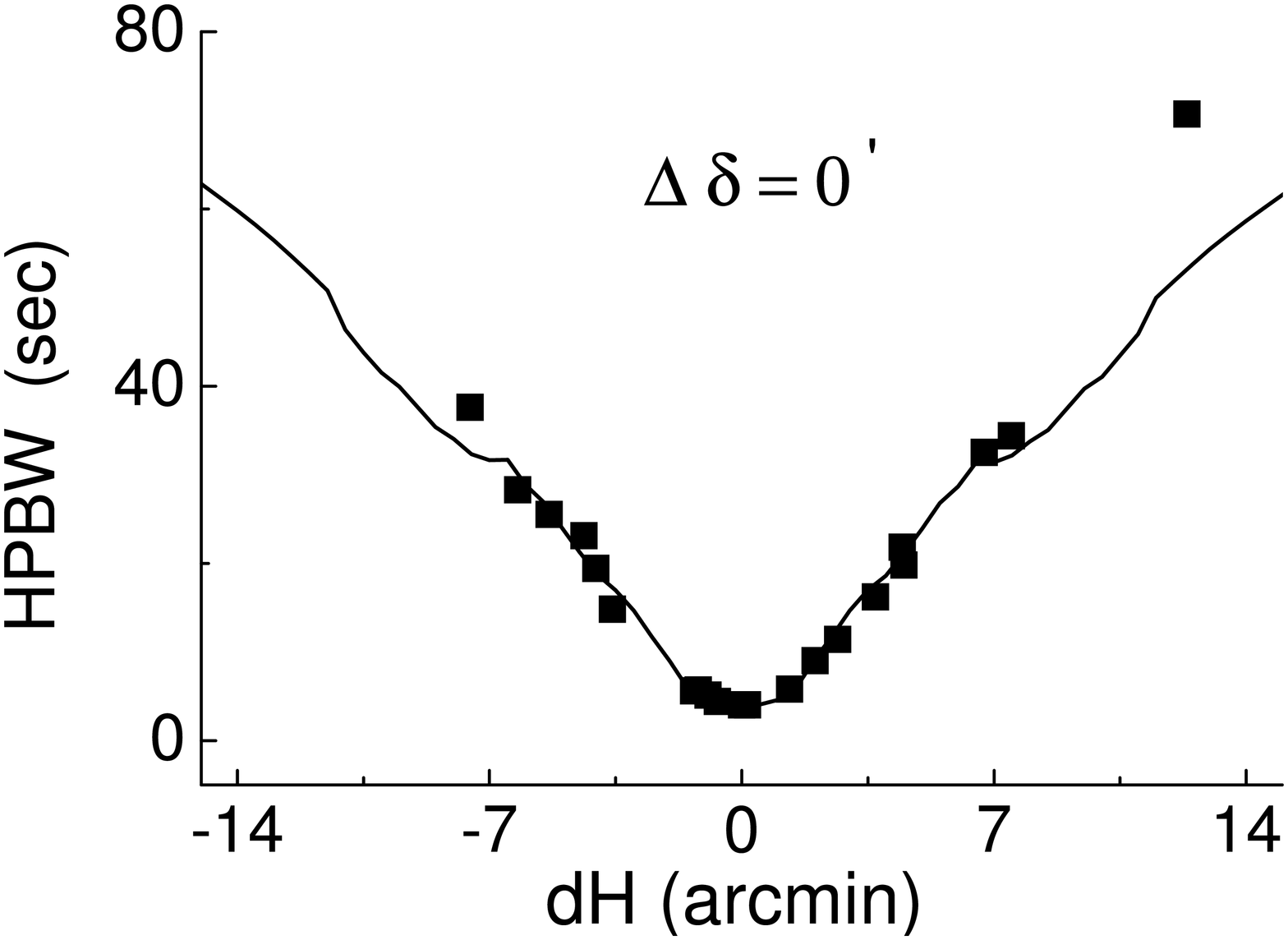}
\includegraphics[angle=-90,width=0.3\textwidth,clip]{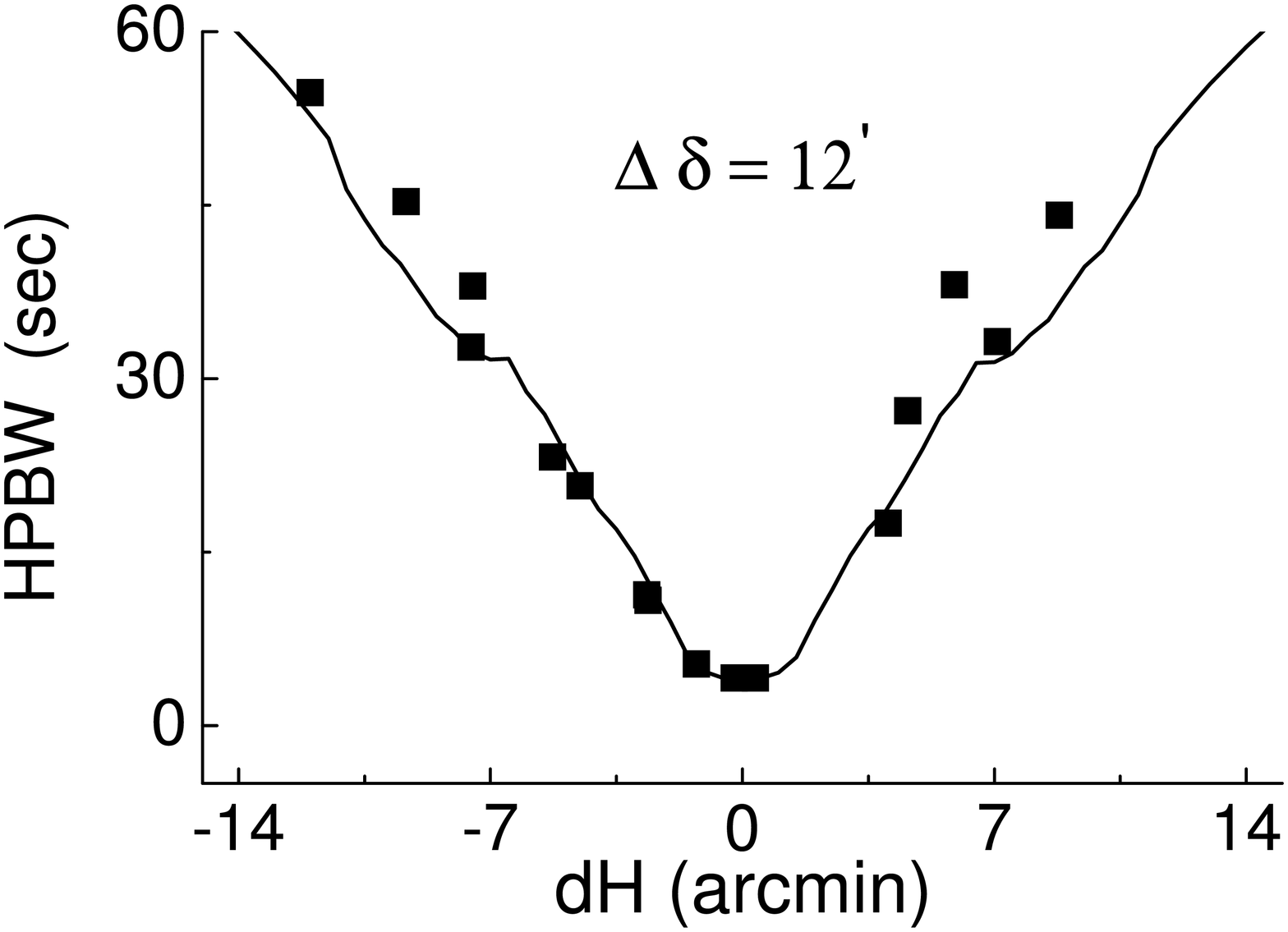}
}
\hbox{
\includegraphics[angle=-90,width=0.3\textwidth,clip]{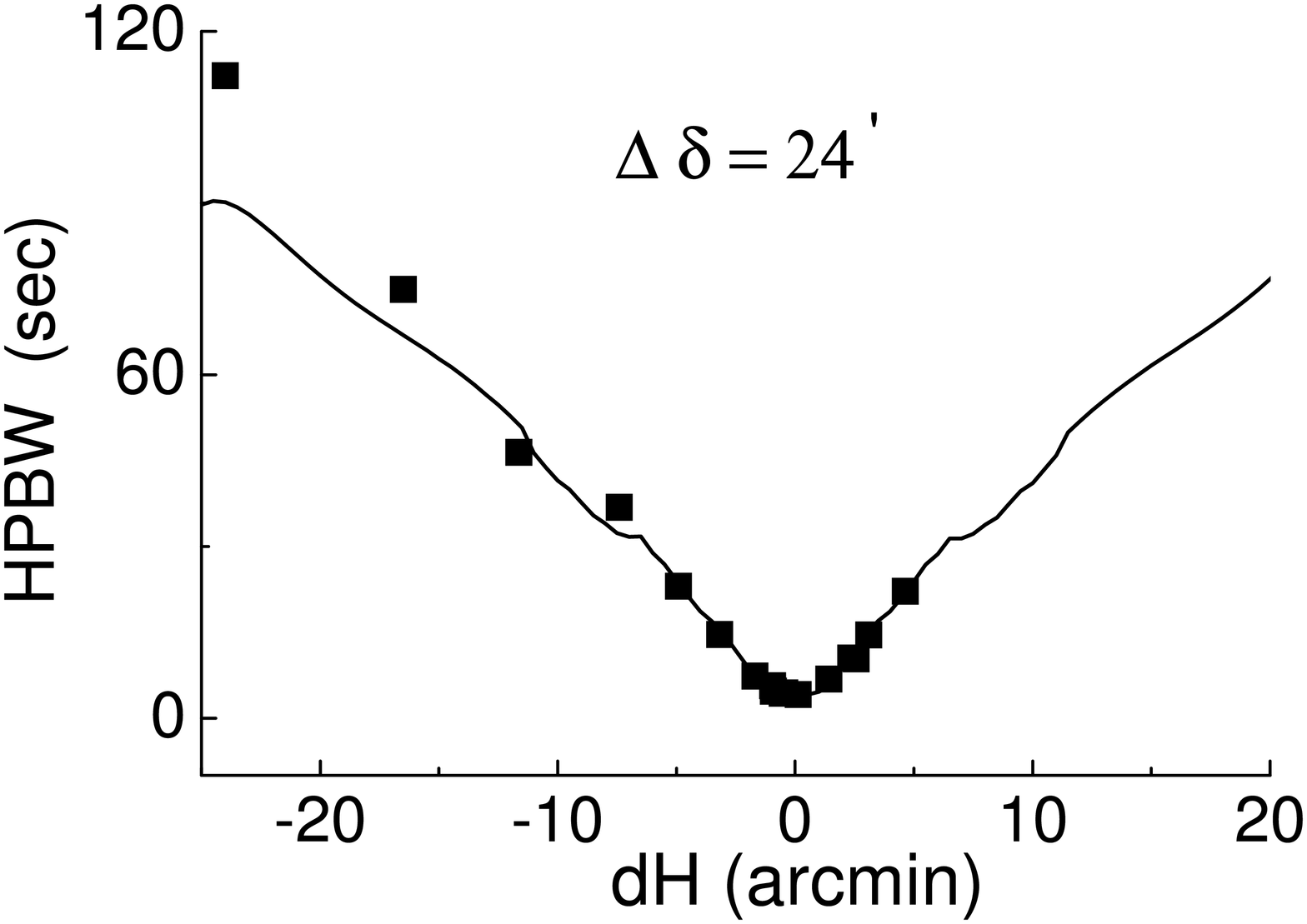}
\includegraphics[angle=-90,width=0.3\textwidth,clip]{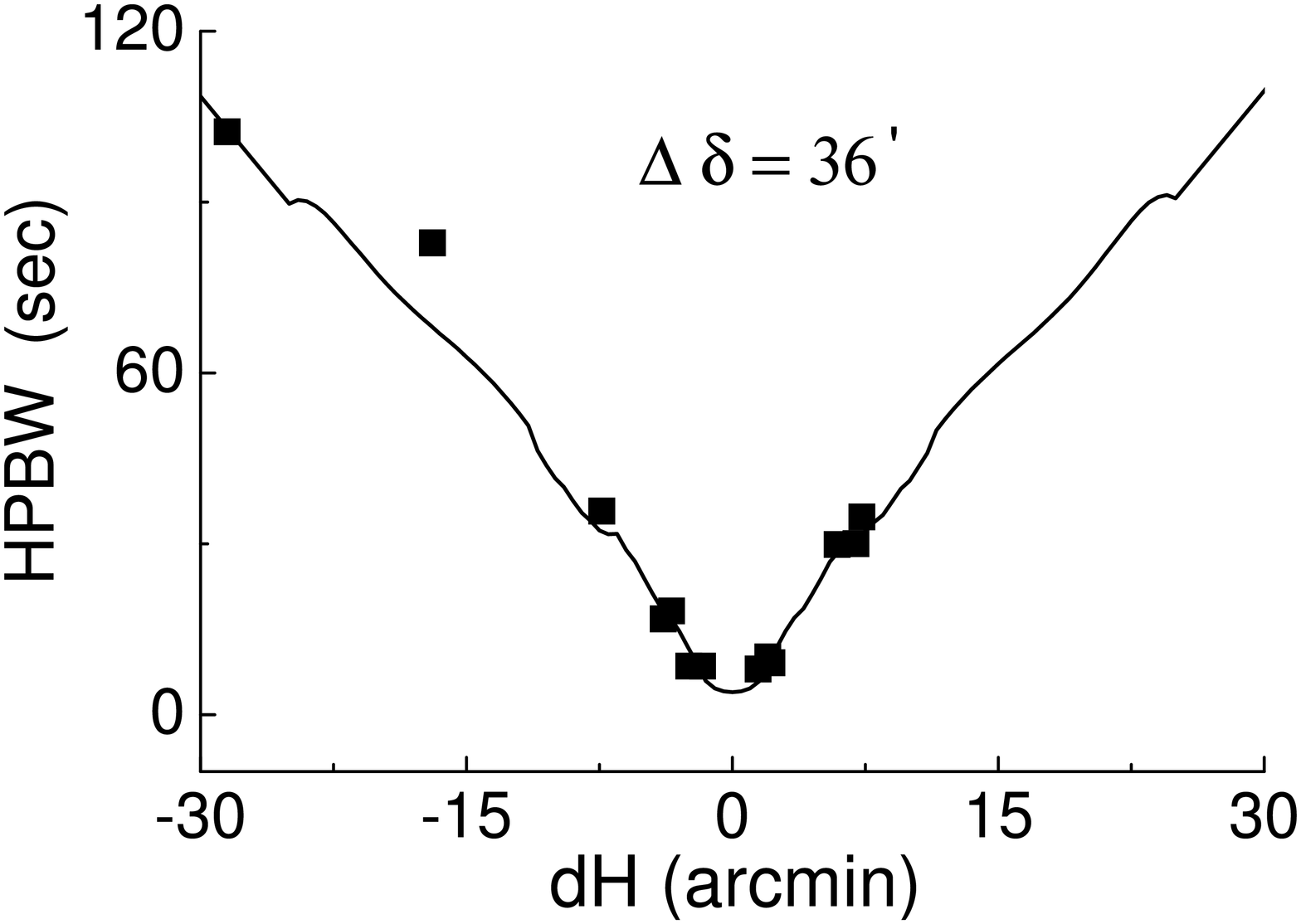}
\includegraphics[angle=-90,width=0.3\textwidth,clip]{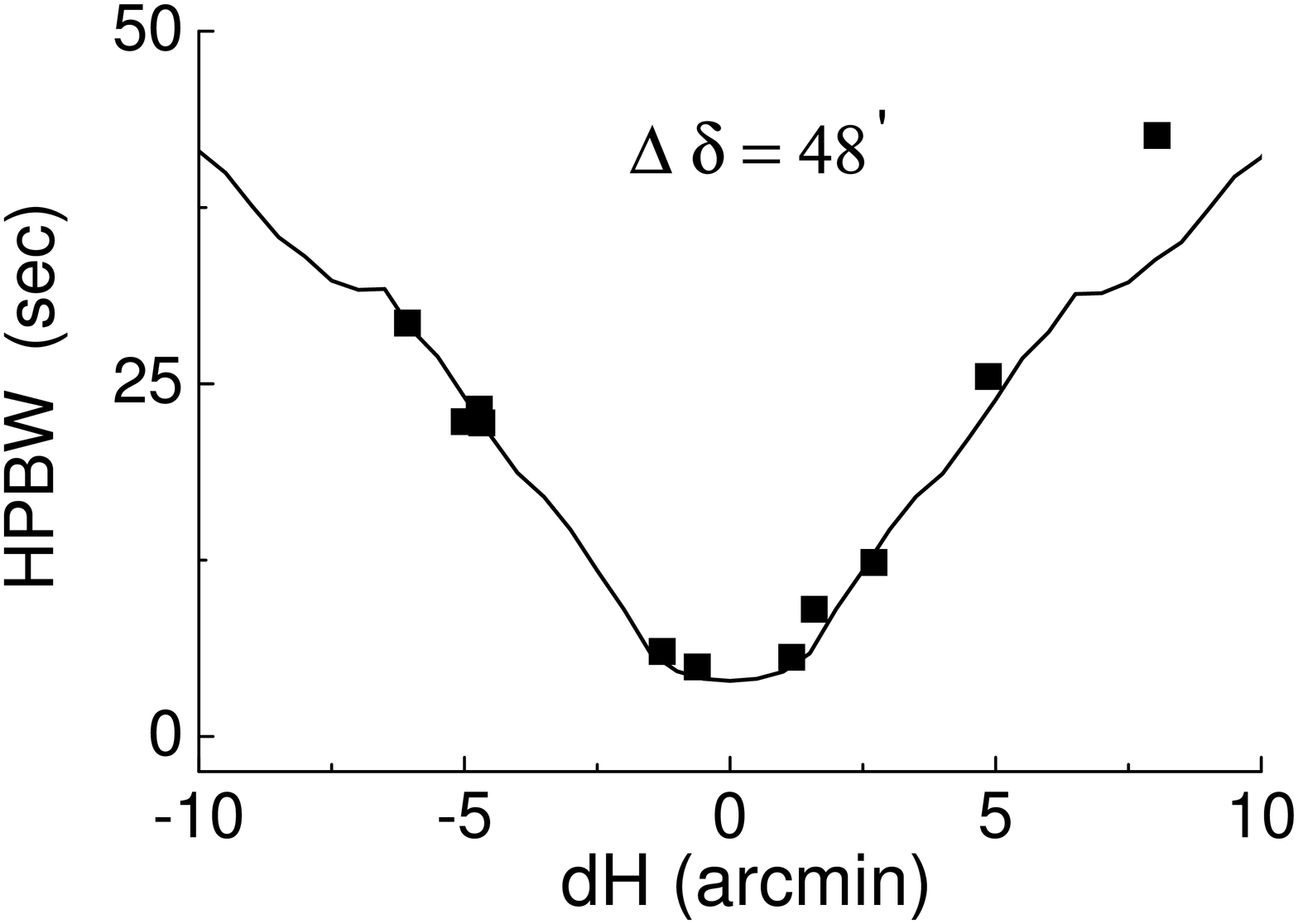}
} } }
%\setcaptionmargin{0mm}
%\captionstyle{normal}
\caption{ Same
as Fig.~\ref{fig4:Majorova_n}, but for the 2003 observing set. }
\label{fig5:Majorova_n}
\end{figure*}

The filled squares in
Fig.~\ref{fig2:Majorova_n},~\ref{fig3:Majorova_n} show the
experimental vertical power  patterns constructed based on the
results of observations of the sources during the 2002 and 2003
sets in each of the nine bands of the survey: in the central band
($\Delta\delta=0$) and in the bands  $\Delta\delta=\pm12',
\pm24', \pm36', \pm48'$ apart from the central band. The solid
lines show the computed power beam patterns.
Figures~\ref{fig4:Majorova_n} and~\ref{fig5:Majorova_n} show the
half-widths of the power beams ($HPBW$) plotted versus $dH$
for the  2002 and 2003 sets, respectively (the filled squares and
solid lines show the experimental  $HPBW$ values and theoretical
$HPBW(dH)$ curves, respectively). As is evident from the plots
shown, experimental data points are, on the whole, close to the
computed curves. Note that the points corresponding to the
experimental half-widths of the PB fit the
theoretical curves better than the experimental data points of
the vertical power  patterns ($F_{v}$). This is most likely
due to the errors of the determination of the fluxes of the
sources studied and also to the variations of the effective
antenna area during observations. Note also that the deviation of
experimental PB from the corresponding
theoretical one differs in different survey bands.
And, finally, in both observing sets experimental $F_{v}$ exceed
the corresponding computed values in the domain of positive $dH$.

\begin{figure*}[]
%\onelinecaptionsfalse
\centerline{
\hbox{
\includegraphics[angle=-90,width=0.4\textwidth,clip]{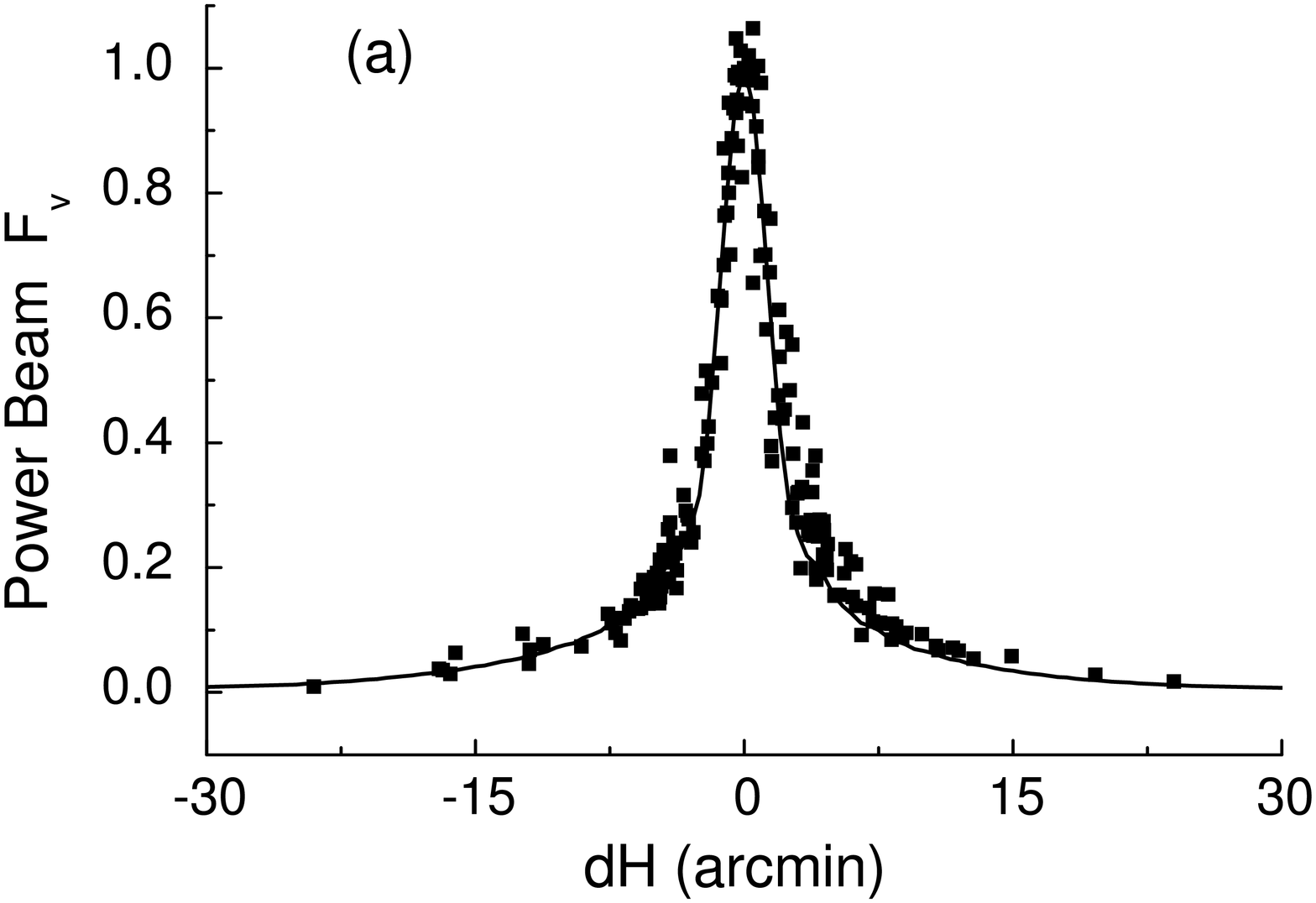}
\includegraphics[angle=-90,width=0.4\textwidth,clip]{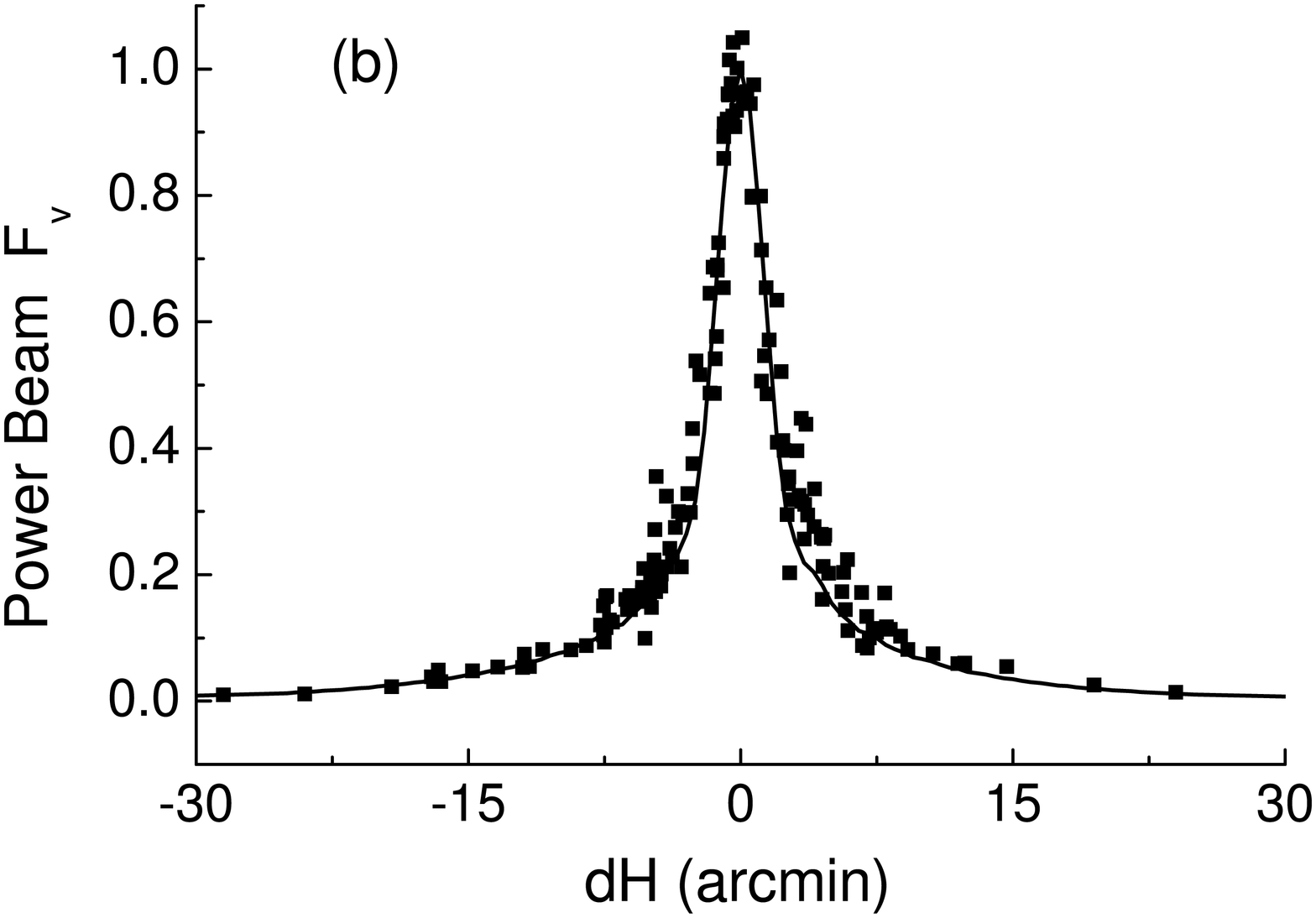}
} }
%\setcaptionmargin{10mm}
%\captionstyle{normal}
\caption{
Vertical PB constructed  using  the results of observations
of  sources in all nine bands of the RZF survey:
$Dec_{2000}=41^o30'42''\pm12'n, ~n=0,1,2,3,4$. (a) the 2002
set and (b) the 2003 set. The filled squares show the
experimental data points of the PB and the solid
lines show the computed PB. }
\label{fig6:Majorova_n}
\end{figure*}

We used the fact that the shape of the power beam pattern at the
$\lambda7.6$ cm wavelength remains virtually unchanged within
the  1$\degr$ interval of source elevation to construct the
dependences  $F_{v}(dH)$ and $HPBW(dH)$ common for all bands. The
filled squares in Fig.~\ref{fig6:Majorova_n} show the
experimental vertical power beam  $F_{v}(dH)$ obtained  from observations of sources out of
 all the
nine bands of the survey and the solid lines show the computed
PB ((a)  the 2002 set and (b)
 the 2003
set). When constructing the power beam  we set the
$(P/T_{a})_{dH=0}$ ratios equal to 2.3 and 2.17 for the  2002 and
2003 sets, respectively. Figure~\ref{fig8:Majorova_n} shows the
experimental   $HPBW(dH)$ dependences obtained from
the results of observations of sources in all
the nine survey bands
(filled squares). The solid lines show the computed
$HPBW(dH)$ dependences.

\begin{figure*}[]
%\onelinecaptionstrue
\centerline{
\hbox{
\includegraphics[angle=-90,width=0.4\textwidth,clip]{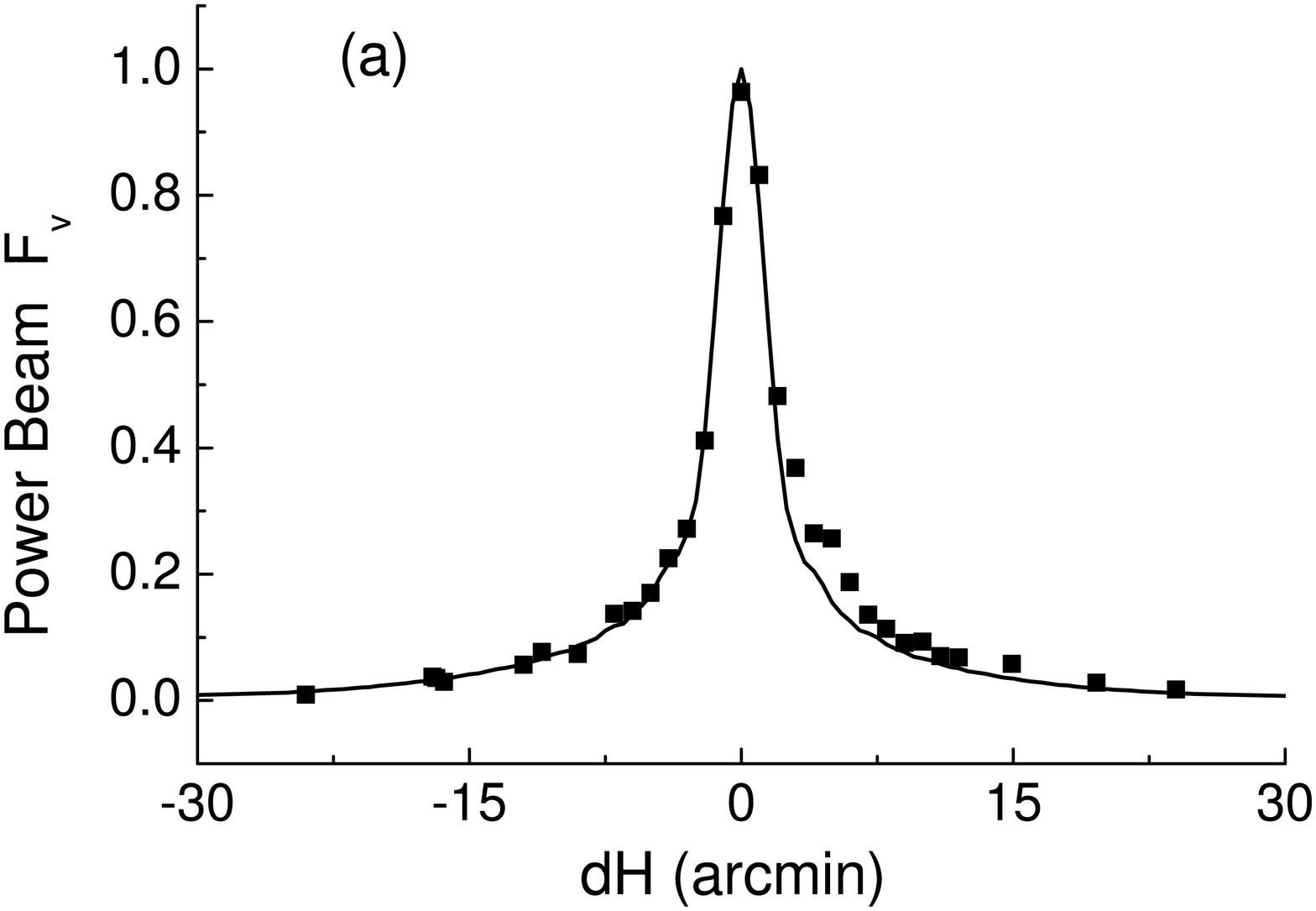}
\includegraphics[angle=-90,width=0.4\textwidth,clip]{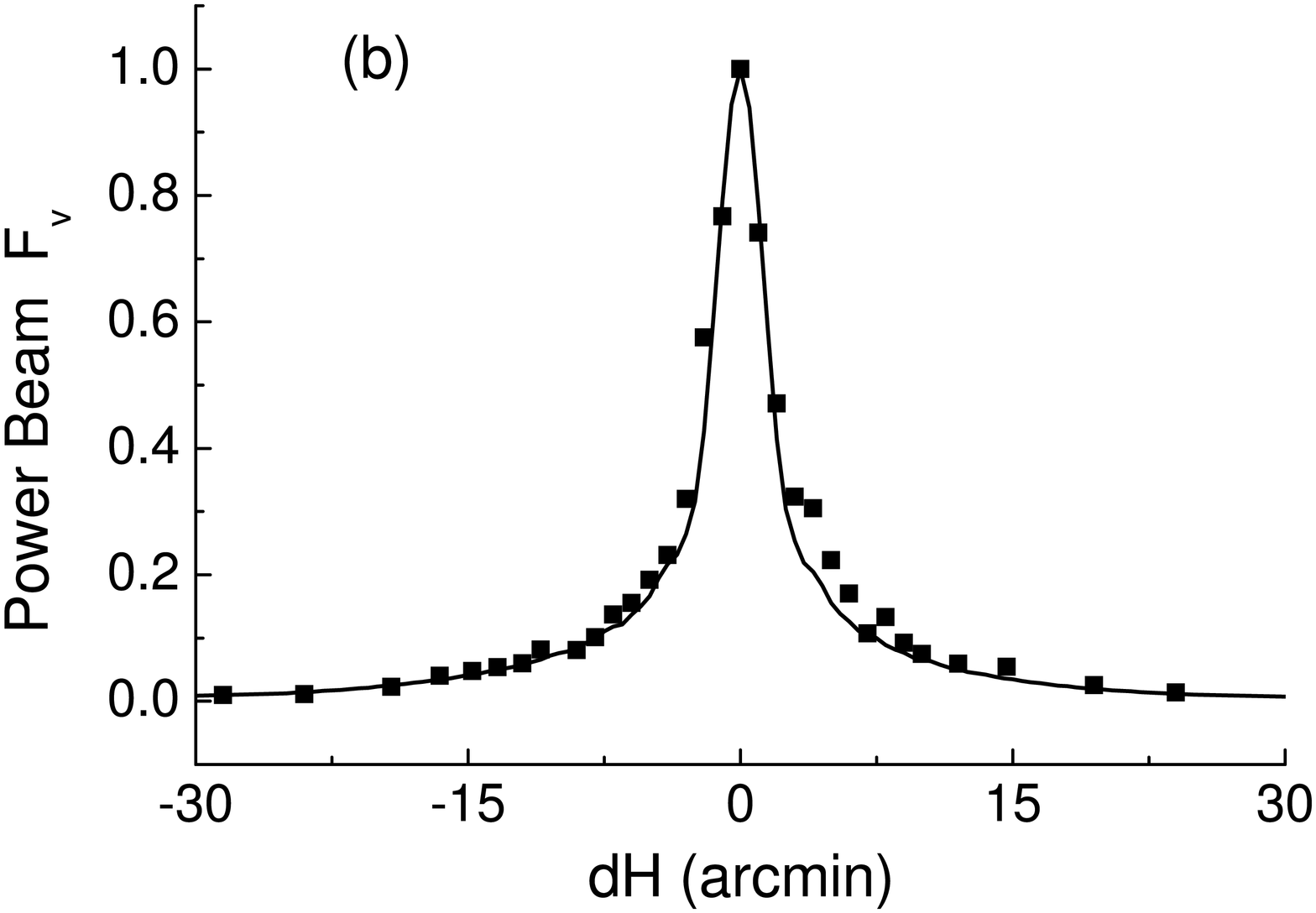}
} }
%\setcaptionmargin{0mm}
%\captionstyle{normal}
\caption{ Same
as Fig.~\ref{fig6:Majorova_n}, but with subsequent
``densification''\, of data points (averaging in the $1'$  $dH$
interval). } \label{fig7:Majorova_n}
\end{figure*}

A rather large scatter of experimental PB data
points is immediately evident from a comparison of the
experimental and theoretical $F_{v}(dH)$ dependences shown in
Fig.~\ref{fig6:Majorova_n}. Moreover, both sets exhibit somewhat
asymmetric power beam. The standard error of the
deviations of the experimental PB from the
computed PB determined over the entire sample of
observed sources is equal to $\sigma=0.071\pm0.005$ and
$\sigma=0.068\pm0.005$ for the 2002 and 2003 sets, respectively.
As for the $HPBW(dH)$ dependences shown in
Fig.~\ref{fig8:Majorova_n}, experimental data points of the 2002
set agree very well with the theoretical curves throughout the
entire interval of  $dH$ values, whereas the experimental data
for the 2003 set agree with the theoretical curves only in the
$-10'< dH < 15'$ interval. To compare the experimental and
computed PB half-widths, we compute the
coefficients $K_{HPBW}$ that are equal to the ratio of the
experimental-to-computed power beam  half-widths. The mean
coefficients $K_{HPBW}$ averaged over the entire sample of
observed sources are equal to $1.00\pm0.08$ and $1.04\pm0.12$ for
the 2002 and 2003 sets, respectively.

Note that the main contributors to the experimental PBs
 shown in Fig.~\ref{fig6:Majorova_n} are the sources
observed in side bands of the survey where measurements were much
less sensitive than in the central band. These very sources
provide the greatest scatter of data points on the experimental
power beam. Recall that no more than $3\div5$ records from
each source were averaged in the bands with
$\Delta\delta=\pm12'n$ ($n=1,2,3,4$), whereas $20 \div 35$
records were averaged for each source in the central band
($\Delta\delta=0$). We therefore constructed, in addition to the
common PBs for the nine bands, the experimental
vertical PBs constructed using observations made in
side bands of the survey exclusively. To reduce the scatter of
data points, we applied the so-called point ``densification''\,
procedure: we averaged the  $F_{v}$ values over some  $dH$
interval. The power beam patterns so constructed are shown in
Fig.~\ref{fig7:Majorova_n} by filled squares ((a)
--- the 2002 set and (b) --- the 2003 set). The standard error of
the deviations of the experimental power beam obtained from the
results of observations made in side bands of the survey with
averaging over $1'$-wide $dH$ intervals is equal to
$\sigma=0.030\pm0.005$ and $\sigma=0.035\pm0.006$ for the 2002
and 2003 sets, respectively. These values proved to be close to
the standard error of the deviations of the power beam pattern
constructed using the results of observations in the central band
of the survey (Fig.~\ref{fig9:Majorova_n})
($\sigma=0.033\pm0.007$ and $\sigma=0.050\pm0.009$ for the 2002
and 2003 sets, respectively).

\begin{figure*}[]
%\onelinecaptionsfalse
\centerline{
\hbox{
\includegraphics[angle=-90,width=0.4\textwidth,clip]{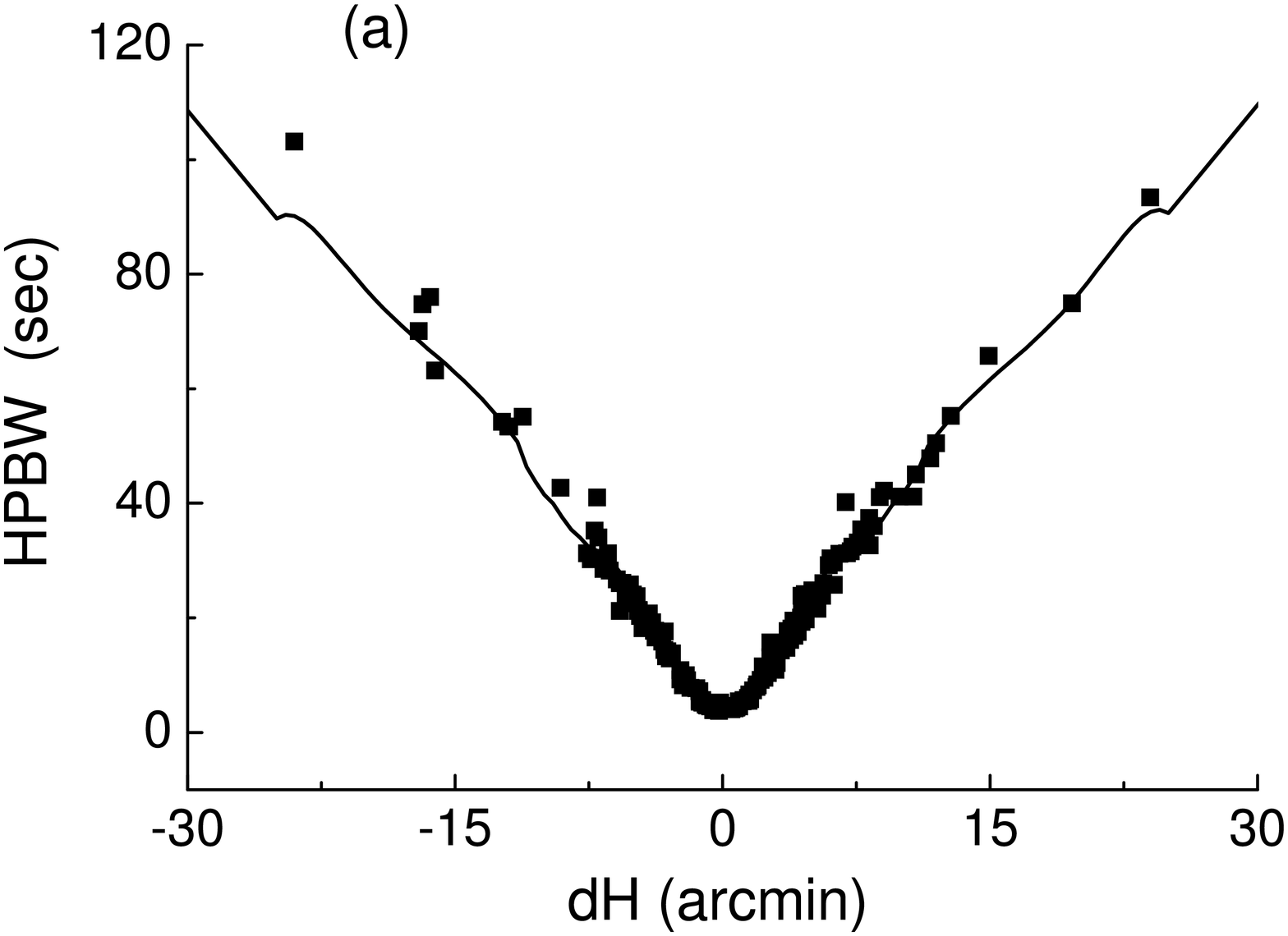}
\includegraphics[angle=-90,width=0.4\textwidth,clip]{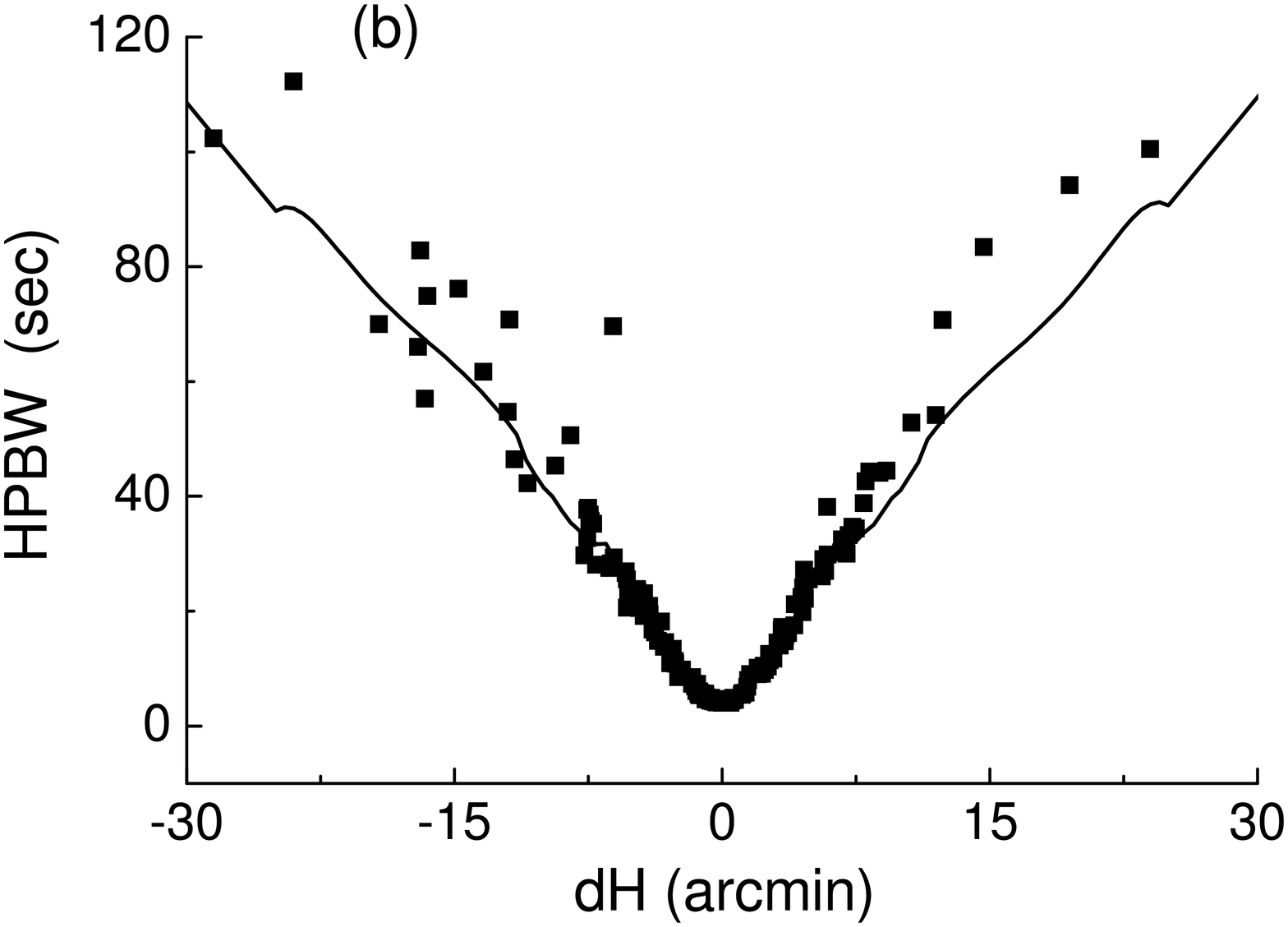}
} }
%\setcaptionmargin{10mm}
%\captionstyle{normal}
\caption{
Dependences of the half-width of the  PB
on  $dH$ obtained from the results of observations of sources in all
the nine RZF survey bands: (a) --- the 2002 set and (b) --- the 2003
set. The filled squares show the experimental data points and the
solid curves show the computed  $HPBW(dH)$ dependences. }
\label{fig8:Majorova_n}
\end{figure*}

In Figs.~\ref{fig9:Majorova_n} and \ref{fig10:Majorova_n} we
show, for comparison, the experimental and vertical PBs obtained from
 the results of observations in the central band and side bands
of the survey. To the latter, we applied the procedure of
``densification''\, of experimental points. The solid lines in the
figures show the computed PBs. For better
visualization, we plot the power beams in the same  $dH$
intervals. It is evident from the plots shown in
Figs.~\ref{fig7:Majorova_n} and \ref{fig10:Majorova_n} that the
experimental PSs obtained in both sets agree well
with the corresponding theoretical curves over the
entire range of  $dH$ values except for the $dH=3'\div5'$
interval, where the experimental PB runs somewhat
above the computed PB. The results of measurements
in the central band of the survey show similar deviations
(Fig.~\ref{fig9:Majorova_n}). These distortions of the power beam pattern
may be due to a systematic error in the elevation-angle
setting of the reflective elements of the primary mirror and to
large errors (``outliers'') of zero settings in the elevation
coordinate for a number of reflective panels. Such errors
appeared because of the wear of the reflective panel mechanisms
and had the same sign. See [\cite{mm2:Majorova_n}] for a
description of their effect on the power beam pattern.

\begin{figure*}[]
%\onelinecaptionsfalse
\centerline{
\hbox{
\includegraphics[angle=-90,width=0.4\textwidth,clip]{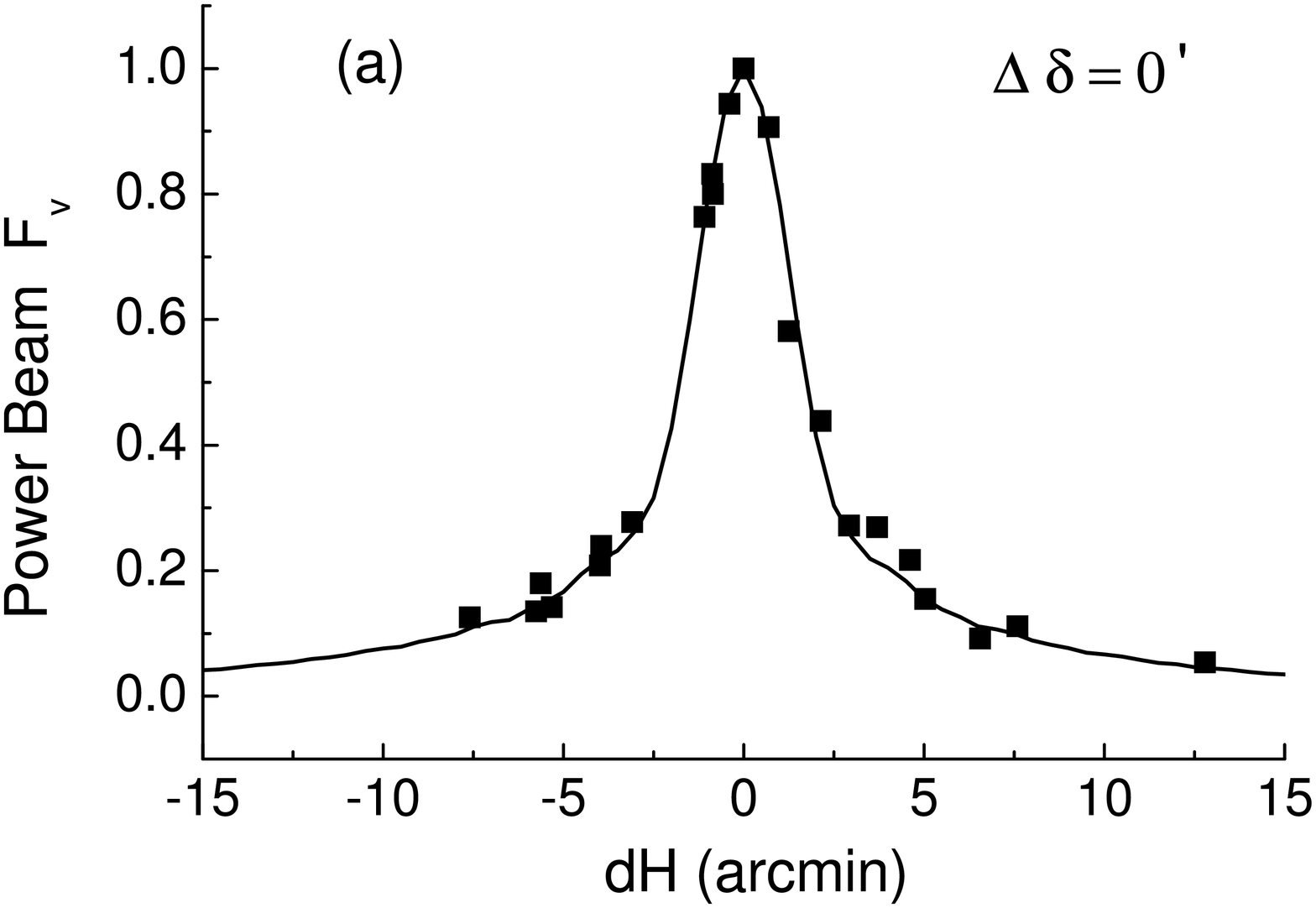}
\includegraphics[angle=-90,width=0.4\textwidth,clip]{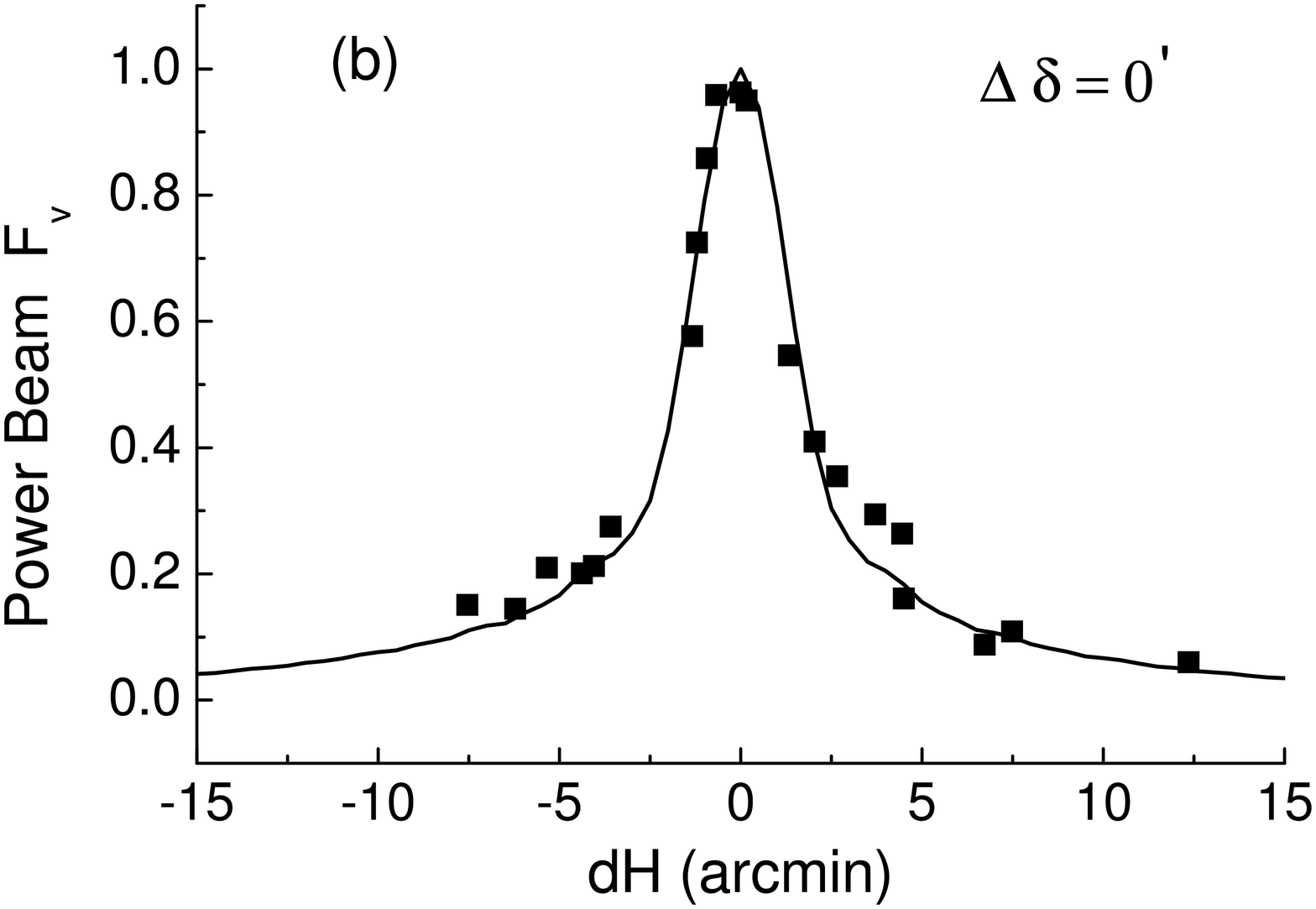}
} }
%\setcaptionmargin{10mm}
%\captionstyle{normal}
\caption{
Vertical PB constructed  using  the results of observations
of sources in the central band of the RZF survey
($Dec_{2000}=41^o30'42''$): (a) the 2002 set and (b) the 2003
set. The filled squares show the data points of the experimental
PB and the solid lines show the computed PB.
 } \label{fig9:Majorova_n}
\end{figure*}

\begin{figure*}[]
%\onelinecaptionsfalse
\centerline{
\hbox{
\includegraphics[angle=-90,width=0.4\textwidth,clip]{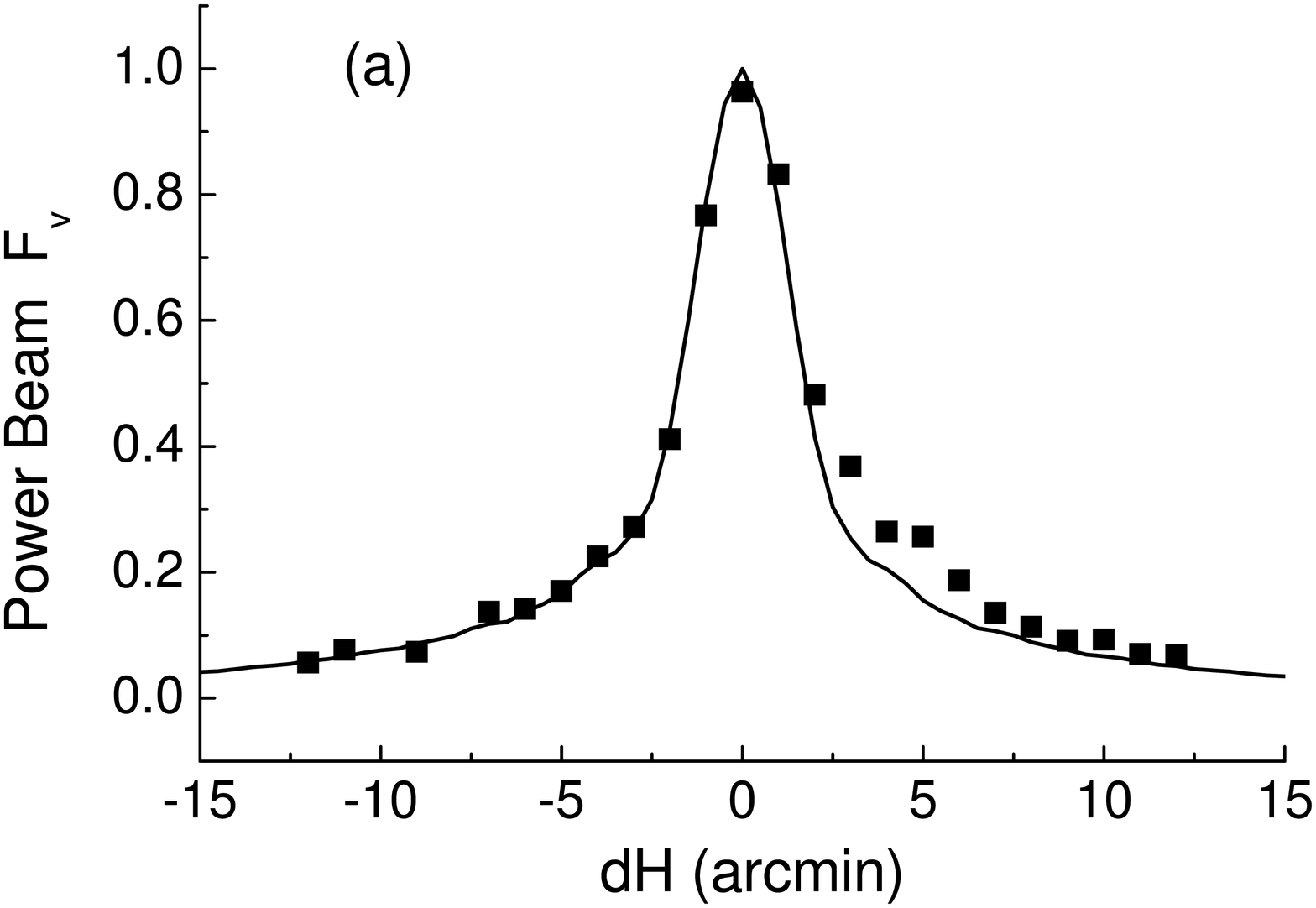}
\includegraphics[angle=-90,width=0.4\textwidth,clip]{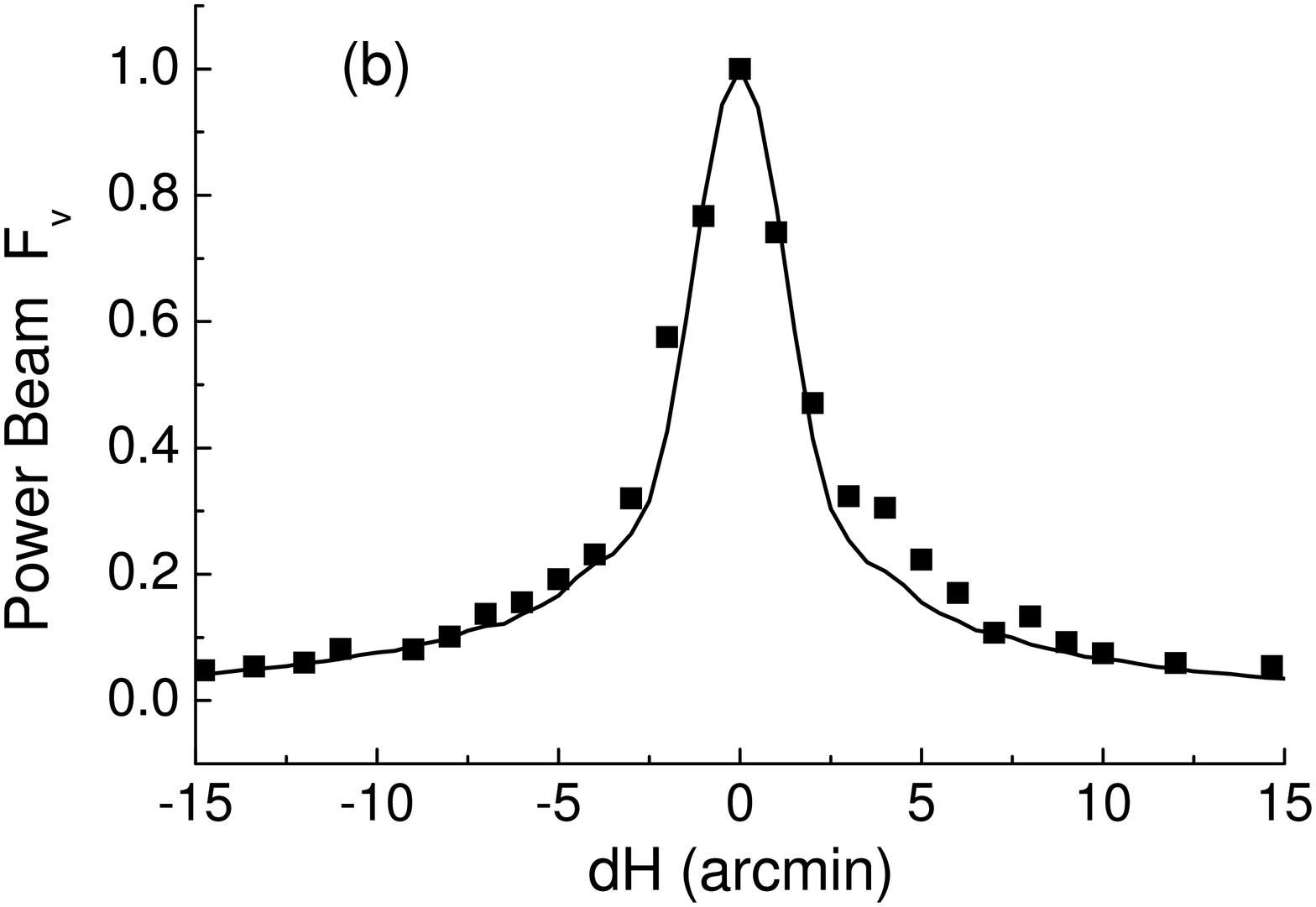}
} }
%\setcaptionmargin{10mm}
%\captionstyle{normal}
\caption{
Vertical PB constructed  using  the results of observations
of sources in the side bands of the RZF survey
($Dec_{2000}=41^o30'42''\pm12'n, ~n=1,2,3,4$) with subsequent
``densification''~ of points: (a) the 2002 set and (b) the 2003
set. The filled squares show the experimental PB
data points and the solid lines show the computed PB.
 } \label{fig10:Majorova_n}
\end{figure*}

\section{RESULTS OF THE POWER BEAM PATTERN MEASUREMENTS FROM  3С84 SOURCE}

\begin{figure*}[]
%\onelinecaptionsfalse
\centerline{
\hbox{
\includegraphics[angle=-90,width=0.4\textwidth,clip]{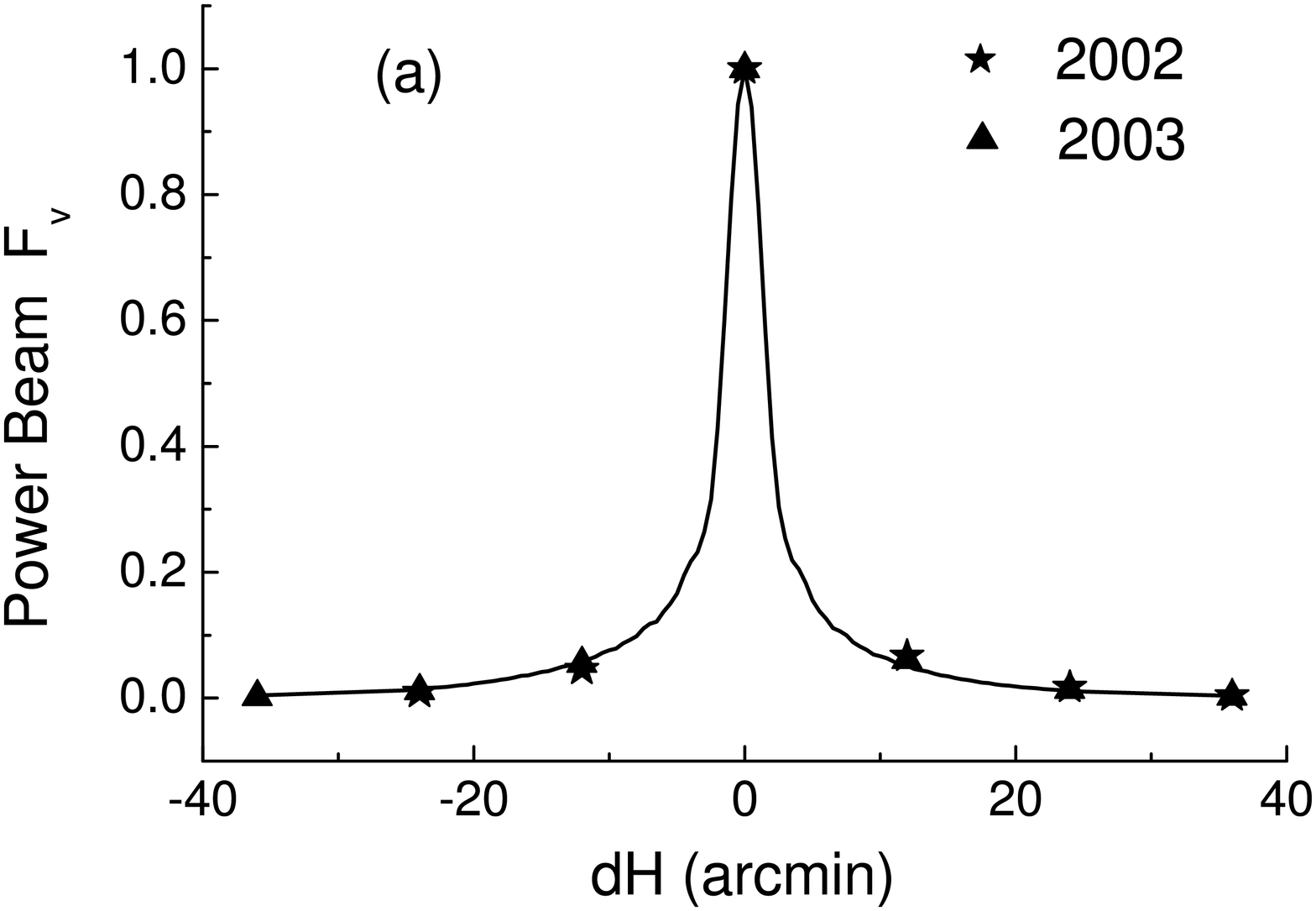}
\includegraphics[angle=-90,width=0.4\textwidth,clip]{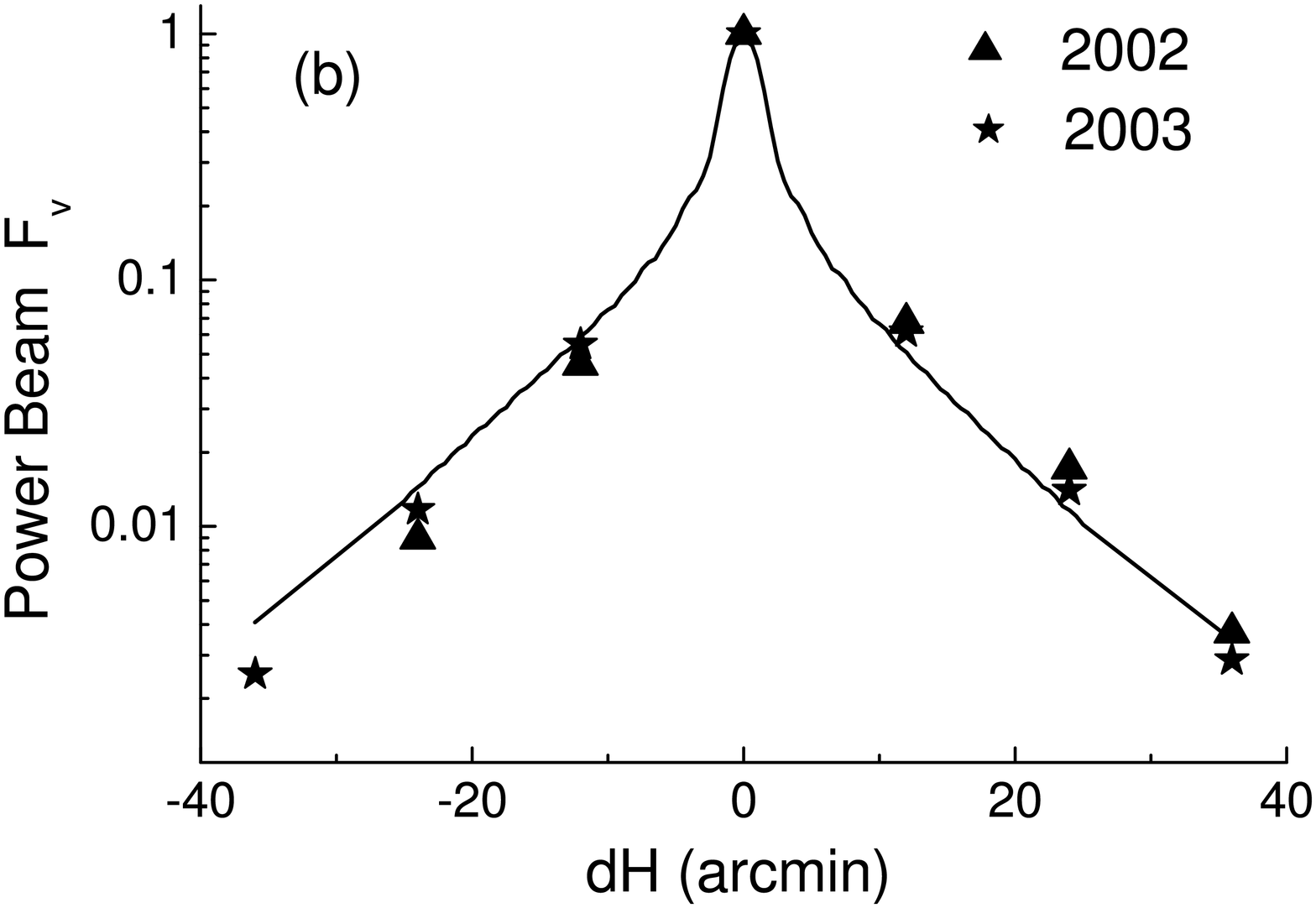}
} }
%\setcaptionmargin{10mm}
%\captionstyle{normal}
\caption{
Vertical PB  constructed  using  the results of observations
of 3С84 in nine sections of the RZF survey
($Dec_{2000}=41^o30'42''\pm12'n, ~n=0,1,2,3,4$) using the
radio-astronomical method described by Kuz'min and Solomonovich
[\cite{ks:Majorova_n}]. The asterisks and triangles show the results
of the measurements made during the 2002 and 2003 sets,
respectively, and the solid lines show the computed PB.
 The power beam patterns are shown in linear (a) and
logarithmic (b) scale. } \label{fig11:Majorova_n}
\end{figure*}

\begin{figure}[]
%\onelinecaptionsfalse
\centerline{
\hbox{
\includegraphics[angle=-90,width=0.4\textwidth,clip]{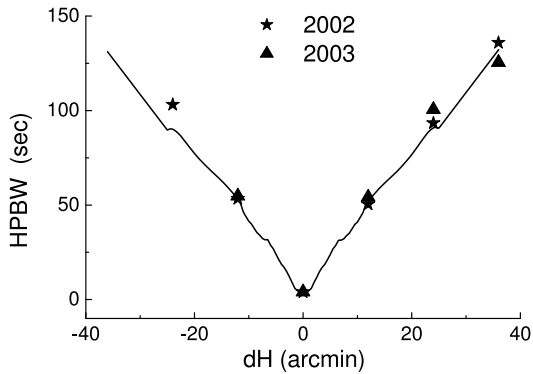}
} }
%\setcaptionmargin{10mm} \captionstyle{normal}
\caption{
Dependences of the half-width of the  PB
$dH$ determined by applying the radio-astronomical method
described by Kuz'min and Solomonovich [\cite{ks:Majorova_n}] to the
results of observations of 3С84 made in nine sections of the RZF
survey ($Dec_{2000}=41^o30'42''\pm12'n, ~n=0,1,2,3,4$). The
asterisks and triangles show the results of the measurements made
during the 2002 and 2003 sets, respectively, and the solid lines
show the computed PBs.} \label{fig12:Majorova_n}
\end{figure}

We measured the vertical power  patterns of the telescope
independently of the data for the 2002 and 2003 sets using the
bright ``point''\, source 3С84 and the technique described in
[\cite{mm2:Majorova_n,ks:Majorova_n}]. This source was observed in
all nine bands of the survey. We normalized the curves of its
transits across the sections with $\Delta\delta=\pm12'n$
($n=0,1,2,3,4$) to the antenna temperature of the source in the
central section of the survey ($\Delta\delta=0$). We computed the
vertical power  pattern values by formula
(\ref{2:Majorova_n}). Figure~\ref{fig11:Majorova_n} shows the
results of the measurements of the vertical power beam.
Figure~\ref{fig12:Majorova_n} shows how the half-widths of the
PBs depend on $dH$. The asterisks and triangles
show the results obtained in the 2002 and 2003 sets,
respectively. The solid lines show the computed PBs
 and the computed half-widths of the PBs.
Figures~\ref{fig11:Majorova_n}(а) and~\ref{fig11:Majorova_n}(b)
feature the vertical power beam patterns in linear and logarithmic
scale, respectively.

As is evident from Figs.~\ref{fig11:Majorova_n} and
\ref{fig12:Majorova_n}, the experimental data points obtained
from the measurements of  3С84 agree very well with the
corresponding computed curves. The standard error of the
deviations of the experimental power beam  from the
computed power beam  is equal to $\sigma=0.010\pm0.004$ and
$\sigma=0.005\pm0.002$ for the 2002 and 2003 sets, respectively.
This is  $3\div10$ times less than the standard error obtained
from the sample of sources from NVSS catalog. Note, however, that
the experimental PB obtained from the results of
observations of 3С84 consists of only nine data points $12'$
apart and has gaps in the  $dH=3'\div5'$ intervals, where the
greatest discrepancies are found between the computed PB
and the experimental  PB constructed using
observations of NVSS sources. The average coefficients $K_{HPBW}$
obtained from the results of observations of 3С84 are equal to
$1.02\pm0.05$ and $1.05\pm0.08$ fore the 2002 and 2003 sets,
respectively, and virtually coincide with the $K_{HPBW}$ values
obtained from the sample of NVSS sources.

\begin{figure*}[]
%\onelinecaptionsfalse
\centerline{
\hbox{
\includegraphics[angle=-90,width=0.33\textwidth,clip]{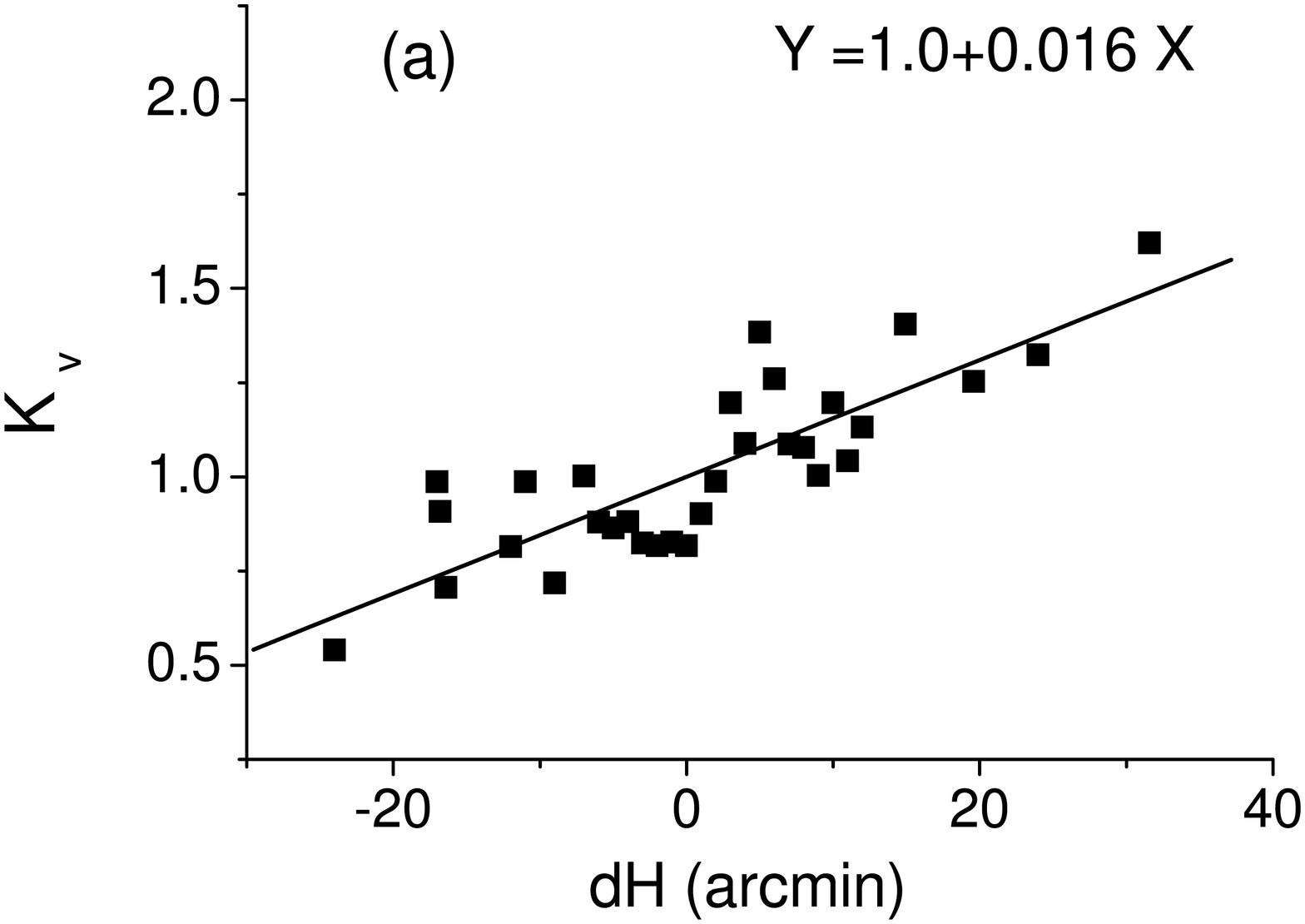}
\includegraphics[angle=-90,width=0.33\textwidth,clip]{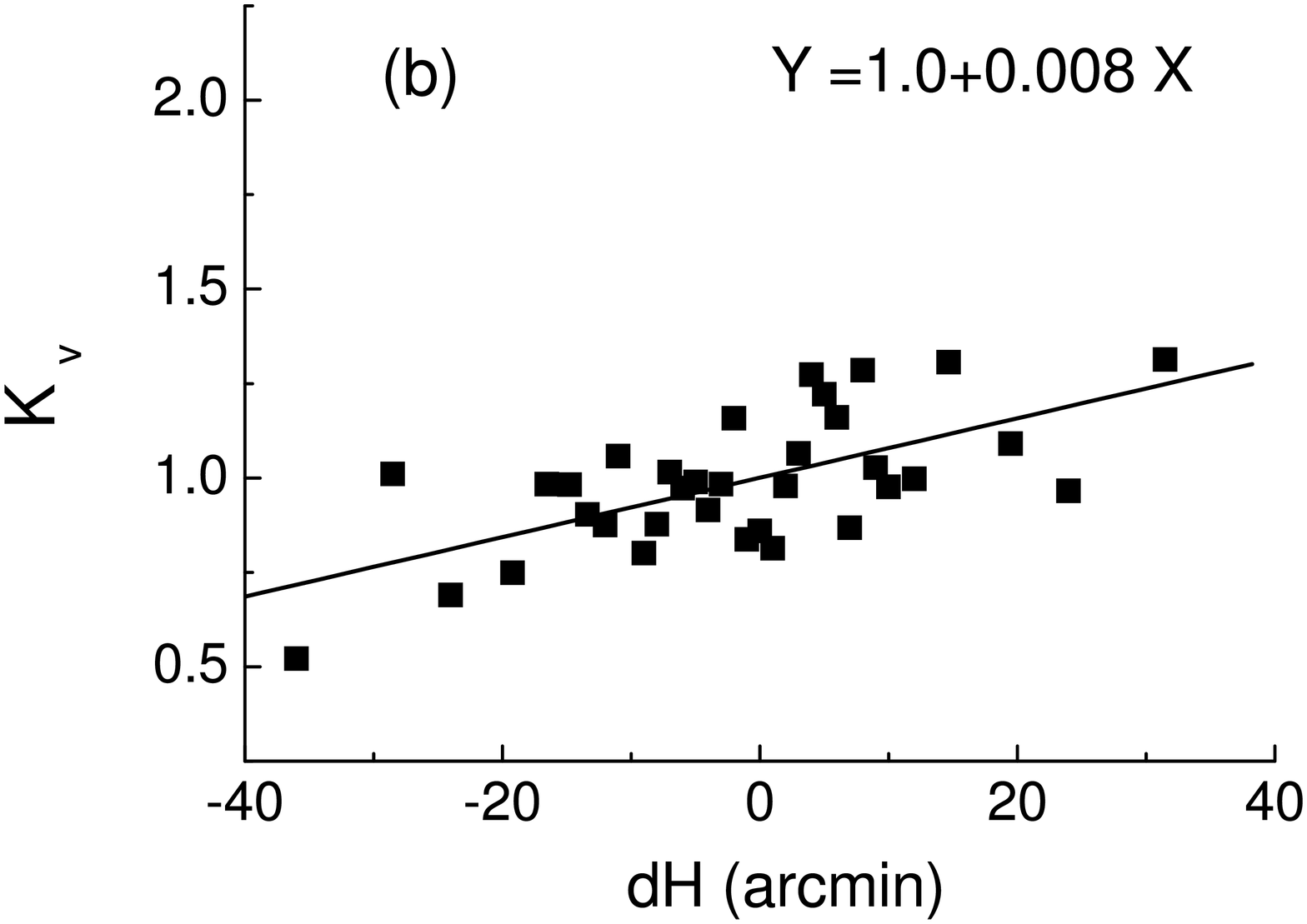}
} }
%\setcaptionmargin{5mm} \captionstyle{normal}
\caption{
Dependences of the ratio  ${K_{v}}$ of the vertical PB
to the computed PB  constructed  using  observations
of sources in the side bands of the RZF survey
($Dec_{2000}=41^o30'42''\pm12'n, ~n=1,2,3,4$) with subsequent
``densification''~ of points against dH: (a)
 the 2002 set and  (b) the 2003 set. }
\label{fig13:Majorova_n}
\end{figure*}

\begin{figure*}[]
%\onelinecaptionsfalse
\centerline{
\hbox{
\includegraphics[angle=-90,width=0.33\textwidth,clip]{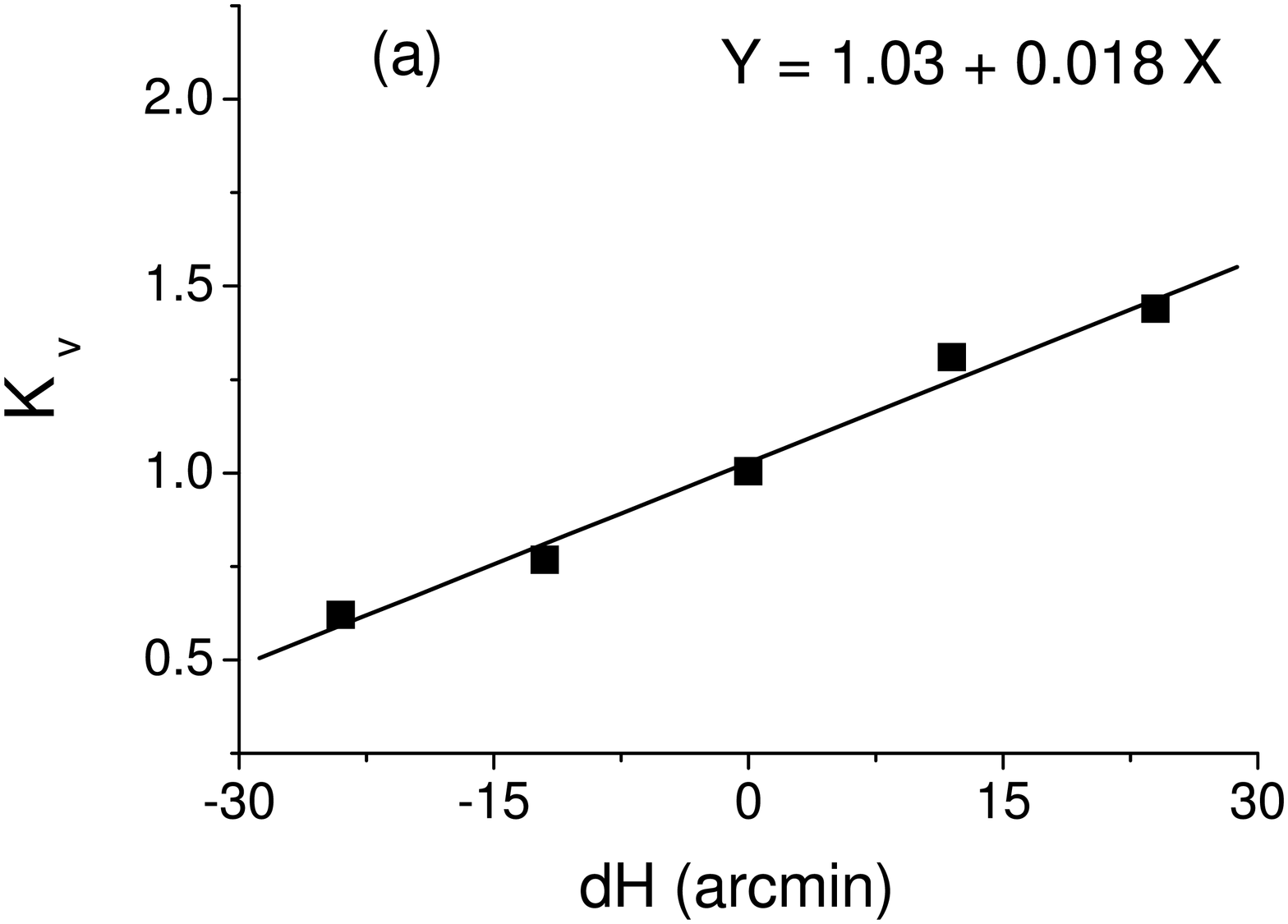}
\includegraphics[angle=-90,width=0.33\textwidth,clip]{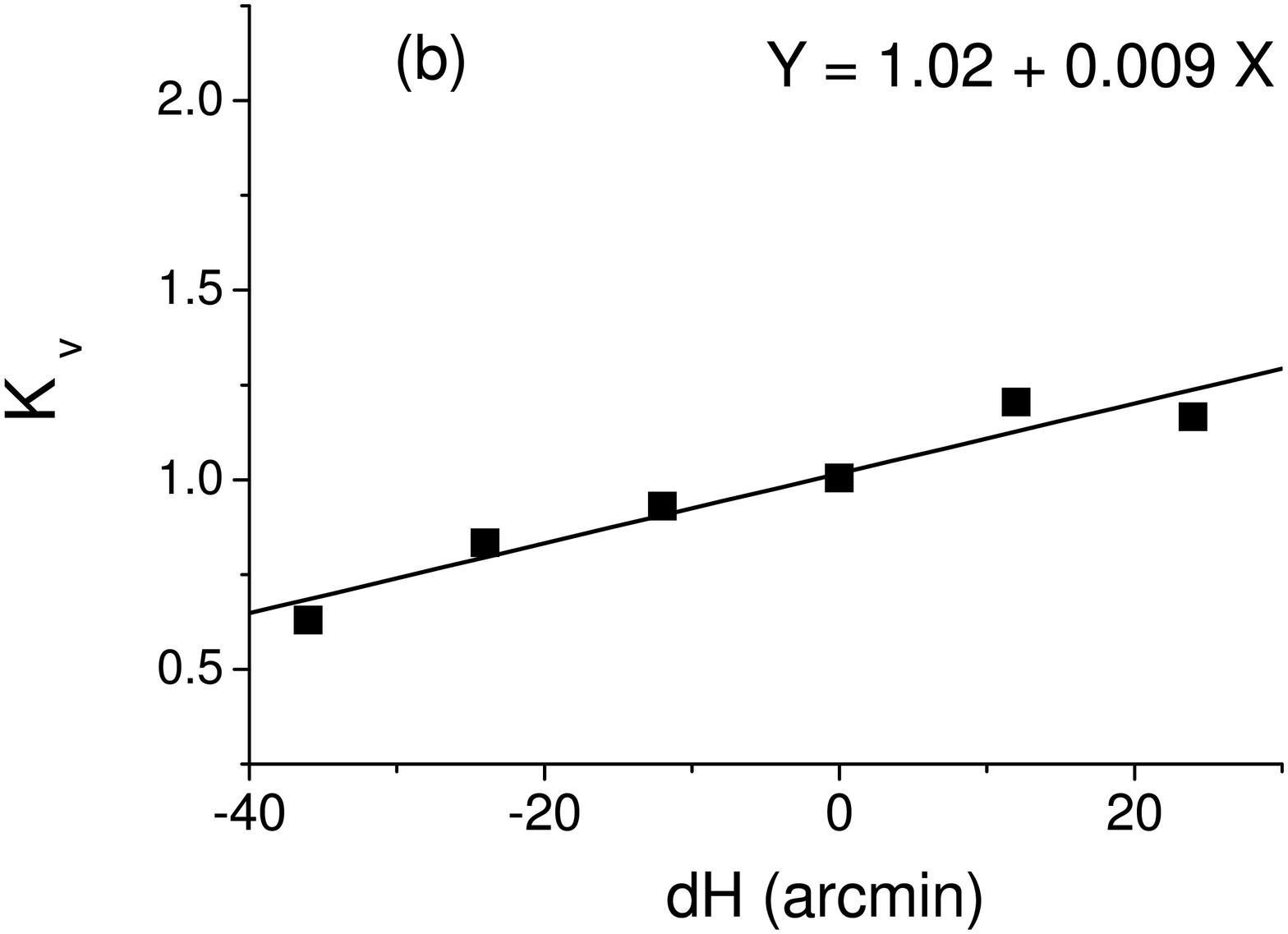}
} }
%\setcaptionmargin{5mm} \captionstyle{normal}
\caption{
Dependences of the ratio ${K_{v}}$ of the vertical PB
to the computed PB constructed  using  the results of
observations of 3С84 in nine bands of the RZF survey
($Dec_{2000}=41^o30'42''\pm12'n, ~n=0,1,2,3,4$) against dH: (a) the 2002
set and  (b) the 2003 set. } \label{fig14:Majorova_n}
\end{figure*}

We constructed the dependences $K_{v}(dH)$ for a more detailed
comparison of the experimental power beam patterns obtained using
two independent methods. Coefficient $K_{v}$ is equal to the
ratio of the experimental vertical PB to the
computed one at the same  $dH$.
Figure~\ref{fig13:Majorova_n} shows the dependences  $K_{v}(dH)$
obtained from the results of observations of a sample of NVSS sources
in the side bands of the survey with subsequent
``densification''\, of data points and
Fig.~\ref{fig14:Majorova_n} shows the corresponding dependences
based on the results of observations of 3С84 ((a) --- the 2002
set and (b) --- the 2003 set.). Approximating linear dependences
$y=a+bx$ are fit to the family of points obtained.

It is immediately apparent that the lines approximating the
dependences $K_{v}(dH)$ constructed using the sample of sources and on
3С84  have virtually the same slope for the same observing set.
It is equal to  $b=0.016$ and $b=0.018$ for the results based on
the sample of sources and 3C84, respectively, in the case of the
2002 observing set, and to $b=0.008$ and $b=0.009$, respectively,
in the case of the 2003 observing set. The slope of the
approximating line indicates a turn of the experimental PB
relative to the computed PB. This turn
shows up conspicuously in Fig.~\ref{fig11:Majorova_n}(b). The
values of the coefficients characterizing the slope of the
approximating curve show that despite their different accuracy
both methods of the measurement of the PB yield
the same PB turn, which is almost twice greater in
the 2002 set compared to the turn of the power beam pattern during
the 2003 set.

\section{COMPARISON OF EXPERIMENTAL POWER BEAM PATTERNS OBTAINED
IN THE 1998--2003 OBSERVING SETS}

During the 2002 and 2003 sets observations were made in nine
survey bands shifted by  $\Delta\delta=12'$ with respect to each
other that alow the experimental BP to be constructed
in the interval $ -36' < dH < 36'$. During the  1998
(1998.10 -- 1999.02) and 2000  (1999.12 -- 2000.03) sets the
sources were observed only in the central band of the survey
 at the declination of  3С84
($Dec_{2000}=41^{o}30'42''$). To compare the experimental
power beam patterns obtained in these observations, we used a
sample of NVSS sources (about 80 objects) with $\lambda$7.6-cm
fluxes greater than 5\,mJy. The number of records
averaged for each source was equal to about  60 for the 1998 and
2000 observations; about 40 records for the 2002 observations, and
about 20 for  the 2003 observations. The  $dH$ values for this
sample lied in the interval $-6' < dH < 6'$, i.e., we studied
mostly the central part  of the power beam
pattern (main beam).

\begin{figure*}[]
%\onelinecaptionsfalse
\centerline{
\vbox{
\hbox{
\includegraphics[angle=-90,width=0.35\textwidth,clip]{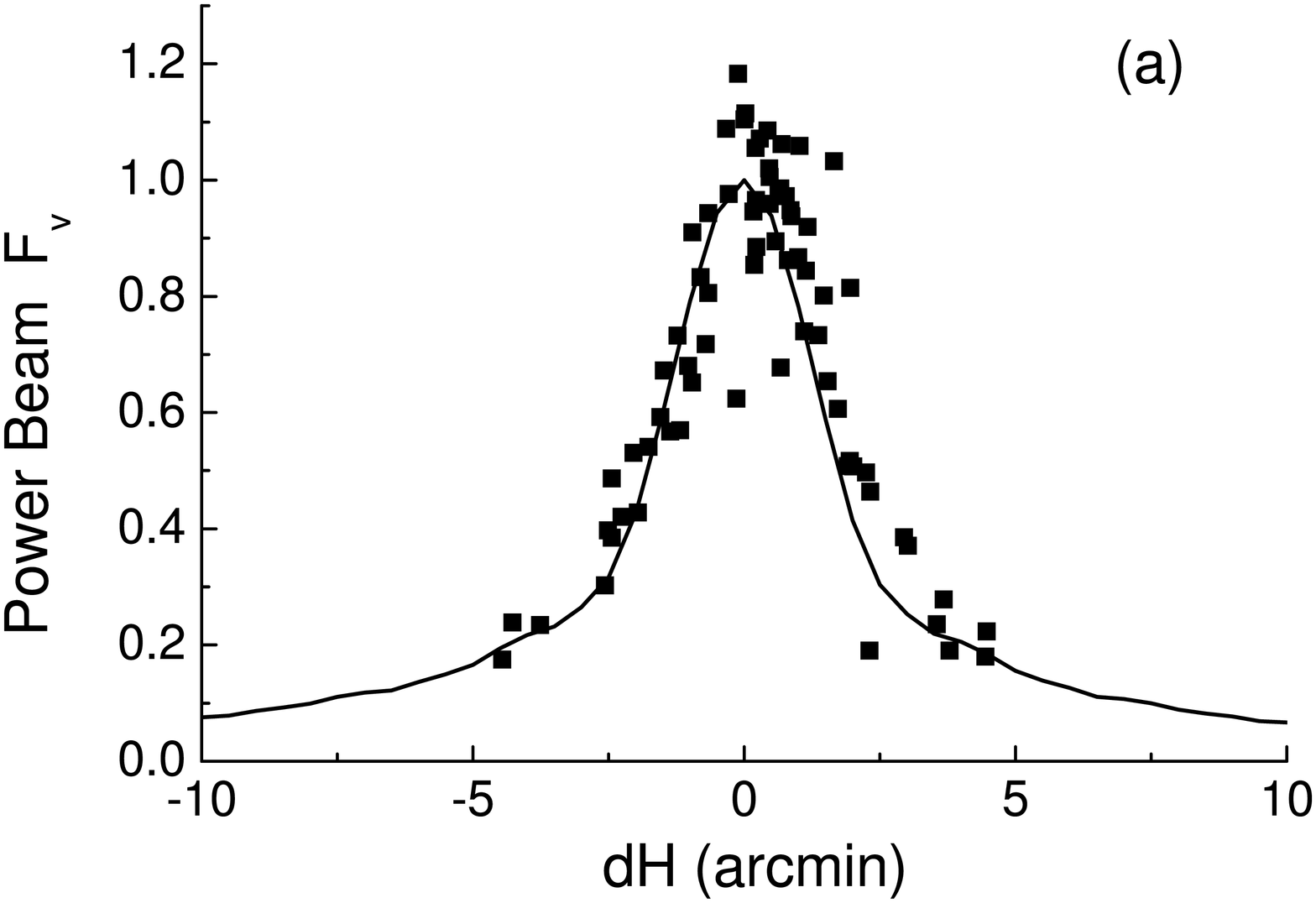}
\includegraphics[angle=-90,width=0.35\textwidth,clip]{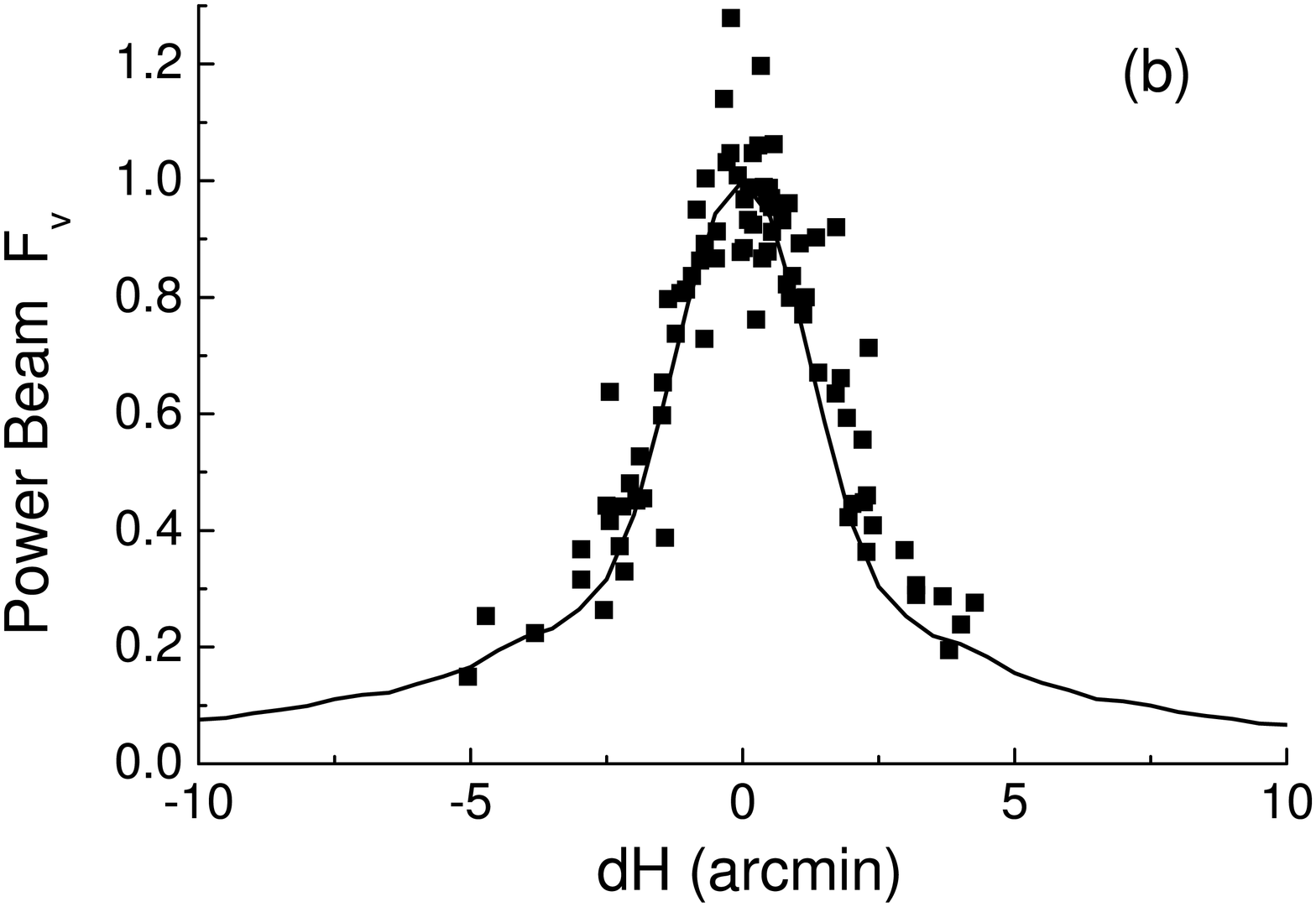}
}
\hbox{
\includegraphics[angle=-90,width=0.35\textwidth,clip]{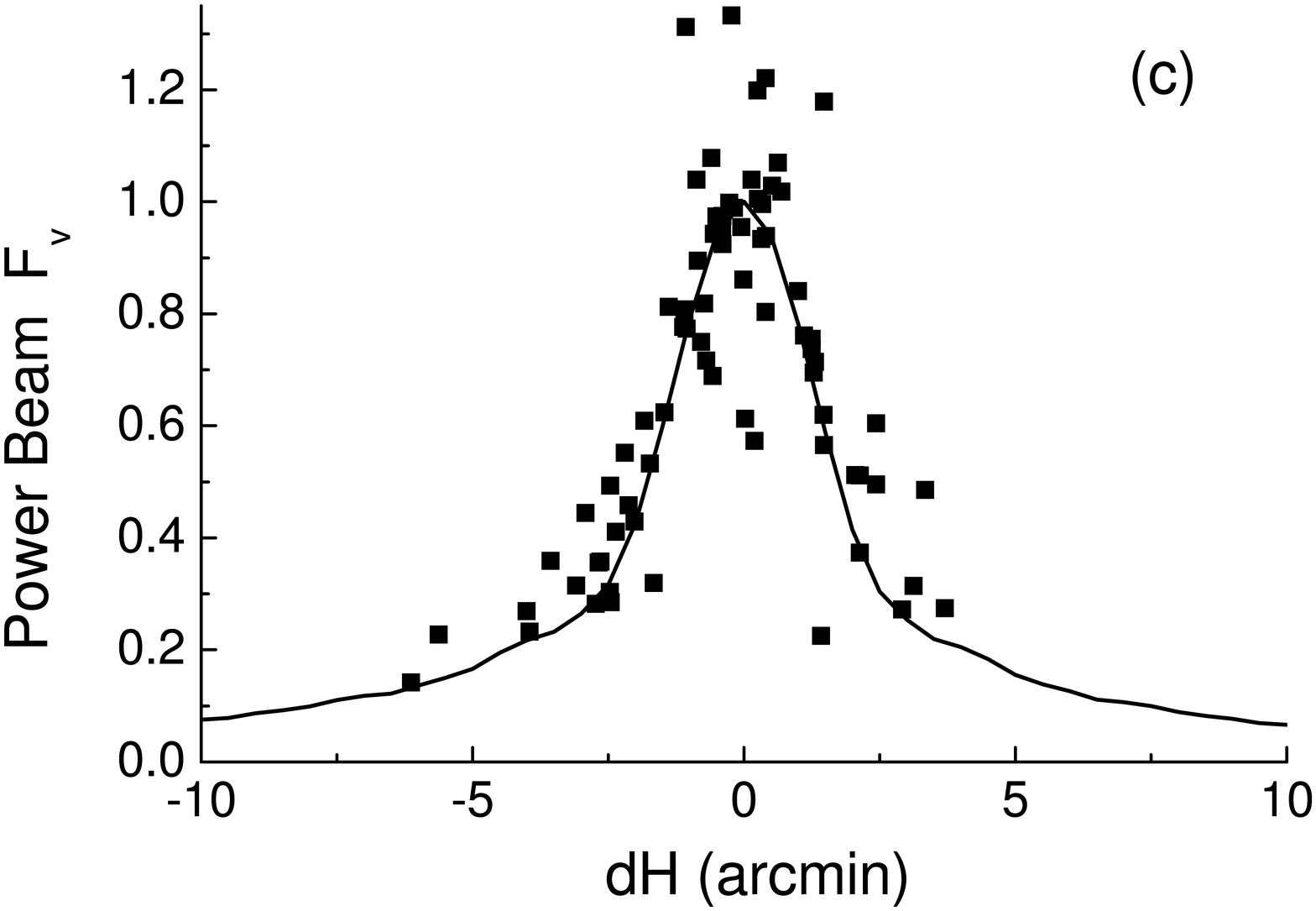}
\includegraphics[angle=-90,width=0.35\textwidth,clip]{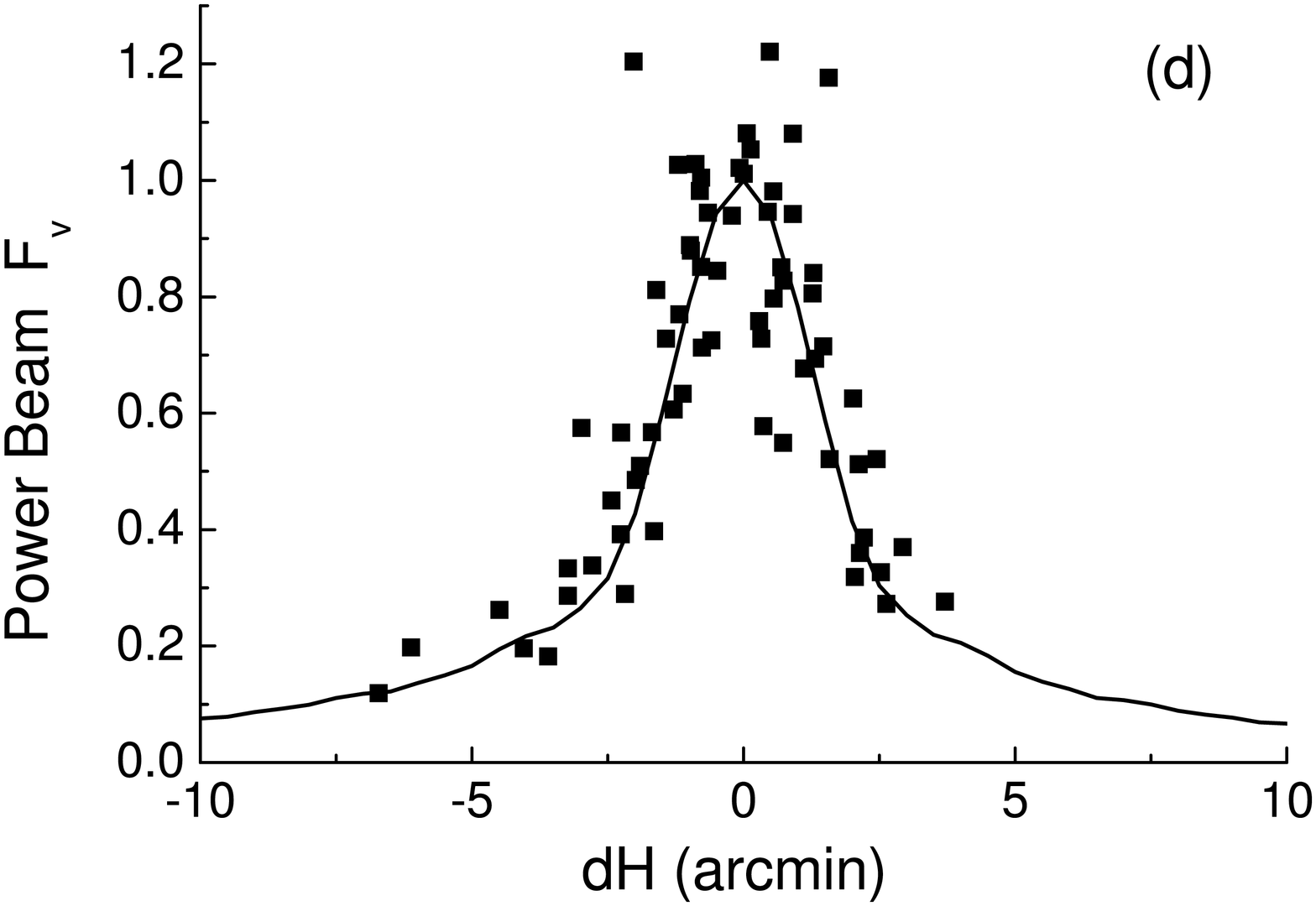}
} } }
%\setcaptionmargin{5mm} \captionstyle{normal}
\caption{
Vertical PB constructed  using the results of observations
of a sample of  NVSS sources with the $\lambda7.6$-cm fluxes  $P
> 5$mJy in the central band of the RZF survey
($Dec_{2000}=41^o30'42''$): (a) --- the 1998 set; (b) --- the
2000 set; (c) --- the 2002 set, and (d) --- the 2003 set. The
filled squares show the experimental data points of the PB
and the solid lines show the computed PB.
} \label{fig15:Majorova_n}
\end{figure*}

\begin{figure*}[]
%\onelinecaptionstrue
\centerline{
\vbox{
\hbox{
\includegraphics[angle=-90,width=0.35\textwidth,clip]{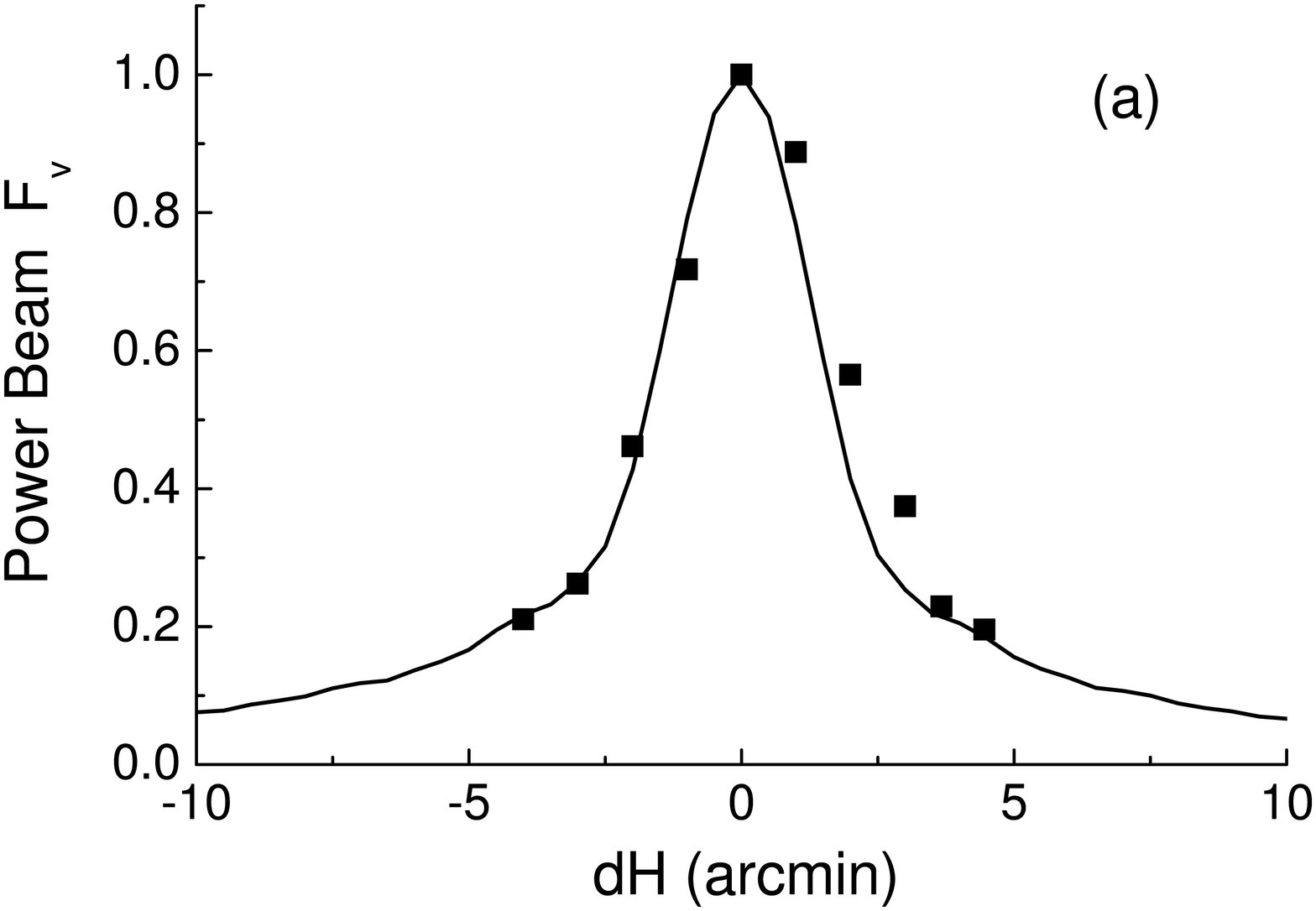}
\includegraphics[angle=-90,width=0.35\textwidth,clip]{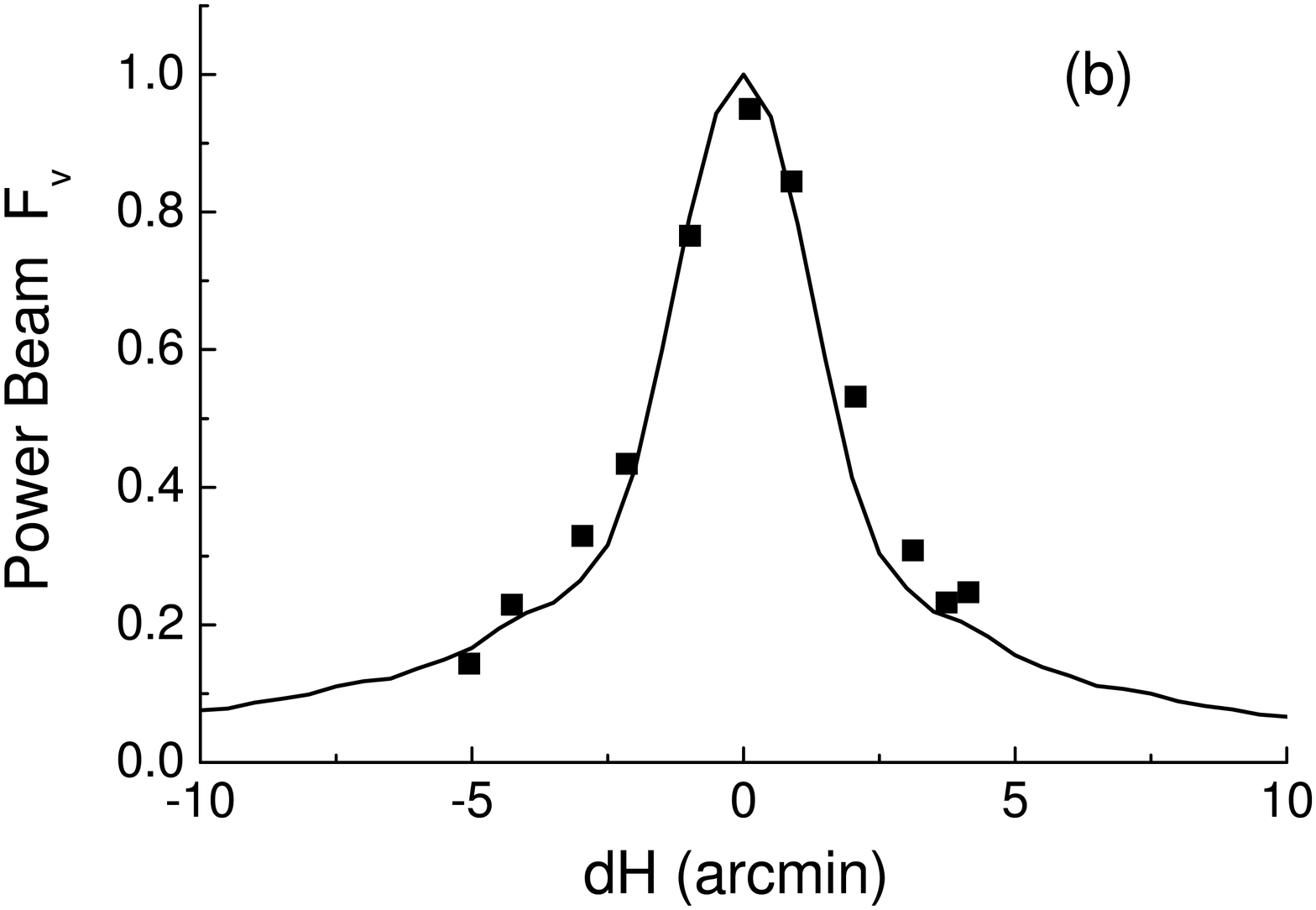}
}
\hbox{
\includegraphics[angle=-90,width=0.35\textwidth,clip]{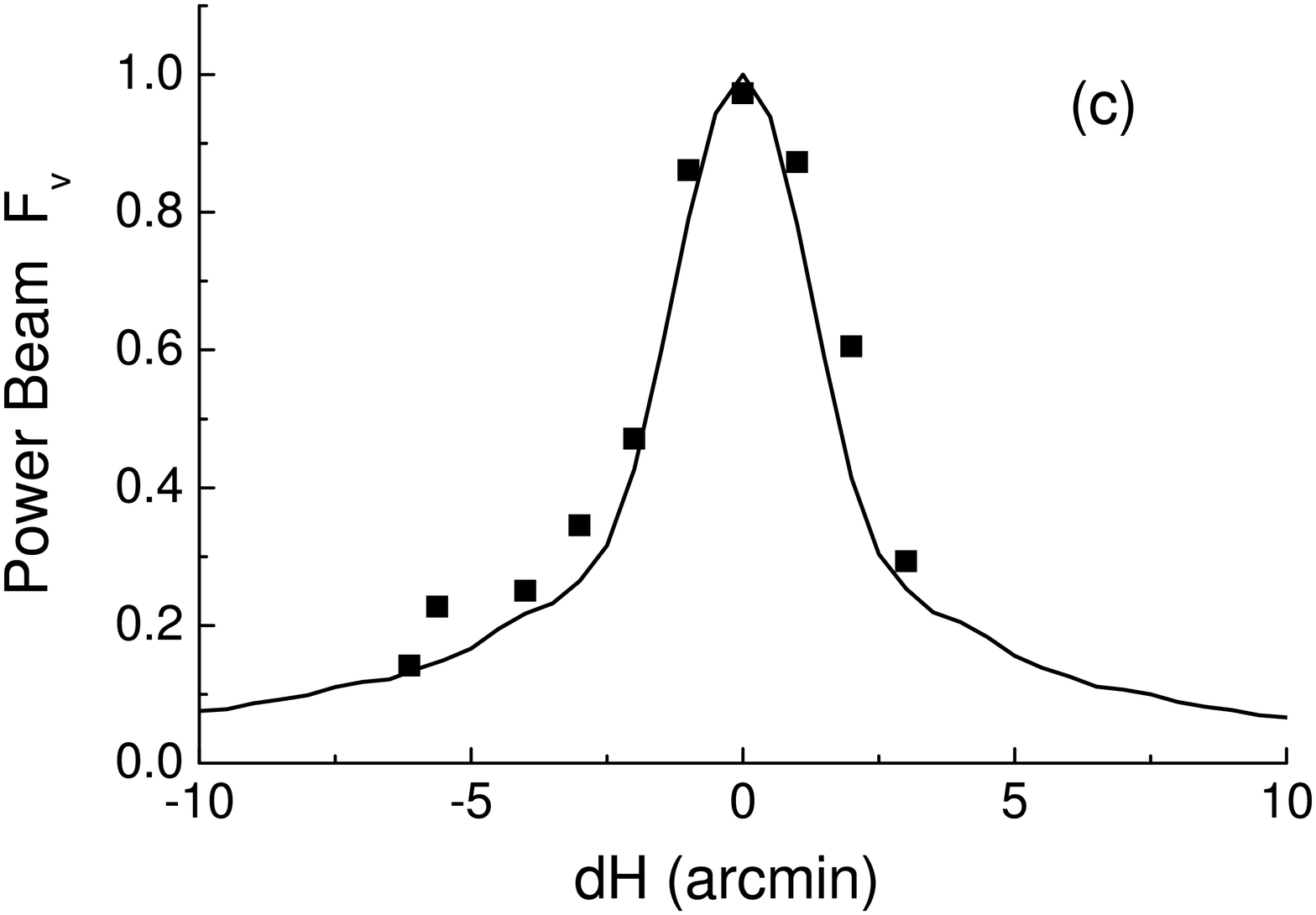}
\includegraphics[angle=-90,width=0.35\textwidth,clip]{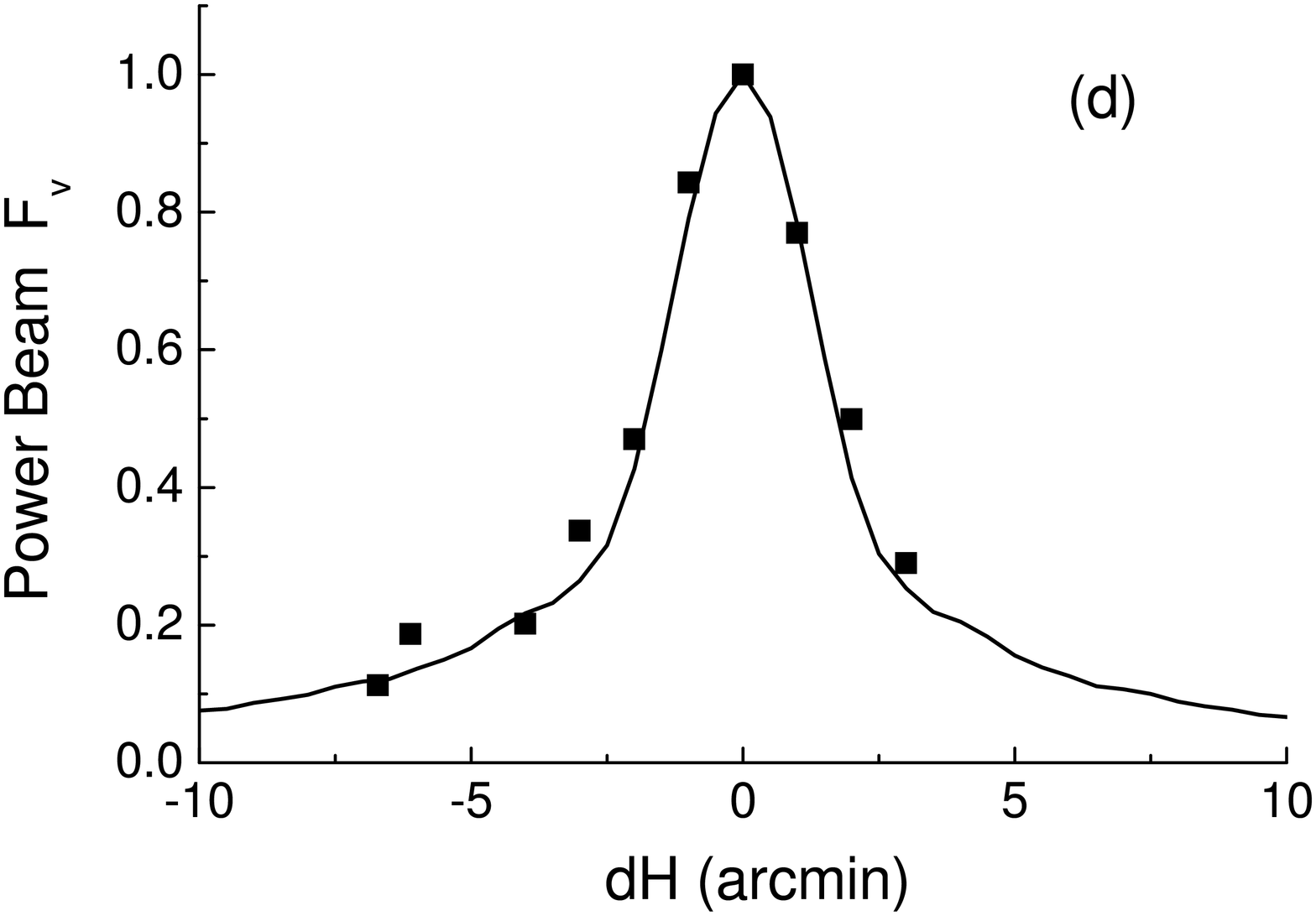}
} } }
%\setcaptionmargin{0mm} \captionstyle{normal}
\caption{ Same
as Fig.~\ref{fig15:Majorova_n}, but with subsequent
``densification''~ of data points. } \label{fig16:Majorova_n}
\end{figure*}

We used the technique described in Section 3 of this paper to
construct the vertical power beams. The filled squares in
Fig.~\ref{fig15:Majorova_n} show the experimental vertical
PB obtained from the results of observations made in
1998, 2000, 2002, and 2003, and the solid lines show the computed
PBs. The standard error of the deviations of the
experimental PB from the computed PB
is equal to  $\sigma=0.13\pm0.02$, $\sigma=0.11\pm0.01$,
$\sigma=0.16\pm0.02$, and $\sigma=0.20\pm0.02$ for the 1998,
2000, 2002, and 2003 sets, respectively. The procedure of
``densification''\, of data points reduced the standard error
down to  $\sigma=0.06\pm0.02$, $\sigma=0.05\pm0.01$,
$\sigma=0.06\pm0.02$, and $\sigma=0.04\pm0.01$ for the 1998,
2000, 2002, and 2003 sets, respectively.The filled squares in
Fig.~\ref{fig16:Majorova_n} show the experimental vertical
PBs obtained after applying the procedure of
``densification''.

\begin{figure*}[]
%\onelinecaptionsfalse
\centerline{
\vbox{
\hbox{
\includegraphics[angle=-90,width=0.35\textwidth,clip]{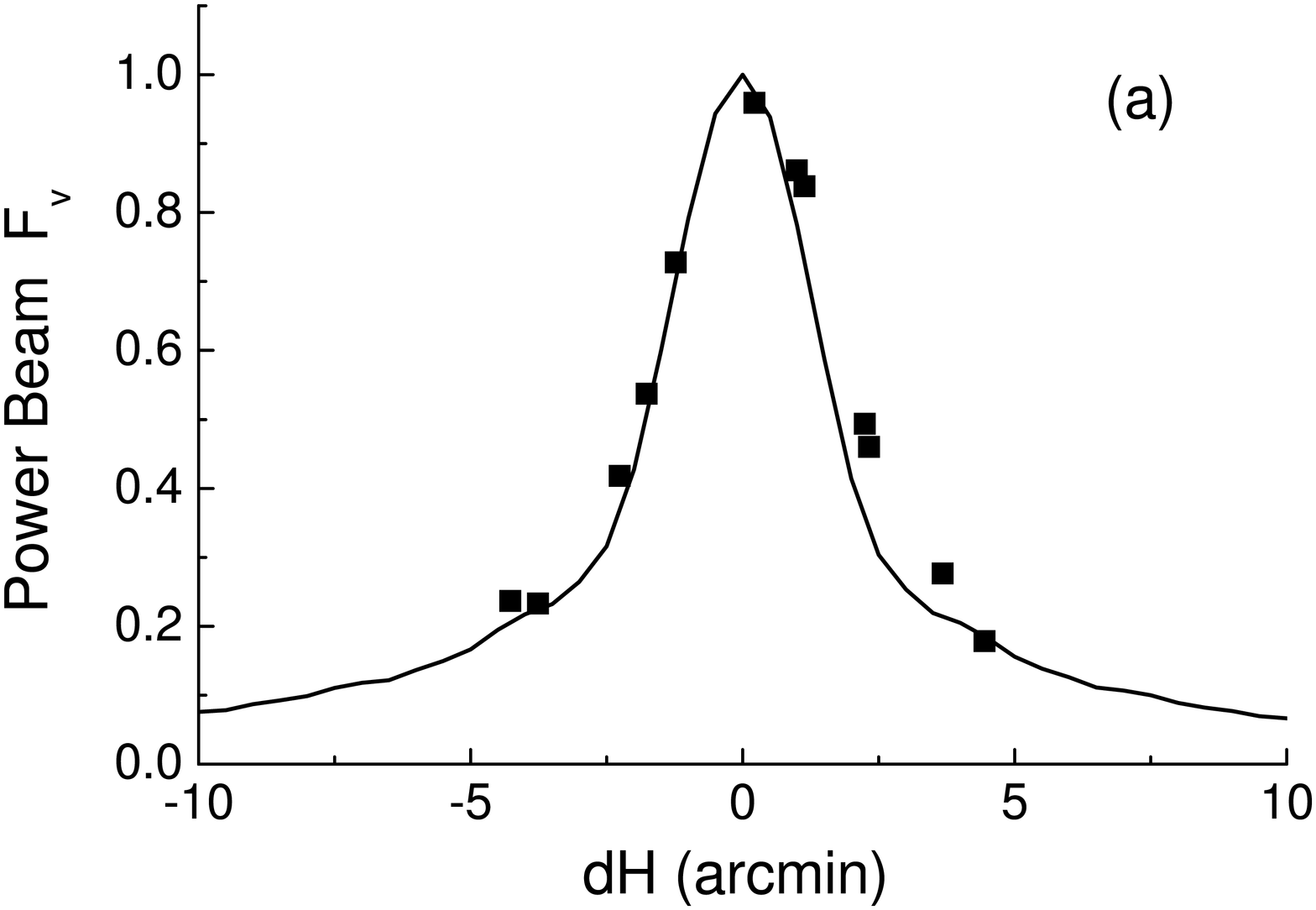}
\includegraphics[angle=-90,width=0.35\textwidth,clip]{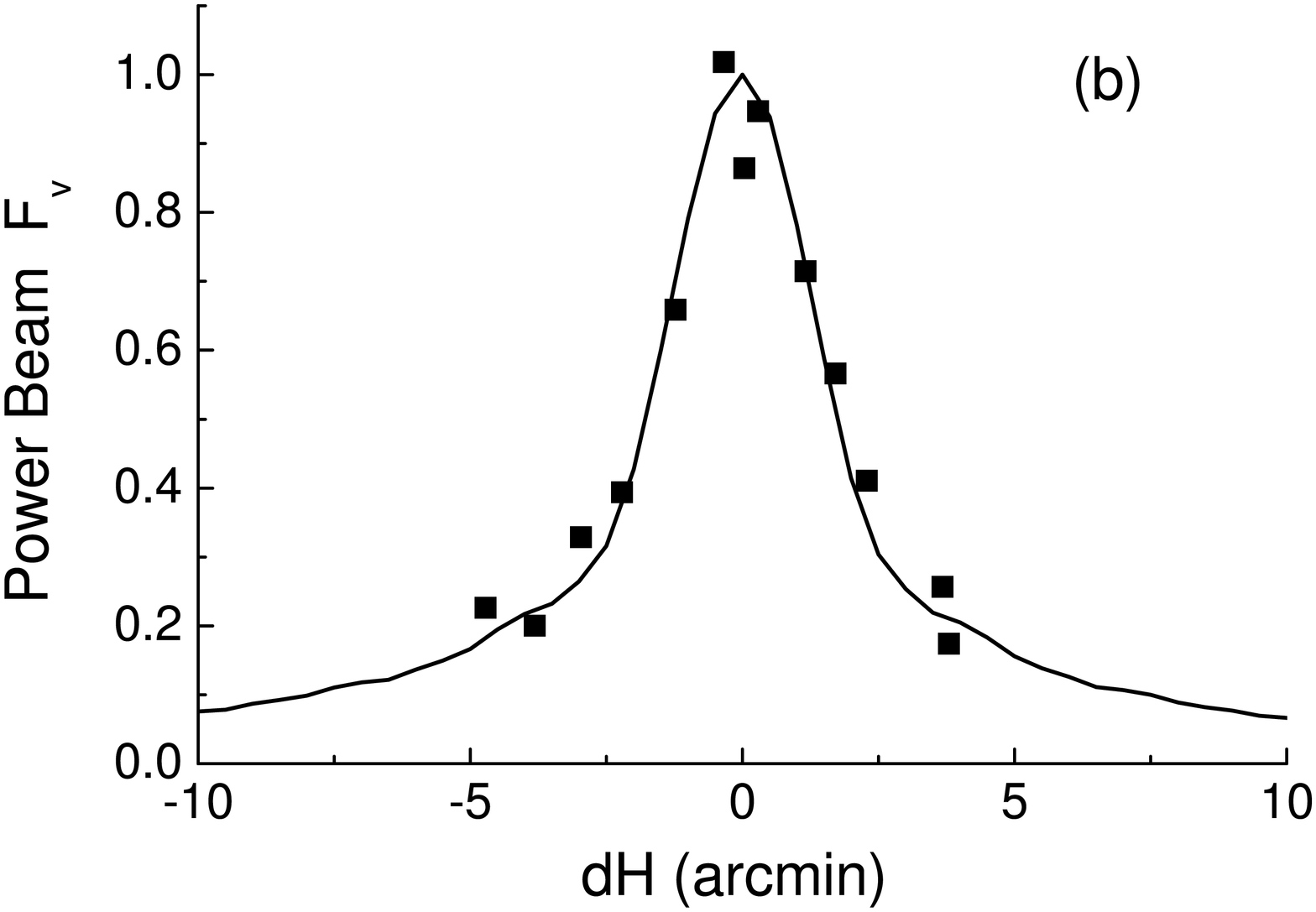}
}
\hbox{
\includegraphics[angle=-90,width=0.35\textwidth,clip]{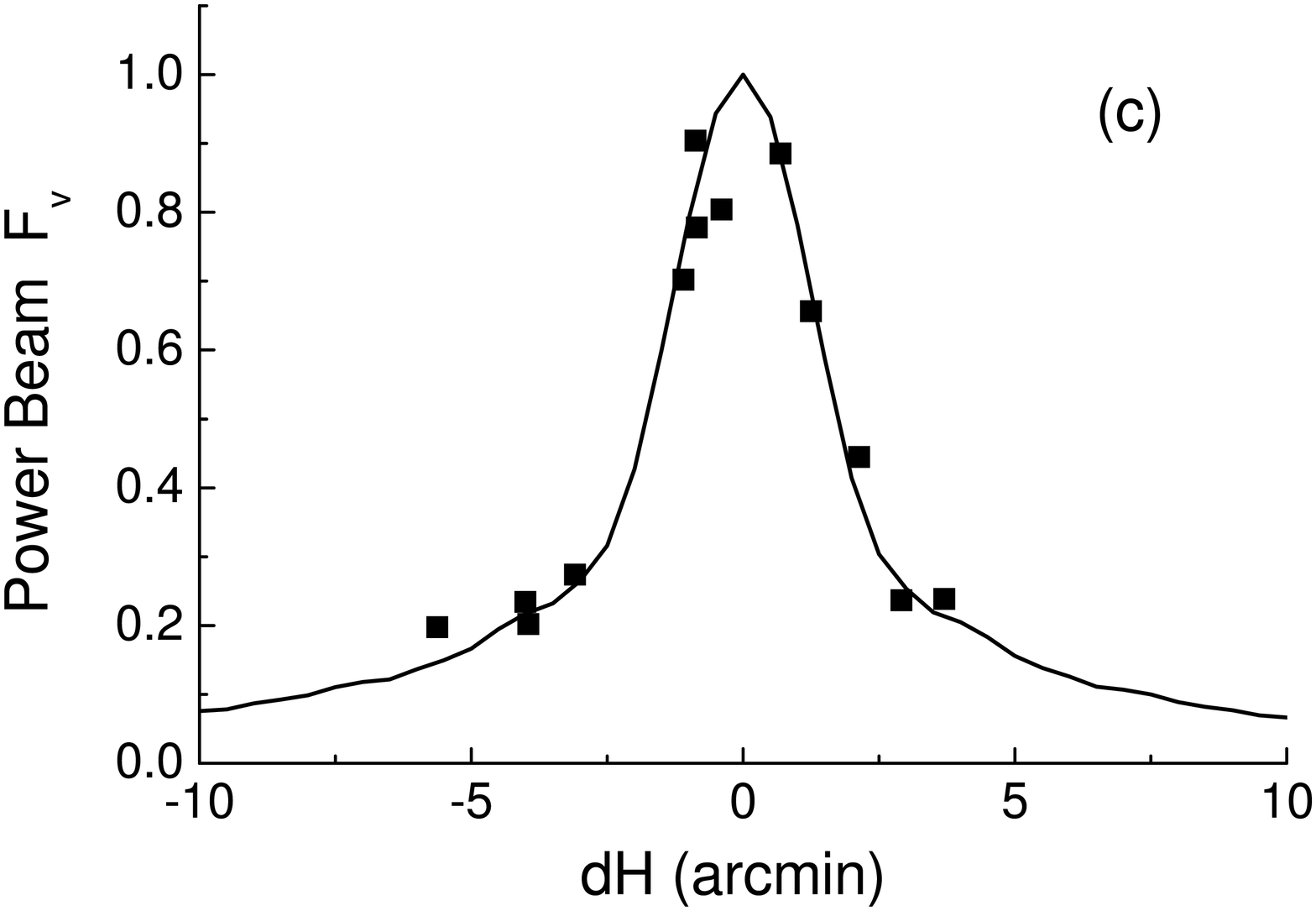}
\includegraphics[angle=-90,width=0.35\textwidth,clip]{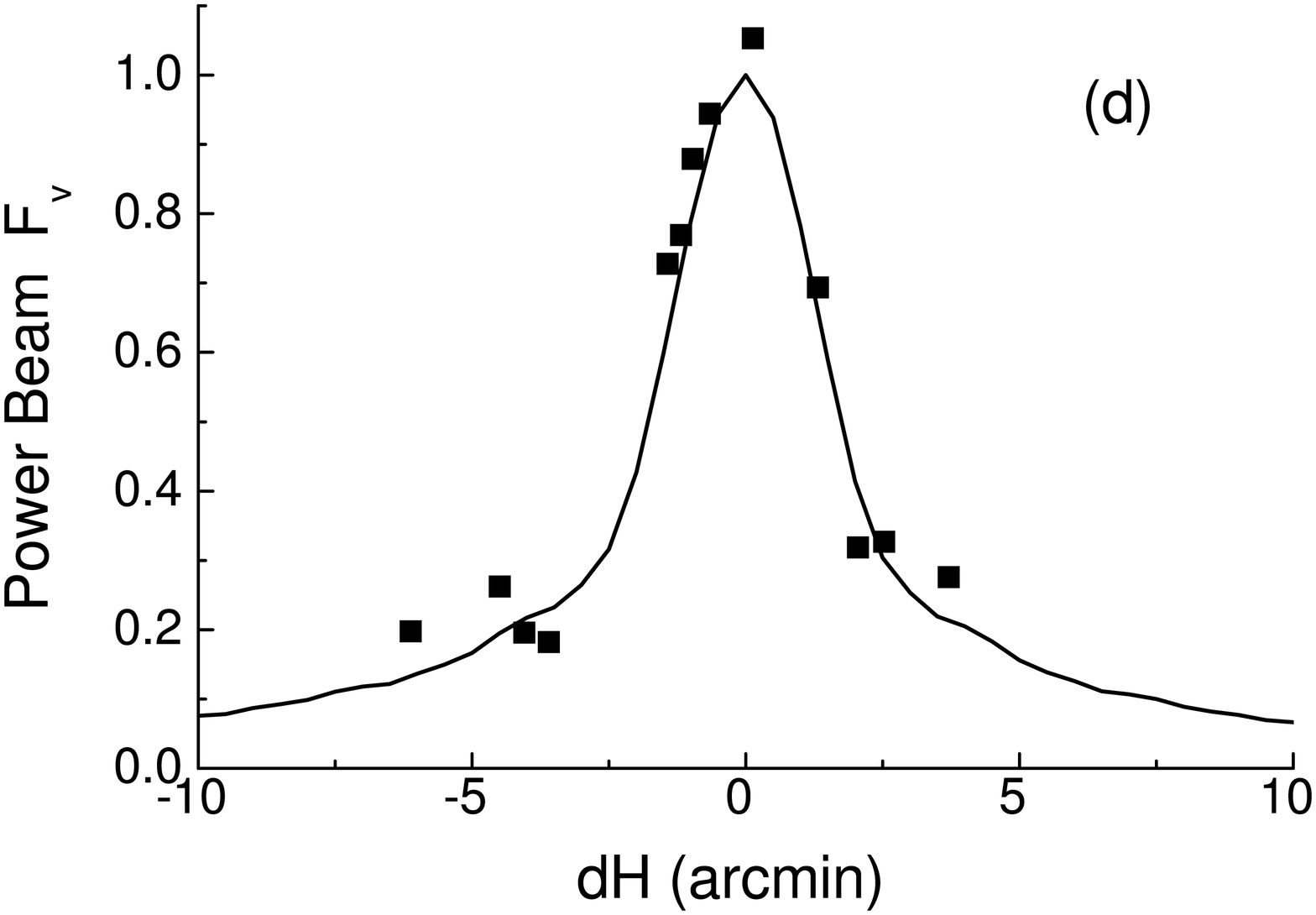}
} } }
%\setcaptionmargin{10mm} \captionstyle{normal}
\caption{
Vertical PB constructed  using  the results of observations
of 13 NVSS sources with the $\lambda7.6$-cm fluxes $P > 150$\,mJy
in the central band of the RZF survey ($Dec_{2000}=41^o30'42''$): (a)
--- the 1998 set; (b) --- the 2000 set; (c) --- the 2002 set, and
(d) --- the 2003 set. The filled squares show the experimental
data points of the PB and the solid lines show the
computed PB.} \label{fig17:Majorova_n}
\end{figure*}

Figure~\ref{fig17:Majorova_n} shows the experimental vertical
power beam patterns obtained from observations of 13  bright
``point''\, sources from the NVSS catalog with the
$\lambda$7.6-cm fluxes greater than  150\,mJy. The standard error
of the deviations of the experimental PB based on
these bright sources from the computed PB is equal
to $\sigma=0.06\pm0.02$ for the  1998, 2000, and 2002 sets and
$\sigma=0.05\pm0.01$ for the 2003 set. These values virtually
coincide with the standard errors of the deviations of the
experimental PBs constructed using the entire sample of
sources with subsequent ``densification''  of data points. Note
that to construct these  PBs, we used sources
 more than half of which  had fluxes in the  5\,mJy<P<50\,mJy
interval. It follows from this that given a large number of
relatively weak sources one can achieve the same accuracy for the
constructed experimental PB as in the case of
using a few strong sources. However, the weaker are the sources,
the bigger must be the sample and the more records are to be
averaged for each source.

We constructed the $K_{v}(dH)$ dependences for the experimental
PB obtained from the entire sample of sources. We fit
approximating lines to the resulting families of data points. The
entire sample yields the following values for the slope of the
approximating line: $b=0.021$, $b=0.023$, $b=0.012$, and
$b=0.006$ for the 1998, 2000, 2002, and 2003 sets, respectively.
These values are close to those obtained from the results of
observations of 3С84 and sources in the side bands of the survey.
The coefficient that characterizes the slope of the approximating
line for the dependences $K_{v}(dH)$ based on observations of 13
bright sources is close to zero.

To compare the vertical PBs obtained for
observations made in different sets, we plot least-square-fit
Gaussians  approximating the experimental power beam patterns
shown in Figs.~\ref{fig15:Majorova_n}, \ref{fig16:Majorova_n},
and \ref{fig17:Majorova_n}. We compare the half-widths of the
Gaussians,  $G_{0.5}$, and the parameters $\Delta_{max}$
characterizing the shift of the summit of the Gaussian along the
$dH$-axis with respect to the coordinate origin ($dH=0$).
Parameter $\Delta_{max}$ characterizes the shift of the maximum
of the vertical PB with respect to the central
section. We found the parameters $G_{0.5}$ and $\Delta_{max}$ to
remain unchanged from on set to another: $G_{0.5}=4.06'\pm0.10$
and $\Delta_{max}=-0.024'\pm0.005$ irrespectively of whether we
fitted the Gaussians to the experimental power beams
obtained from the entire sample of sources or to those constructed using
the sample of bright sources exclusively. Fitting Gaussians to
the computed PBs yielded the same $G_{0.5}$ and
$\Delta_{max}$ values --- $G_{0.5}=4.06'\pm0.10$ and
$\Delta_{max}=-0.024'\pm0.005$ --- thereby confirming the good
agreement between experimental and computed power beam patterns.

Given that the standard errors of the deviations of the
experimental PBs from the corresponding computed
PBs obtained from either the bright sources or from the
entire sample virtually coincide for different observing sets, we
can conclude that the experimental power beam pattern was
sufficiently stable during the RZF survey
(1998--2003) and agreed well with the computed power beam pattern
in its central part ($\pm6'$).

\section{ESTIMATING THE MEASUREMENT ERRORS OF SOURCE FLUXES IN THE RZF  SURVEY}

The data obtained when constructing the experimental PBs
 for the  1998--2003 observing sets can be used to
estimate the measurement errors for the fluxes of the sources
that were observed during the RZF survey at  7.6 cm
and were included into the  RZF catalog [\cite{p1:Majorova_n}].

Given the antenna temperature of a source crossing the section
that is $dH$ apart from the central section, one can convert the
antenna temperature of the source to its value in the central
section. We can then use formula (\ref{1:Majorova_n}), where
$k/S_{eff}=(P/T_{a})_{dH=0}$, to determine the  $\lambda7.6$-cm
source flux and interpolate (or extrapolate) it to $\lambda21$.
The source fluxes so determined can be compared to the fluxes
given by the  NVSS catalog. It is evident that this chain is the
inverse of the one used to construct the experimental PB
using the sample of sources with known fluxes. The
ratio of the flux of a source with given $dH$ (or declination)
given in the NVSS catalog to the observed flux is equal to the
ratio of the experimental PB obtained from the results
of observations of this source to the computed PB
in the $dH$ section crossed by the source. The standard errors of
the determination of the source fluxes in the sample considered
are equal to the standard errors of the deviation of  $K_{v}$
from the mean value averaged over the same sample of sources.

We computed the ratios $K_{v}$ both for the sample of sources
observed in nine survey bands during the 2002 and 2003 sets and
for the sample of sources observed in the central band of the
survey during the  1998-- 2003 sets. The standard errors
$\sigma_{K_{v}}$ of the deviation of  $K_{v}$ from its mean value
computed from observations of sources in nine survey bands are
equal to $\sigma_{K_{v}}=0.27\pm0.03$ and
$\sigma_{K_{v}}=0.26\pm0.03$ for the 2002 and 2003 sets,
respectively. These standard errors are computed for a sample of
140 NVSS sources with the 7.6-cm fluxes exceeding 80\,mJy. We did
not ``densify''\, the data points of the experimental PB.
 Removing the bias in the distribution of $K_{v}(dH)$ due
to the turn of the experimental PB relative to the
computed one reduces the standard error down to
$\sigma_{K_{v}}=0.24\pm0.02$ and $\sigma_{K_{v}}=0.25\pm0.02$ for
the 2002 and 2003 sets, respectively.

The standard error for the sample of 80 sources observed in the
central survey band during the 1998--2003 sets is equal to
$\sigma_{K_{v}}=0.23\pm0.03$ for the 1998 and 2000 sets and
$\sigma_{K_{v}}=0.27\pm0.03$ for the 2002 and 2003 sets. The
7.6-cm fluxes of these sources were no lower than 5\,mJy.
Restricting the sample exclusively to objects with fluxes above
100\,mJy results in a standard error of
$\sigma_{K_{v}}=0.15\pm0.03$ for the 1998 and 2000 sets and
$\sigma_{K_{v}}=0.23\pm0.05$, for the 2002 and 2003 sets. The
standard error for the sample of 13 bright sources is
$\sigma_{K_{v}}=0.15\pm0.04$, $\sigma_{K_{v}}=0.14\pm0.04$,
$\sigma_{K_{v}}=0.16\pm0.04$, and  $\sigma_{K_{v}}=0.18\pm0.04$
for the 1998, 2000, 2002, and 2003 sets, respectively.

The standard errors for the observations of sources whose
declinations lie in the $Dec_{2000.0}=41^o30'42'' \pm2'$ (the
band of the RZF catalog) interval are equal to $0.15 \div 0.24$.
They virtually agree with the source flux estimates obtained by
Bursov et al.~[\cite{p1:Majorova_n}].

To sum up the results obtained, it can be concluded that the
standard error of the fluxes of sources observed during the
RZF survey at 7.6~cm lies in the $(0.15\div0.27)\pm0.05$
interval and depends both on the brightness of the source and on
its distance from the central section of the power beam pattern.

\section{CONCLUSIONS}

We propose a method for constructing the experimental power beam pattern
for RATAN-600 radio telescope using NVSS sources with
well-known coordinates and fluxes that cross various sections of
the power beam  during the deep sky survey.

During the 2002 and 2003 observational sets the RZF survey was
conducted in nine survey bands $\Delta\delta=\pm12'$ apart. We
used a rather preliminary sample consisting of 140 sources
observed during these sets to construct the experimental and vertical
PBs of the radio telescope, which we then compared
to the experimental PBs obtained from observations of 3С84
 using the
traditional method [\cite{ks:Majorova_n}].
 The standard error of the deviations of the experimental
PB from the computed one for the entire sample of
observed sources is equal to $0.070\pm0.005$. The application of
the procedure of ``densification''\, of data  points (averaging
the $F_{v}$ values in a $1'$ interval of $dH$ values) reduced the
standard error of the deviations of the experimental PB
from the computed one down to
$0.033\pm0.006$.  For sources observed in the central band of the
survey, where the sensitivity of measurements was much higher
than in the side bands of the survey, the standard error is equal
to $0.033\pm0.007$ for the 2002 set and $0.050\pm0.009$, for the
2003 set (without ``densification '').

A comparison of the experimental and computed power beams showed that
experimental values exceed the computed ones in the $dH=3'\div5'$
interval for both sets.
Besides,
a turn of the experimental PB
relative to the computed PB takes place.
 These distortions of the PB
may be due to a systematic error of the elevation setting
of the reflective elements of the primary mirror and to large
deviations ``outliers''  of the zero points in a number of
reflective panels. These ``outliers'' in the elevation coordinate
had the same sign and were due to the wear of the moving
mechanisms of the reflective panels during observations.

A comparison of the power beam constructed using observations of
the sample consisting of NVSS sources with the experimental
power beam obtained from observations of 3С84 in the same
sections of the survey shows that the accuracy of the proposed
method is at least three times lower than the accuracy of
measuring the PB from observations of  3С84.
However, although the two methods have different accuracy of
measurement of the vertical PB, the estimated
turns of the experimental PB relative to the
computed one agree well with each other. Moreover,
the two methods yield virtually coincident average ratios of half-widths of
the experimental and computed PBs obtained from
the entire sample of sources ($K_{HPBW}=1.03\pm0.06$) and from
3С84 ($K_{HPBW}=1.02\pm0.10$). The latter  fact indicates that
 the error of the determination of the
fluxes of the observed sources is dominant when constructing
 the experimental PB using the method considered.

We studied the stability of the PB during the
zenith field survey using a sample of 80 NVSS sources observed in
the central band of the survey during the  1998, 2000, 2002, and
2003 sets. The parameters of the experimental PBs
consructed using this sample of sources proved to remain virtually
unchanged throughout the sets considered. Thus the half-width
$G_{0.5}$ of the Gaussian fitted to the experimental PB
and the parameter $\Delta_{max}$ that characterizes the
shift of the maximum of the experimental PB
relative to the central section do not change from one set to
another and are equal to the  $G_{0.5}$ and $\Delta_{max}$ values
for the Gaussian fitted to the computed PBs
($G_{0.5}=4.06'\pm0.10$ and $\Delta_{max}=-0.024'\pm0.005$).
Moreover, the standard errors of the deviations of the
experimental PBs from the computed ones
 proved to be very accurately the same for all
observational sets. The mean standard deviation averaged over all
sets is $\sigma=0.055\pm0.020$. The differences between the
$\sigma$ values for different sets are within the corresponding
measurement errors. We can thus conclude that the experimental
PB of RATAN-600 was sufficiently stable during the
RZF survey and agreed well with the computed PB
in its central part ($\pm6'$).

We use the data obtained when constructing the experimental
PB for the  1998--2003 sets to estimate the flux
measurement errors for the sources that were observed at 7.6 cm
during the zenith field survey and were included into the RZF
catalog [\cite{p1:Majorova_n}]. The standard error of the measured
fluxes for the entire sample of sources considered  varies in the
interval $(0.15\div0.27)\pm0.05$ depending on how bright is the
source and how far it is from the central section of the PB.
 The RZF catalog includes only objects located within
$-2' < dH <2'$ from the central section of the survey. The
estimate of the flux error ($\pm15\%\div\pm25\%$) obtained by
Bursov et al.~[\cite{p1:Majorova_n}] agrees with the results of
this study implying flux errors of $0.15 \div 0.24$ in the $-2' <
dH <2'$ interval.

The foregoing leads us to the following conclusion. Despite the
possibly large scatter of data points on the experimental
PBs constructed using the technique proposed in
this paper, these power beam patterns, if subjected to the
procedure of ``densification'', fully reproduce the features of
the PB measured using the more accurate
radio-astronomical method [\cite{ks:Majorova_n}]. Our technique can
be used not only to measure the power beams in the process
of the survey, but also to draw sufficiently reliable conclusions
concerning the state of the antenna of the radio telescope and
the measurement errors of the survey proper.

The data on the two-dimensional power beam pattern of RATAN-600
 are already used to interpret the
two-dimensional sky survey in the declination band
$\pm1^{\circ}$: $40^{o}30'42'' < Dec_{2000} < 42^{o}30'42''$,
$0^h\le R.A.< 24^h$ and for dee\-per ``cleaning'' of discrete radio
sources located far from the axis of the PB.

\begin{acknowledgements}
We are grateful to Yu.~N.~Parijskij for supporting the work and
useful comments.

This work made use of the CATS database developed by
O.~V.~Verkhodanov, S.~A.~Trushkin, H.~Andernach, and
V.~N.~Chernenkov.

This work was supported by the Russian Foundation for Basic
Research(grant no.~05-02-1751) and the Program of President of
the Russian Academy of Sciences for the Support of Leading
Scientific Schools (``The School of S.E.Khaikin'').

\end{acknowledgements}

%\columntovsizefalse

%\onecolumngrid
%\newpage

\end{document}